\documentclass[sigplan,10pt]{acmart}

\newif\ifdoubleblind
\doubleblindfalse     %

\def\enableComments{0}  %
\def\isReadyToSubmit{1} %

\usepackage{adjustbox} %
\usepackage{algorithm} %
\usepackage[noend]{algpseudocode} %
\usepackage{anyfontsize} %
\usepackage{array} %
\usepackage{bm} %
\usepackage{calc} %
\usepackage{courier} %
\usepackage{dsfont} %
\usepackage{enotez} %
\setenotez{backref=true} %
\usepackage{ifthen} %
\usepackage{listings} %
\usepackage{multirow} %
\usepackage{makecell} %
\usepackage{paralist} %
\usepackage{pbox} %
\usepackage{pythonhighlight} %
\usepackage{soul} %
\usepackage{subcaption} %
\usepackage{tabularx} %
\usepackage{tikz} %
\usetikzlibrary{positioning, arrows.meta, shapes, shadows, matrix}
\usepackage{flushend} %
\usepackage{verbatim} %
\usepackage{wrapfig} %
\usepackage{xspace} %
\usepackage[utf8]{inputenc}
\usepackage[T1]{fontenc} %

\newcommand{\ie}{{\em i.e.,~}}
\newcommand{\eg}{{\em e.g.,~}}

\makeatletter
\renewcommand{\ALG@name}{Alg.}
\makeatother

\def\Def{Def.~}
\def\F{Fig.~}
\def\T{Tab.~}
\def\Alg{Alg.~}
\def\Thm{Thm.~}
\def\Eq{Eq.~}
\def\Lem{Lem.~}
\def\Line{Line~}

\newcommand{\heading}[1]{{\vspace{1pt}\noindent\bf{#1}}} %

\makeatletter
\newcommand{\rigidsection}{\@startsection{section}{1}{\z@}%
  {-.75\baselineskip}%
  {.25\baselineskip}%
  {\ACM@NRadjust\@secfont}}
\newcommand{\rigidsubsection}{\@startsection{subsection}{2}{\z@}%
  {-.75\baselineskip}%
  {.25\baselineskip}%
  {\ACM@NRadjust\@subsecfont}}
\makeatother

\algblock{Input}{EndInput}
\algnotext{EndInput}
\algblock{Output}{EndOutput}
\algnotext{EndOutput}
\algblock{Config}{EndConfig}
\algnotext{EndConfig}

\newcommand{\graycomment}[1]{\textcolor{gray}{// \em #1}}
\algrenewcommand\algorithmicindent{0.5em} %

\newcommand{\code}[1]{\lstinline|#1|}

\newcommand{\circled}[1]{%
  \tikz[baseline=(char.base)] \node[draw, circle, inner sep=1pt](char) {#1};%
}
\newcommand{\step}[2][blue]{{\color{#1} \circled{#2}}}

\ifnum\enableComments=1
\newcommand{\asaf}[1]{\textcolor{orange}{\{Asaf: #1\}}}
\newcommand{\ml}[1]{\textcolor{purple}{\{Mathias: #1\}}}
\newcommand{\pierre}[1]{{\color{blue}{Pierre: #1}}}
\newcommand{\pierreinternal}[1]{{\color{blue}{Pierre: #1}}}
\newcommand{\pierrerm}[1]{{\color{blue}{\st{#1}}}}
\newcommand{\pierrerp}[2]{{\color{blue}{\st{#1}{#2}}}}
\newcommand{\ben}[1]{{\color{green}{Ben: #1}}}
\newcommand{\roxana}[1]{{\color{cyan}{Roxana: #1}}}
\newcommand{\giorgio}[1]{{\color{green!70!black}{Giorgio: #1}}}
\newcommand{\tosupport}[1]{{\uline{#1}}}
\newcommand{\todo}[1]{{\color{red} TODO(#1). }}
\newcommand{\todomsg}[2]{{\color{red} TODO(#1): #2}}
\newcommand{\mc}[1]{{\color{orange}{Mark: #1}}}
\newcommand{\mcrm}[1]{{\color{orange}{\st{#1}}}}
\newcommand{\mcrp}[2]{{\color{orange}{\st{#1}{#2}}}}
\newcommand{\ac}[1]{{\color{pink}{Alison: #1}}}

\else

\newcommand{\asaf}[1]{}
\newcommand{\ml}[1]{}
\newcommand{\pierre}[1]{}
\newcommand{\pierreinternal}[1]{}
\newcommand{\ben}[1]{}
\newcommand{\kelly}[1]{}
\newcommand{\roxana}[1]{}
\newcommand{\pierrerm}[1]{}
\newcommand{\pierrerp}[2]{}
\newcommand{\todo}[1]{}
\newcommand{\todomsg}[2]{}
\newcommand{\mc}[1]{}
\newcommand{\mcrm}[1]{}
\newcommand{\mcrp}[2]{}
\newcommand{\ac}[1]{}
\newcommand{\giorgio}[1]{}
\newcommand{\jason}[1]{}
\newcommand{\xiaotian}[1]{}
\newcommand{\jo}[1]{}
\newcommand{\tosupport}[1]{}
\renewcommand{\endnote}[1]{}
\fi

\newtheoremstyle{TheoremNum}
{\topsep}{\topsep}              %
{\itshape}                      %
{}                              %
{\bfseries}                     %
{.}                             %
{ }                             %
{\thmname{#1}\thmnote{ \bfseries #3}}%
\theoremstyle{TheoremNum}

\newtheorem{numberedthm}{Theorem}

{\makeatletter
 \gdef\xxxmark{%
   \expandafter\ifx\csname @mpargs\endcsname\relax
     \expandafter\ifx\csname @captype\endcsname\relax
       \marginpar{\textcolor{red}{xxx~}}%
     \else
       \textcolor{red}{xxx~}%
     \fi
   \else
     \textcolor{red}{xxx~}%
   \fi}
 \gdef\xxx{\protect\xxx@inner}
 \gdef\xxx@inner{\@ifnextchar[\xxx@lab\xxx@nolab}
 \long\gdef\xxx@lab[#1]#2{{\bf [\xxxmark \textcolor{red}{#2} ---{\sc #1}]}}
 \long\gdef\xxx@nolab#1{{\bf [\xxxmark \textcolor{red}{#1}]}}
 \ifnum\isReadyToSubmit=1
   \long\gdef\xxx@lab[#1]#2{}\long\gdef\xxx@nolab#1{}%
 \fi
}

\acmConference[SOSP’26]{The 32nd ACM Symposium on Operating Systems Principles}{September 29--October 2, 2026}{Prague, Czechia}
\newcommand{\ppa}{Attribution\xspace}
\newcommand{\ppas}{Attribution's\xspace}

\newcommand{\convsite}{conversion site\xspace}

\newcommand{\impsite}{impression site\xspace}
\newcommand{\impsites}{\impsite{}s\xspace}

\newcommand{\ex}[1]{{\em #1.ex}\xspace}

\newcommand{\persiteFilterVarName}{\textnormal{querier}\xspace}

\newcommand{\persiteFilterEps}{\epsilon_{\textnormal{querier}}}
\newcommand{\epsQuerier}{\persiteFilterEps}
\newcommand{\globalFilter}{\textnormal{global budget}\xspace}
\newcommand{\globalFilterVarName}{\textnormal{global}\xspace}

\newcommand{\globalFilterEps}{\epsilon_{\textnormal{global}}}

\newcommand{\epsGlobal}{\globalFilterEps}

\newcommand{\impQuota}{\textnormal{impression-site quota}\xspace}
\newcommand{\impQuotaVarName}{\textnormal{imp-quota}\xspace}
\newcommand{\impQuotaVar}{\code{imp-quota}\xspace}

\newcommand{\impQuotaEps}{\epsilon_{\textnormal{imp-quota}}}

\newcommand{\convQuotaVarName}{\textnormal{conv-quota}\xspace}
\newcommand{\convQuotaVar}{\code{conv-quota}\xspace}

\newcommand{\convQuotaEps}{\epsilon_{\textnormal{conv-quota}}}
\newcommand{\quotacount}{\textnormal{per-action domain cap}\xspace}
\newcommand{\quotacountVarName}{per-action domain cap\xspace}
\newcommand{\mathquotacountVar}{\kappa_{\textnormal{action}}}
\newcommand{\quotacountVar}{$\mathquotacountVar$\xspace}
\newcommand{\quotacountValue}{\textnormal{per-action domain cap}}

\newcommand{\enc}{\persiteFilterEps}
\renewcommand{\c}{\globalFilter}

\newcommand{\ec}{\globalFilterEps}

\newcommand{\qimp}{\impQuota}

\newcommand{\eqimp}{\impQuotaEps}

\newcommand{\eqconv}{\convQuotaEps}

\newcommand{\continue}{\textsf{CONTINUE}\xspace}

\newcommand{\true}{\textsf{TRUE}\xspace}
\newcommand{\false}{\textsf{FALSE}\xspace}
\newcommand{\pass}{\text{pass}}

\newcommand{\N}{\mathbb{N}}
\newcommand{\Z}{\mathbb{Z}}
\newcommand{\R}{\mathbb{R}}
\newcommand{\D}{\mathbb{D}}
\newcommand{\E}{\mathbb{E}}
\newcommand{\Ev}{\mathbb{F}}

\newcommand{\cA}{\mathcal{A}}
\newcommand{\cB}{\mathcal{B}}
\newcommand{\cC}{\mathcal{C}}
\newcommand{\cD}{\mathcal{D}}
\newcommand{\cE}{\mathcal{E}}
\newcommand{\cF}{\mathcal{F}}
\newcommand{\cG}{\mathcal{G}}

\newcommand{\cI}{\mathcal{I}}

\newcommand{\cK}{\mathcal{K}}
\newcommand{\cL}{\mathcal{L}}

\newcommand{\cN}{\mathcal{N}}
\newcommand{\cM}{q}
\newcommand{\cP}{\mathcal{P}}
\newcommand{\cQ}{\mathcal{Q}}
\newcommand{\cR}{\mathcal{R}}
\newcommand{\cS}{\mathcal{S}}
\newcommand{\cT}{\mathcal{T}}
\newcommand{\cU}{\mathcal{U}}

\newcommand{\cX}{\mathcal{X}}

\newcommand{\lap}{\operatorname{Lap}}

\newcommand{\bfb}{\mathbf{b}}

\newcommand{\bfF}{\mathbf{F}}

\newcommand{\chal}{\chi}

\newcommand{\tmax}{t_{\max}}
\newcommand{\emax}{e_{\max}}

\newcounter{evalClaimCounter}

\ifdoubleblind
  \newcommand{\sysname}{Oscar\xspace}
\else  %
  \newcommand{\sysname}{Big~Bird\xspace}
\fi  %
\newcommand{\pdslib}{pdslib\xspace}  %
\newcommand{\pdslibbf}{\textbf{pdslib}\xspace}   %

\title{\sysname: Resilient Privacy Budgeting Across Untrusted Web Domains}

\ifdoubleblind
  \author{Paper \#981}
  \affiliation{\country{}}
\else
  \author{Pierre Tholoniat}
  \authornote{Contributions made at Columbia University. P.T. is now at Google. A.C. and R.G. were also affiliated with Microsoft and Meta, respectively.}
  \affiliation{\institution{Columbia University}\country{USA}}

  \author{Alison Caulfield}
  \authornotemark[1]
  \affiliation{\institution{Columbia University}\country{USA}}

  \author{Giorgio Cavicchioli}
  \affiliation{\institution{Columbia University}\country{USA}}

  \author{Mark Chen}
  \affiliation{\institution{Columbia University}\country{USA}}

  \author{Benjamin Case}
  \affiliation{\institution{Meta Platforms, Inc.}\country{USA}}

  \author{Asaf Cidon}
  \affiliation{\institution{Columbia University}\country{USA}}

  \author{Roxana Geambasu}
  \authornotemark[1]
  \affiliation{\institution{Columbia University}\country{USA}}

  \author{Mathias L\'ecuyer}
  \affiliation{\institution{University of British Columbia}\country{Canada}}

  \author{Martin Thomson}
  \affiliation{\institution{Mozilla}\country{Australia}}
\fi  %

\usepackage{mdframed}
\AtBeginDocument{%
  \let\origtheorem\theorem
  \let\endorigtheorem\endtheorem
  \renewenvironment{theorem}
    {\begin{mdframed}[
        linewidth=0.4pt,
        roundcorner=1pt,
        skipabove=0pt,
        skipbelow=0pt,
        innerleftmargin=1pt,
        innerrightmargin=1pt,
        innertopmargin=0pt,
        innerbottommargin=0pt
      ]\begingroup\footnotesize\origtheorem}
    {\endorigtheorem\endgroup\end{mdframed}}%
}

\begin{document}

\acmYear{2026}\copyrightyear{2026}
\setcopyright{cc}
\setcctype[4.0]{by}
\acmConference[SOSP '26]{Symposium on Operating Systems Principles}{September 29--October 2, 2026}{Prague, Czech Republic}
\acmBooktitle{Symposium on Operating Systems Principles (SOSP '26), September 29--October 2, 2026, Prague, Czech Republic}
\acmDOI{10.1145/3830418.3843897}
\acmISBN{979-8-4007-2585-2/26/09}



\settopmatter{printfolios=false}

\begin{abstract}

    The W3C Attribution API is an emerging standard for privacy-preserving advertising measurement. Its current privacy architecture enforces individual differential privacy (IDP) independently for each domain (e.g., an advertiser) issuing queries. We show that this guarantee is unsound under realistic system behavior: it fails under cross-querier data adaptivity and can also fail when shared limits are enforced across queriers. The issue is not the on-device accounting model itself -- device-epoch IDP -- but treating each querying domain in isolation.
    We propose \sysname, a privacy-budget manager that makes global device-epoch IDP -- enforced jointly across all domains -- both sound and deployable for Attribution. \sysname addresses the main obstacle to global enforcement in open multi-querier systems: denial-of-service depletion of a shared global budget by Sybil web domains. Its key insight is that benign Attribution workloads have a stock-and-flow structure: impressions create potential privacy loss, conversions realize it, and meaningful budget consumption should be tied to genuine user actions across distinct web domains. \sysname enforces this structure with privacy-loss-based quotas on impression and conversion sites and a per-user-action cap on how many quotas can be activated, ensuring that adversarial impact scales with genuine user interactions rather than with the number of Sybil domains.
    We implement \sysname in Rust, integrate it into Firefox's Attribution prototype, and evaluate it theoretically and empirically on real ad-tech data. We show that \sysname provides rigorous global device-epoch IDP, formal resilience to depletion attacks, and utility for benign queriers under attack.

\end{abstract}

\makeatletter
\let\countrywithdisplay\country
\renewcommand{\country}[1]{\global\@ACM@countrypresenttrue}
\makeatother
\maketitle
\renewcommand{\shortauthors}{Tholoniat et al.}
\let\country\countrywithdisplay
\pagestyle{empty}  %

\vspace{-0.2cm}
\section{Introduction}
\label{sec:introduction}

An important new standard is taking shape at the W3C: the Attribution API, designed to enable privacy-preserving advertising measurement at web scale~\cite{ppa, TKM+24}. Advertising measurement is a core economic function of the web, yet today it relies heavily on cross-site tracking mechanisms such as third-party cookies and fingerprinting, which undermine user privacy and have proven difficult to regulate or replace. Attribution aims to provide a principled alternative: a browser-mediated interface for cross-site measurement that replaces raw data collection with structured, differentially private aggregation.

At a high level, Attribution enables tasks such as conversion attribution measurement -- linking ads viewed on content sites to purchases on seller sites -- while ensuring that only noisy aggregate statistics are released. Browsers mediate all raw-data flows, store user activity locally, and release encrypted reports that are aggregated with differential privacy (DP), eliminating the need for cross-site tracking or centralized user-level data collection.

A key feature of Attribution is device-side privacy accounting based on individual differential privacy (IDP), which we developed for Attribution in prior work, Cookie Monster~\cite{TKM+24}. Each browser enforces a per-epoch privacy budget for each querier (e.g., an advertiser), deducting privacy loss before releasing reports. This design offers strong utility by charging privacy loss only when a user meaningfully contributes to a query, and it has been adopted as the foundation of the emerging standard -- including its per-querier enforcement scope. The standard is being developed with participation from all major browsers; as of August 2026 it is in Wide-Area Review, en route to Candidate Recommendation.

Given its role as a {\em web standard} -- not merely an implementation, and likely to be deployed across billions of devices -- scrutinizing Attribution's privacy guarantees is essential and urgent. We therefore ask: are its guarantees sound, and can they be enforced robustly in practice?

{\em Contribution 1:} We show that Attribution's central privacy guarantee -- device-epoch IDP enforced per querier -- is fundamentally unsound under realistic system behavior. The issue is not IDP or the device-epoch unit, but per-querier enforcement. In multi-querier systems, privacy guarantees must account for interactions across queriers. We identify two failure modes: (1) cross-querier data adaptivity in streaming settings causes privacy loss to exceed per-querier bounds, even without collusion; and (2) introducing shared limits across queriers, a common defense against collusion, also violates per-querier guarantees. We prove these results in a general DP system model, showing that the problem is not specific to Attribution but reflects a fundamental mismatch between per-querier enforcement and deployment realities.

Focusing on Attribution, we therefore advocate global device-epoch IDP -- enforced across all queriers -- as the sound privacy foundation. However, global enforcement introduces a key systems challenge. In open environments where any domain can invoke the API, a shared global privacy budget becomes vulnerable to denial-of-service depletion: adversarial domains can exhaust users' budgets, preventing benign measurement.

    {\em Contribution 2:} We present \sysname, a global privacy-budget manager for Attribution that provides both rigorous global IDP enforcement and resilience to depletion attacks by untrusted web domains. \sysname is based on a key structural insight: legitimate Attribution workloads exhibit a stock-and-flow pattern of privacy consumption, where impressions create potential privacy loss (stock) and conversions realize it (flow), tied to genuine user actions across sites. Attacks violate this structure by generating stock and flow without corresponding user activity.

\sysname restores this structure using coordinated, IDP-aligned quotas on impression sites, conversion sites, and per-user-action domain participation. These constraints ensure that privacy-budget consumption -- whether benign or adversarial -- scales with genuine user interactions rather than with the number of attacker-controlled (Sybil) domains.

We implement \sysname in Rust as a generic IDP budget management library, {\em \pdslib}, integrate it into Firefox's Attribution prototype, analyze it theoretically, and evaluate it empirically on real ad-tech data. Our results show that \sysname enforces rigorous global device-epoch IDP, achieves formal resilience to depletion attacks, and preserves utility for benign workloads under attack.

Both contributions are now part of the Attribution draft.\footnote{B.C. and M.T. are specification editors \cite{ppa}. All changes described here were proposed, discussed, and adopted through the W3C public process.}
Following the recommendation in \S\ref{sec:analysis-positive-result}, the specification added a section recording our results as limitations of per-querier semantics and discussing their implications~\cite{ppa-spec-formal-analysis}.
The draft also adopted the Big Bird algorithm from \S\ref{sec:contribution2} (Alg. 2) as its budget-management mechanism~\cite{ppa-spec-budget-store}.
The latter resolved an open issue~\cite{charlie-issue}---DoS depletion of the global safety limits---that the editors had flagged as a blocker to the standard's reaching the Wide-Area Review stage, which it has since reached. We also release our source code as a reference implementation for the standard as well as for broader use cases: \pdslib at \url{https://github.com/columbia/pdslib}, Firefox integration at \url{https://github.com/columbia/pdslib-firefox}, and experimental infrastructure at \url{https://github.com/columbia/bigbird}.
Indeed, while we focus on Attribution, our results have broader implications for DP systems in untrusted querier settings: (1) multi-granularity enforcement is fundamentally unsound, and (2) global budget management requires depletion defenses aligned with the structure of benign workloads. We view \sysname as one step toward this broader design space.
\section{Background}
\label{sec:background}

\begin{figure}[t]
  \centering
  \includegraphics[width=.95\linewidth]{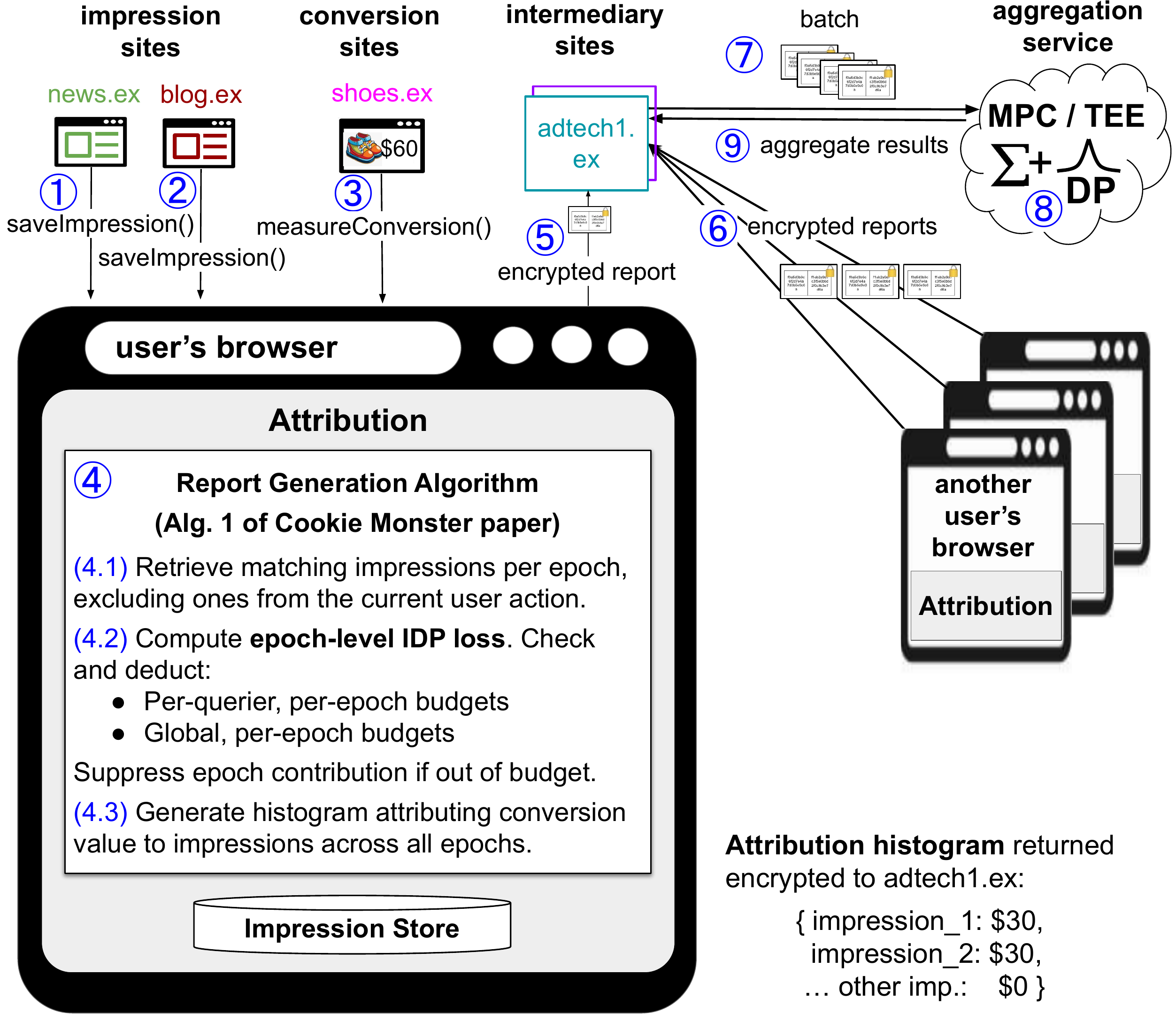}
  \caption{{\bf \ppa API architecture.}}
  \label{fig:ppa-overview}
\end{figure}

\subsection{\ppa API Architecture}
\label{sec:ppa-architecture}

W3C \ppa is a privacy-preserving, browser-based interface for cross-site conversion measurement. Figure~\ref{fig:ppa-overview} shows the workflow, involving four principals, where \mbox{\code{.ex}} denotes fictional example domains:
(1) \emph{impression sites} (e.g., \ex{news}, \ex{blog}) register shown ads via \code{saveImpression()};
(2) \emph{conversion sites} (e.g., \ex{shoes}) call \code{measureConversion()} to request encrypted reports of which ads a user saw;
(3) \emph{intermediaries} (e.g., \ex{adtech}) receive encrypted reports; and
(4) \emph{aggregation services} (e.g., \code{divviup.org}) aggregate reports and add DP noise using trusted MPC or TEE.

The browser mediates all information flows: impressions are stored locally, conversions trigger encrypted reports, and aggregators release only DP-protected statistics. A typical sequence is:
\step{1}--\step{2} impression sites record ads;
\step{3} a conversion site calls \code{measureConversion()};
\step{4} the browser fetches relevant impressions, checks local privacy budgets, and
\step{5} generates encrypted reports;
\step{6}--\step{9} reports are batched, aggregated, and released with DP noise via an aggregation service.

Critically, \emph{privacy enforcement happens on-device}. Before releasing a report, the browser deducts the corresponding privacy loss from a local \emph{privacy budget store}, returning null reports when budget is exhausted \textcolor{blue}{(4.2)}. This enforcement uses an algorithm presented in the Cookie Monster paper~\cite{TKM+24}.

\vspace{-0.2cm}
\subsection{Cookie Monster Accounting}
\label{sec:cookie-monster}

Our prior work, Cookie Monster~\cite{TKM+24}, proposed using \emph{individual differential privacy (IDP)}~\cite{ESS15,JYC15,FZ21} for efficient, on-device privacy accounting in Attribution, and the accounting it introduced is the one the standard adopted. We summarize it here because both of this paper's contributions build on it: \S\ref{sec:contribution1} revisits the scope of its guarantee, and \S\ref{sec:contribution2} builds a budget manager on top of its accounting.

Under IDP, each browser maintains its own privacy budgets and deducts privacy loss based on the user's actual data contribution (the \emph{individual sensitivity}), rather than a population-wide worst-case bound (the \emph{sensitivity}) as in traditional DP.
This distinction enables a crucial optimization for \ppa. Advertising signals are sparse: many users see ads but never convert; many conversions have no matching impression; and users purchase different products and therefore participate in different queries. Cookie Monster's accounting exploits this sparsity. Each device partitions its data into non-overlapping epochs (e.g., one week), and when a conversion occurs, the browser charges privacy loss \emph{only for epochs containing at least one impression relevant to that conversion}. If an epoch contains no such impression, the conversion incurs \emph{zero} privacy loss for that device-epoch. When a charge is applied, it is bounded by \(\text{(value / maxValue)} \times \epsilon\), using parameters from the \code{measureConversion()} call.

This contrasts with traditional DP, where privacy loss is charged for every query based on population-level sensitivity, even if a device contributed no relevant data. For Attribution, avoiding charges in irrelevant epochs is essential for utility: it reduces privacy loss for typical users and allows many measurements before budgets are exhausted.

\subsection{Queriers and Privacy Guarantees}
\label{sec:ppa-queriers-privacy-guarantees}

Attribution performs privacy accounting \emph{with respect to a querier} -- the entity on whose behalf aggregation occurs. In today's API, the querier is the \emph{advertiser} measuring its own conversions. A planned extension will add \emph{intermediaries} as queriers for cross-advertiser queries (e.g., for training ad-placement models). Each browser therefore maintains \emph{per-epoch budgets per querier}. We showed in Cookie Monster that, {\em when queriers are modeled in isolation}, such per-querier budgets satisfy device-epoch-level IDP~\cite{TKM+24}. %

The \ppa specification acknowledges a well-known limitation: if queriers collude, their privacy losses add. To mitigate this, the API introduces \emph{global safety limits} -- per-epoch budgets spanning all queriers. In the current design, per-querier budgets remain the primary enforcement mechanism, with global limits viewed as defense-in-depth.
\rigidsection{Contribution 1: Per-Querier DP Analysis}
\label{sec:contribution1}

Attribution's focus on per-querier enforcement, with global limits as defense-in-depth, is not unique. Several DP systems have similarly proposed per-querier or per-analyst enforcement~\cite{GHK+18,GHH+25,KVLH25,TKM+24,DProvDB} to align guarantees with organizational boundaries. Such guarantees are intuitively appealing, especially when queriers can reasonably be assumed to act independently and not collude (e.g., competing organizations or scientists under IRB restrictions). Some of these systems also, like Attribution, enforce DP at multiple granularities for defense-in-depth~\cite{GHK+18,KVLH25,DProvDB}.

Our analysis of per-querier guarantees -- developed initially for Attribution but generalized here to broader DP systems -- shows that their limitations extend beyond collusion.
Any DP analysis has three components:
(1) a formal privacy property (definition, privacy unit, enforcement scope),
(2) a model of system behavior specifying how data is generated and queried, and
(3) a proof that the modeled system satisfies the property.
If the behavioral model omits deployment-relevant behaviors that affect data generation or querying, the proof may still be correct \emph{relative to the model}, while the stated privacy property fails to capture the system's actual semantics.

We identify two recurring modeling gaps in analyses of per-querier DP. First, existing models typically treat each querier in isolation, neglecting cross-querier influences on data generation that naturally arise in streaming systems. Second, when per-querier budgets and shared limits coexist, prior work has generally modeled them separately, overlooking how their joint enforcement shapes system behavior.

When these omitted behaviors are modeled, per-querier guarantees become unsound: even without collusion, a single querier's view can reveal more about the dataset than the per-querier budget permits. This does not render per-querier budgets useless -- they remain valuable engineering constraints -- but it does mean their formal guarantees rest on stronger assumptions than previously recognized. More broadly, these results reinforce a key takeaway: rigorous privacy analysis relies as much on \emph{faithful system modeling} as on the mathematics of the DP proof itself.

\vspace{-0.2cm}
\subsection{Multi-Querier System Model}
\label{sec:contribution1:formal-model}

We define a generic model for a multi-querier DP system, not specific to \ppa. \Alg\ref{alg:adaptivity_model_simplified} shows a system with $K$ queriers ($Q^1, \dots, Q^K$), each with its own privacy budget $\cF^k$, interacting with a shared database $D$ over discrete time steps $t$. The model accommodates both streams and static databases: a data generation process $\cG$ may append new data to $D$ at each step, while in the static case $\cG$ outputs nothing after the initial step. It supports both per-querier enforcement, via $\cF^k$, and an abstract shared limit $\cL$, which can represent a global budget. At each step, querier $k$ selects a query $Q^k_t$ based only on its \emph{own} past results, thereby explicitly {\em excluding intentional collusion}. The query executes only if both $\cF^k$ and $\cL$ approve, in which case the querier receives result $V_t^k$; otherwise it receives a data-independent result $\bot$.

To analyze DP, we introduce a {\em challenge bit} $\chal$ that governs whether a particular record $x_0$ is added to the database at time $t_0$. Neither the queriers nor $\cG$ see $\chal$, but as queriers issue queries, their views may reveal information about it, modeling the leakage of an individual's presence through DP query results. Privacy loss quantifies how hard it is for an adversary to distinguish between $\chal=0$ and $\chal=1$, whether the adversary controls a single querier (per-querier guarantees) or all queriers (global guarantees).

Formally, \Alg~\ref{alg:adaptivity_model_simplified} satisfies $\epsilon$-DP for querier $k \in [K]$ if for $\forall x_0 \in \cX, \forall t_0 \in [\tmax], \forall \cG, \forall (\cQ^1, \dots, \cQ^{n+1}),\forall S \subseteq \cR^{t_{\max}}$:
{\footnotesize
\[
    \left| \ln \left( \frac{\Pr[V^{k,(\chal=0)} \in S]}{\Pr[V^{k,(\chal=1)} \in S]} \right) \right| \bm{\le} \epsilon.
    \vspace{-0.2cm}
\]
}
By contrast, a global guarantee would bound $\frac{\Pr[V^{(\chal=0)} \in S]}{\Pr[V^{(\chal=1)} \in S]}$, if for example $\cL$ were implemented as a global budget.

We next analyze per-querier guarantees and show two orthogonal cases in which they degrade linearly with the number of queriers. These correspond to conditions \textbf{(A)} and \textbf{(B)} in \Alg~\ref{alg:adaptivity_model_simplified}: adaptive data generation in streaming settings, and the presence of a shared limit in addition to per-querier budgets. By toggling these conditions on and off, we isolate the effect of each on per-querier guarantees. Appendix~\ref{sec:appendix:extension_idp} extends these results to IDP budgets, as used in \ppa.

\algrenewcommand\algorithmicfunction{def}
\algrenewcommand\textproc{}
\begin{algorithm}[t]
    \footnotesize
    \caption{\footnotesize \bf Formal model of a generic multi-querier, pure-DP system}
    \label{alg:adaptivity_model_simplified}
    \begin{algorithmic}[1] %
        \Input
        \State Challenge bit $\chal \in \{0,1\}$
        \State In-or-out record $x_0$ to be inserted at time $t_0$
        \State Data generation process $\cG$ \label{line:data-generation}
        \State Queriers $\cQ^1, \dots, \cQ^K$
        \State Per-querier budgets (modeled as pure-DP filters) $\cF^1, \dots, \cF^K$
        \State Shared limit $\cL$
        \EndInput

        \For{$t \in [t_{\max}]$}
        \Statex \ \ \graycomment{Data generation model:}
        \Statex \ \ \textcolor{red!75!black}{\textbf{// (A) Cross-querier data adaptivity}}
        \State $D_t \gets \cG(\textcolor{red!75!black}{V^1_{<t}, \dots V^K_{<t}})$ \ \graycomment{\footnotesize new data depends on {\bf all} past results} \label{line:data-adaptivity}
        \If{$t = t_0$ and $\chal = 1$} \ \graycomment{\footnotesize insert in-or-out record on challenge}
        \State $D_t \gets D_t + x_0$
        \EndIf

        \Statex \ \ \graycomment{Query execution model:}
        \For{$k \in [K]$}
        \State $\cM_t^k \gets \cQ^k(V^k_{<t})$ \ \graycomment{\footnotesize $Q^k$ selects query based on {\bf its own} past results}
        \Statex \ \ \ \ \textcolor{red!75!black}{\textbf{// (B) Shared-limit enforcement}}
        \If{$\cF^k(\cM_1^k, \cdots, \cM_t^k) \textcolor{red!75!black}{ \wedge \cL(\cM_1^1, \cdots, \cM_1^K, \cdots \cM_t^k)}$} \label{line:shared-limit}
        \State $V^k_{t} \gets \cM_t^k(D_{\le t})$
        \Else
        \State $V^k_t \gets \bot$ \label{line:v_bot}
        \EndIf
        \EndFor
        \EndFor

        \State \Return $V^1, \dots, V^K$
    \end{algorithmic}
\end{algorithm}

\vspace{-0.2cm}
\subsection{Failure Mode 1: Adaptive Data Generation}
\label{sec:analysis-per-querier-data-adaptivity}

We first consider the case where data is generated over time with no shared limits ($\cL$ always outputs \true). Adaptivity in data generation -- where new data depends on past DP query results -- has been studied in both streaming~\cite{GS13, DMR+22, JRSS23} and interactive~\cite{sage, dpf, TKM+24, GHH+25} systems. Modeling such adaptivity is essential because it both directly affects DP analyses, as prior work shows, and is inevitable in streaming systems. For instance, a movie recommender may use prior DP measurements to suggest a film, and the user's decision to watch it then becomes new data for future measurement.

Line~9 of \Alg~\ref{alg:adaptivity_model_simplified} captures data adaptivity for multi-querier systems: $D_t \gets \cG(V_{<t}^1, \dots, V_{<t}^K)$, meaning data can depend not only on one querier's past results but also on others'. This cross-querier dependency arises even without collusion, through user behavior or mechanisms like auctions. For example, \code{news.ex} may show an impression for either \code{shoes.ex} or \code{hats.ex}, depending on bids shaped by prior DP results from both sites.

We prove that under this model, per-querier budgets (formalized as privacy filters~\cite{filters_rogers}) with capacity $\epsilon$ do not suffice:

\begin{theorem}[Per-querier guarantees are unsound under adaptive data generation]
    \label{thm:data-adaptivity}

    Consider \Alg\ref{alg:adaptivity_model_simplified}, with $K = n+1$ queriers,
    where each $\cF^k$ is a pure-DP filter with capacity $\epsilon > 0$,
    and where $\cL$ always outputs \true (\ie no shared limit).
    Denote by $V^{n+1,(\chal)}$ the view of $\cQ^{n+1}$ on challenge bit $\chal$.
    Then, there exists $\epsilon' > 0$ such that $\exists x_0 \in \cX, \exists t_0 \in [\tmax], \exists \cG, \exists \cQ^1, \dots, \cQ^{n+1},
        \exists S \subseteq \cR^{t_{\max}}$:
    \begin{align*}
        \left| \ln \left( \frac{\Pr[V^{n+1,(\chal=0)} \in S]}{\Pr[V^{n+1,(\chal=1)} \in S]} \right) \right| \bm{\ge} \epsilon + n\epsilon'.
    \end{align*}
\end{theorem}

\Thm\ref{thm:data-adaptivity} shows that with per-querier budgets of capacity $\epsilon$, privacy loss can exceed $\epsilon$, with {\em excess loss growing linearly in the number of queriers}. This arises not from collusion but from worst-case analysis that DP must account for. The proof, given in Appendix~\ref{sec:appendix:analysis-per-querier-data-adaptivity}, works by constructing an example of data generation and querying processes that exhibits that much excess loss. While the theorem addresses only pure-DP guarantees (yielding a linear increase in excess loss), the result likely extends to approximate DP (with excess loss growing with the square root of the number of queriers).

Prior analyses overlooked this point. ARA's implementation supports multiple adtechs~\cite{Goo22}, but its proof models only a single querier~\cite{GHH+25}, effectively defining data generation as $\cG(V^k_{<t})$ for a fixed $k$ rather than our more general $\cG(V_{<t}^1, \dots, V_{<t}^K)$. %
Cookie Monster, by contrast, allowed multiple queriers but modeled a static dataset in its proof~\cite{TKM+24}.

\vspace{-0.2cm}
\subsection{Failure Mode 2: Shared Limits}
\label{sec:analysis-per-querier-shared-limits}

We next consider the case where a shared limit is added to a system with per-querier enforcement (line~\ref{line:shared-limit} in \Alg\ref{alg:adaptivity_model_simplified}). Such limits have been proposed in both static~\cite{GHK+18, DProvDB} and streaming~\cite{KVLH25} DP systems as defense-in-depth against collusion. We instantiate $\cL$ as a pure-DP filter shared across all queriers, which lets us quantify the excess privacy loss. The same reasoning, however, applies to other shared-limit mechanisms, such as caps on the total number of queries or rate limiters or quotas spanning subsets of queriers.

We show that even without data adaptivity, per-querier privacy loss can exceed $\epsilon$ in the presence of a shared budget. To isolate this effect from \Thm\ref{thm:data-adaptivity}, we restrict attention to static databases ($\cG$ returns $\emptyset$ after the first time step). In this setting, the shared budget acts as a cross-querier leakage channel, enabling excess loss that, as before, grows linearly with the number of queriers covered by the shared limit:

\begin{theorem}[Per-querier guarantees are unsound under adaptive query budgets with shared limits]
    \label{thm:shared-limit}

    Take $n + 1$ queriers, with per-querier filters with capacity $\epsilon$, and a shared filter with capacity $\epsilon_g < (n + 1)\epsilon$.
    There exists $\epsilon' >0$ such that
    for all $n \ge \frac{4\ln2}{(1-e^{-\epsilon/2})^2}$,
    there exists $\cG$ that returns $\emptyset$ after the first time step,
    such that $\exists x_0 \in \cX, \exists t_0  \in [\tmax], \exists \cQ^1, \dots, \cQ^{n+1}, \exists S \subseteq \cR^{t_{\max}}$:
    \begin{align*}
        \left| \ln \left( \frac{\Pr[V^{n+1, (\chal=0)} \in S]}{\Pr[V^{n+1,(\chal=1)} \in S]} \right) \right| \bm{\ge} \epsilon + n\epsilon'.
    \end{align*}
\end{theorem}
The proof constructs an attack that induces this excess loss. While it may appear to rely on collusion, the same leakage can arise inadvertently. At a high level, multiple queriers spend part of their budgets on an initial measurement and issue a follow-up query only when the result is high (e.g., measuring ad exposure and conditionally computing a histogram). The shared filter then encodes this signal: it is more likely to be exhausted when the true value is high, since more queriers issue follow-up queries. A new querier can thus infer this information from whether its own query is accepted or rejected, exceeding the privacy bound of its local budget. For IDP, rejection is not directly observable, requiring a more subtle argument (see \S\ref{sec:appendix:analysis-per-querier-shared-limits} for DP and IDP proofs).

This effect has been missed in prior work on multi-granularity enforcement, either because it was not formalized at all~\cite{GHK+18} or because per-querier and global limits were analyzed in isolation or under non-adaptive queriers~\cite{KVLH25, DProvDB}. Conceptually, this is like treating the two conditions in $\cF^k(\cM_1^k, \cdots, \cM_t^k) \wedge \cL(\cM_1^1, \dots, \cM_1^K, \dots, \cM_t^k)$ as if they acted independently, when in real systems they are necessarily enforced jointly.

\vspace{-0.2cm}
\subsection{Siloing Assumption and Takeaways}
\label{sec:analysis-positive-result}

It is natural to ask whether any conditions -- realistic or not -- could make per-querier enforcement sound. Appendix~\ref{sec:appendix:siloing} presents one such condition (Thm.~\ref{thm:siloing-assumption}), the \emph{siloing assumption}: each querier operates on a completely disjoint data stream, and shared limits are disabled or so loose that they never block queries. Under these constraints, per-querier DP is sound in our general model (\Alg\ref{alg:adaptivity_model_simplified}).

These constraints are almost certainly unattainable in practice. Worse, attempting to engineer them -- for example by weakening or removing shared limits -- undermines the goal of strong user privacy. We therefore strongly recommend {\em against} doing so.

    {\bf Instead, the takeaways are three-fold.}
First, \ppa should explicitly acknowledge the semantic limitations of per-querier enforcement.
Second, per-querier budgets should be complemented with global enforcement; our second contribution -- and the remainder of this paper -- shows how to do so robustly and efficiently. Attribution has since adopted both: the specification now records the semantic limitations of per-querier enforcement~\cite{ppa-spec-formal-analysis},
and specifies global enforcement managed by the algorithm of \S\ref{sec:contribution2}~\cite{ppa-spec-budget-store}.
Third, a broader lesson for the research community: rigorous privacy analysis depends as much on \emph{faithful system modeling} as on DP mathematics. While this is well-understood in formal methods, the DP literature often relies on oversimplified models to keep proofs simple. Our results highlight why modeling fidelity must become a first-class concern in DP systems research.
\rigidsection{Contribution 2: Global IDP with \sysname}
\label{sec:contribution2}

Global enforcement escapes the failures of \S\ref{sec:contribution1} because both are failures of scope, not of DP accounting. Thm. 3.1 and Thm. 3.2 exhibit leakage that a per-querier budget cannot see: information routed through the data stream or through the shared limit, and thus outside the budget meant to bound it. A global budget has no analogous outside channel---it accounts for the joint information released across all queriers, so cross-querier channels are inside the quantity being bounded, not a way around it. This section therefore claims a global guarantee only, and shows that sound global device-epoch IDP is achievable under a system model we make as comprehensive as we can (Thm. 4.2). Per-querier budgets remain in the design as engineering constraints protecting utility, not privacy; the caveats of §3 apply to them here as anywhere.

Global enforcement does, however, introduce a new systems challenge: in open, multi-querier settings with untrusted participants, a shared global budget becomes vulnerable to {\em depletion attacks}---a denial-of-service (DoS) on utility rather than privacy.
In Attribution, {\em any} web domain can invoke the API, allowing an attacker controlling many Sybil domains to act as distinct queriers, exhaust users' budgets, and block benign measurement. Although Attribution introduces global safety limits for privacy, it leaves their management unspecified and identifies DoS depletion as a key roadblock to deployment~\cite{charlie-issue}.

We present {\em \sysname}, a global budget manager for Attribution that makes device-epoch IDP both sound and deployable. Its key idea is to align budget consumption with the structure of legitimate workloads: impressions create potential privacy loss, conversions realize it, and meaningful consumption should scale with genuine user actions across distinct sites. \sysname enforces this via IDP-aligned quotas on impression and conversion sites and a per-user-action cap on domain participation, while preserving device-epoch IDP.

\vspace{-0.2cm}
\subsection{Threat Model}  %
\label{sec:threat-model}

\begin{figure}[t]
  \centering
  \begin{subfigure}{0.48\linewidth}
    \centering
    \includegraphics[width=\linewidth]{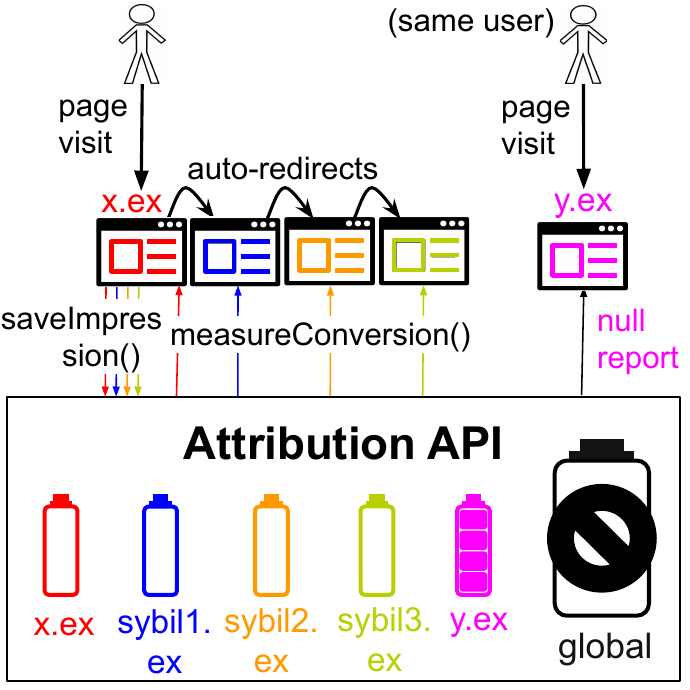}
    \caption{Sybils as conversion sites}
    \label{fig:example-attack-conversion-sites}
    \vspace{0.1cm}
  \end{subfigure}
  \hfill
  \begin{subfigure}{0.48\linewidth}
    \centering
    \includegraphics[width=\linewidth]{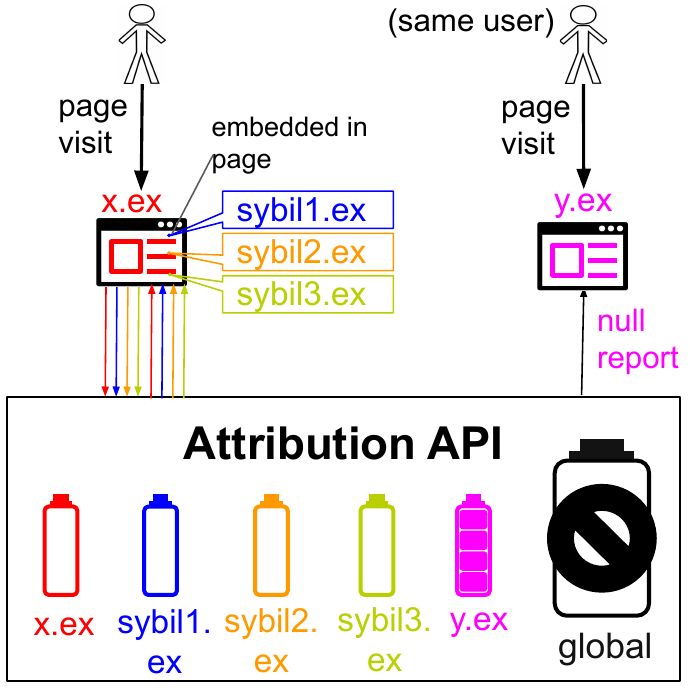}
    \caption{Sybils as intermediaries}
    \label{fig:example-attack-intermediaries}
    \vspace{0.1cm}
  \end{subfigure}
  \caption{{\bf Example attacks.} Sybils deplete the global budget from a single user action by acting as (a) conversion sites or (b) intermediaries. Per-querier and global budgets are shown as batteries (filled = available, empty = drained).}
  \label{fig:example-attacks}
\end{figure}

Global budgets are a {\em shared resource}, and on the web, attacker identities are weak (Attribution uses eTLD+1 domain names) and browser API use is generally open. An attacker can exploit this by registering many {\em Sybil domain names} and coordinating them to deplete a device's global budget. If this occurs on many devices, benign queriers are locked out of measurement, yielding a DoS based on the privacy layer.

\heading{Example attacks.}
\F\ref{fig:example-attacks} shows two example DoS depletion attacks effective against current \ppa.
In \F\ref{fig:example-attack-conversion-sites}, a malicious site \ex{x} uses aggressive redirection: when a user visits \ex{x}, it registers impressions under many Sybil domains posing as conversion sites, then auto-redirects the user through them. Each Sybil calls \code{measureConversion()}, consuming global budget. Since Attribution does per-querier checks, it can spend up to $\epsilon_{\text{querier}}$; with $\epsilon_{\text{global}}/\epsilon_{\text{querier}}$ Sybils, \ex{x} can exhaust a user's global budget from a single visit. When the user later visits a legitimate merchant (e.g., \ex{y}), that merchant receives null reports -- both useless and biased. If \ex{x} attracts many users even once, it can disable \ppa for benign queriers across the epoch.

Attacks need not be so overt. An attacker can instead embed many Sybil intermediaries (\F\ref{fig:example-attack-intermediaries}), each requesting reports and drawing from distinct per-querier budgets. More subtly, it can pace Sybil activity over time or across repeated visits to gradually drain users' budgets, making the attack harder to detect.

\heading{Generic attack model.}
The preceding examples illustrate specific tactics that an attacker can use to deplete global budgets in \ppa. Our objective, however, is to build defenses at the budget-management layer that withstand a broader class of attackers, regardless of how tactics evolve. We thus model a generic attacker that controls a set of domains visited by genuine users and aims to degrade the accuracy of benign queriers' measurements on genuine user data by exhausting the global budget on users' devices. We characterize the {\em attacker's strength} along three dimensions:
(i) how many domains it controls;
(ii) how many intentional user actions it can induce across those domains; and
(iii) how many distinct genuine users perform these actions.

\heading{Resilience goal.}
The attacker's impact in the \F\ref{fig:example-attacks} examples is {\em disproportionate to its strength}: the attacker is weak along one or more dimensions yet still depletes a device's global budget. Our goal is to {\em restore proportionality} along all three dimensions of strength: depletion should scale not with Sybil count alone, but with genuine user actions and with the number of users it is able to attract, so that an attacker must be strong along all three dimensions to deplete substantially.
We further require that this holds regardless of how the attacker behaves or what it knows, without assuming that benign traffic looks any particular way.

\heading{Assumptions.}
We exclude two classes of threats from our model.
First, we do not address bot-driven {\em poisoning} attacks in which adversaries fabricate impressions or conversions (including null reports) to skew aggregates. These attacks degrade data quality rather than deplete global budgets on genuine devices, and defending against them requires mechanisms orthogonal to global-budget management, such as recently proposed fraud prevention for \ppa~\cite{CATT25}.
Our focus is on preserving measurement availability for genuine-user data -- a prerequisite for meaningful measurement even without the poisoning threat.

Second, we protect only benign sites that do not directly or transitively partner with malicious ones (e.g., by embedding attacker-controlled intermediaries or serving their ads). \sysname operates purely at the global-budget layer and has no visibility into partner relationships or ad-network governance; attacks that exploit such partnerships require ecosystem-level controls (registration, contracts), not budget logic. Our aim is to confine adversarial impact to sites that interact with Sybils, while unrelated benign sites retain utility.
\vspace{-0.2cm}
\subsection{Candidate Approaches}
\label{sec:candidate-approaches}

The research challenge is to add DoS depletion resilience without overconstraining benign workloads, violating device-epoch IDP, or undermining its optimizations. Na\"ive approaches fail on at least one of these fronts.

\heading{Rate-limit API calls.}
One could cap how many impressions a site may log or how many reports it may request per time unit. However, API calls are a poor proxy for privacy loss. Under IDP, consumption depends on a user's actual data contribution, and many events incur zero loss: most impressions never lead to conversions, and many conversions have no attributable impression. These would still count against an API-call limit, disabling IDP's sparsity benefits. Worse, sites cannot know in advance which events will incur loss, so such caps force conservative throttling of legitimate activity and degrade utility. Any fixed limit therefore either suppresses benign workloads or must be set so loosely that attackers operate beneath it.

\heading{Rate-limit total budget consumption.}
A more aligned approach is to cap budget consumption itself, e.g., by limiting how much global budget can be spent over short time windows. This blocks bursty attacks like those in \F\ref{fig:example-attacks}, but fails against paced strategies that spread consumption over time. For example, a site can exploit repeated or prolonged user presence (e.g., an open tab) to gradually drain budget via redirects or intermediary requests. Extending the time window to catch such attacks risks making the limiter itself a DoS target. Moreover, setting the cap below $\epsilon_{\text{querier}}$ harms benign workloads, especially small advertisers that may legitimately consume their budget in bursts.

\heading{Cap total budget consumption per user action.}
A key weakness in \F\ref{fig:example-attacks} is that a single user action can drain a user's entire global budget. A natural fix is to cap budget per user action: each action initializes a budget $\epsilon_{\text{action}}$, shared across all queriers triggered by that action. This ties consumption to a meaningful unit, but still allows a single high-engagement site to deplete budget over many interactions with the same user. Among candidate approaches, this is closest to \sysname.

\heading{Interactions of caps with device-epoch IDP.}
We find it easy to define caps that violate device-epoch IDP by coupling decisions across epochs. For example, \S\ref{appendix:cap-interactions-with-idp} presents a natural but IDP-violating variant of the preceding approach: capping total budget consumption across all epochs following a user action. Any defense must therefore ensure that both checks and consequences are confined to a single epoch; in this paper, we show how to do so.
\vspace{-0.2cm}
\subsection{\sysname Design}
\label{sec:big-bird-overview}

Based on the limitations of alternative approaches, we establish four principles for \sysname's design.
    {\bf P1:} Limits should track \emph{IDP budget consumption}, not proxies such as API-call counts, since IDP budget consumption is not always correlated to the number of API calls.
    {\bf P2:} Defenses should target \emph{minimal, intrinsic properties of legitimate \ppa use}, not specific attack tactics (e.g., redirection chains, embedded intermediaries, fast-fire vs.\ slow-burn attacks) or speculative assumptions about benign workloads (e.g., that benign traffic will never involve redirections or more than $x$ intermediaries). This keeps defenses robust as both attacks and benign workloads evolve.
    {\bf P3:} Global-budget use should be \emph{ultimately tied to intentional user actions across distinct first-party sites}, since that is what \ppa is meant to measure.
    {\bf P4:} \ppas device-epoch IDP must be preserved.

\begin{figure}[t]
    \centering
    \begin{subfigure}{0.57\linewidth}
        \centering
        \includegraphics[width=\linewidth]{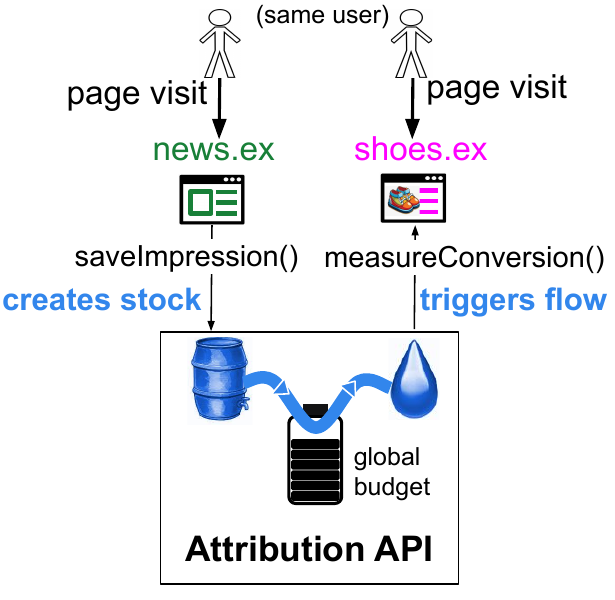}
        \caption{Privacy loss in normal use}
        \label{fig:stock-and-flow-normal-use}
        \vspace{0.1cm}
    \end{subfigure}
    \hfill
    \begin{subfigure}{0.39\linewidth}
        \centering
        \includegraphics[width=\linewidth]{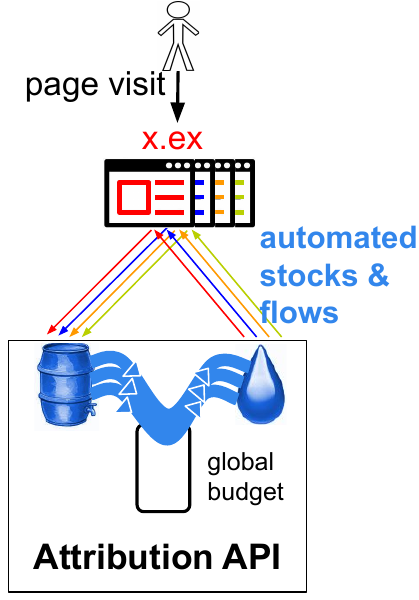}
        \caption{Privacy loss in attack}
        \label{fig:stock-and-flow-attack}
        \vspace{0.1cm}
    \end{subfigure}
    \caption{{\bf Stock-and-flow pattern of budget consumption in \ppa.} (a) Normal use: user actions on impression sites create stock and user actions on conversion sites trigger flows. (b) Attacks break this pattern by automatically creating many stocks and flows from very few user actions.}
    \label{fig:stock-and-flow}
\end{figure}

\heading{Key insight: stock-and-flow.}
\F\ref{fig:stock-and-flow} illustrates this pattern.
The \ppa API is intended to measure {\em user interactions across distinct first-party sites}, such as impression and conversion sites.
In normal \ppa use (\F\ref{fig:stock-and-flow-normal-use}), privacy loss should only be consumed when there is something to measure: a user visits one site (and sees an ad) and later visits a distinct site (and converts there). That creates a stock-and-flow pattern for privacy budget consumption: (1) {\em stock} is created when the user visits an impression site and the site saves an impression; and (2) {\em flow} is triggered when a user visits a conversion site and the site requests a report. Non-zero privacy loss is realized only when {\em both steps occur}, a property afforded by the IDP mechanism. By contrast, attacks disrupt this structure (\F\ref{fig:stock-and-flow-attack}) by coalescing domains or automating stock creation and flow triggering (\eg via automatic redirections or embedded intermediaries), allowing substantial global budget consumption from minimal user actions. \sysname restores this pattern with privacy-loss-based quotas that cap {\em stock creation} and {\em flow triggering} for every caller alike, ultimately tying both to {\em intentional user actions}.

\begin{figure}[t]
    \centering
    \includegraphics[width=0.9\linewidth]{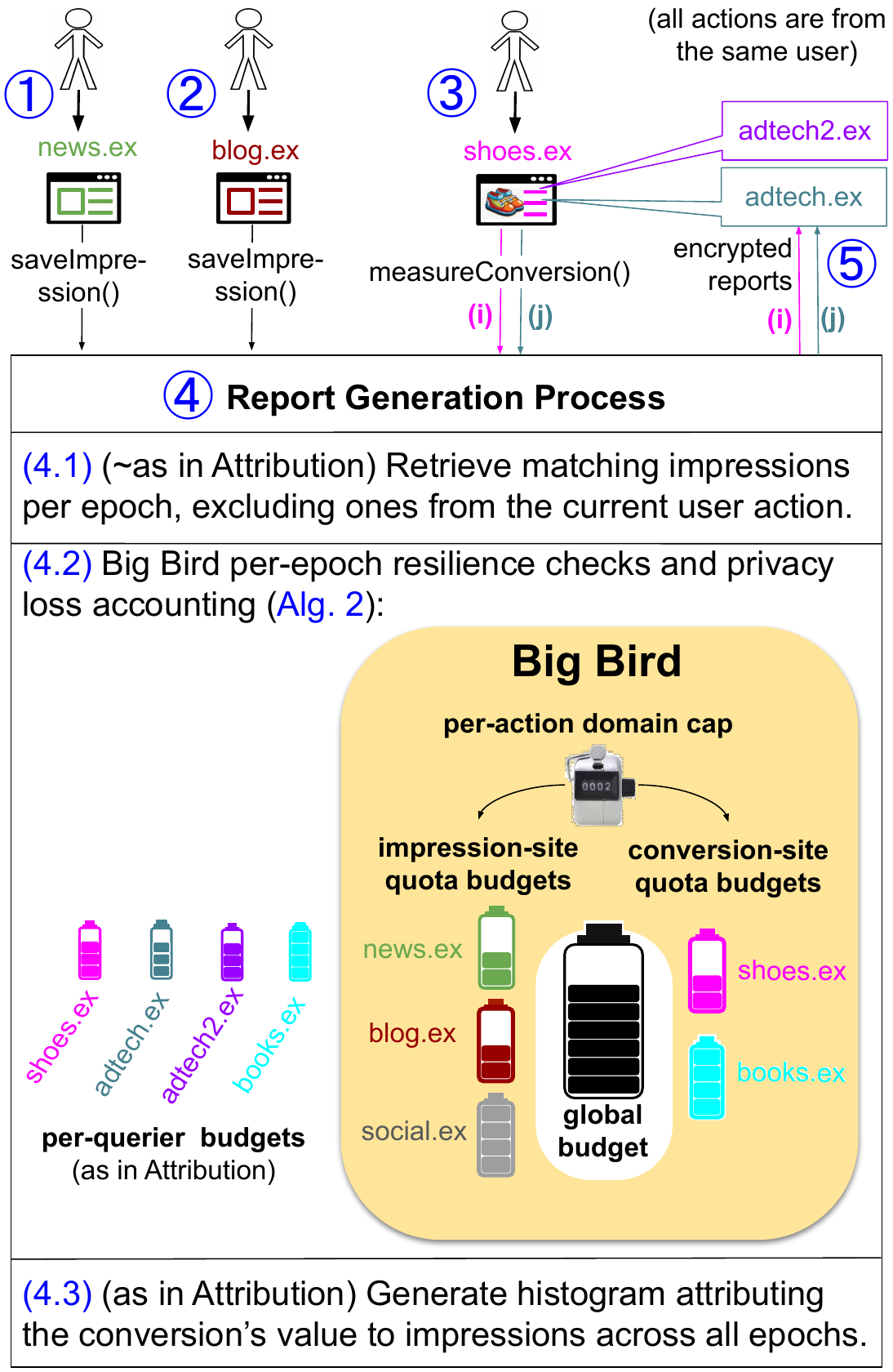}
    \caption{{\bf \sysname architecture}. %
        Depicts budget structure for a single epoch, though in reality privacy and quota budgets exist for each epoch.
    }
    \label{fig:big-bird-architecture}
    \vspace{-0.5em}
\end{figure}

\heading{\sysname architecture.}
\F\ref{fig:big-bird-architecture} shows the \sysname architecture. It sits as a management layer around \ppas global budgets and protects them from depletion by enforcing quotas that restore the expected stock-and-flow pattern, in each epoch: (1) {\em impression-site quotas} limit how much stock any one domain can contribute; (2) {\em conversion-site quotas} limit how much flow a domain can trigger; and (3) a {\em \quotacountVarName} limits how many new domains can create quotas in the context of a single user action.

The first two quotas are privacy budgets implemented as IDP filters, just like global and per-querier budgets. However, their purpose is not privacy; they protect the global budget from depletion. These quotas account for individual privacy loss in a manner aligned with the global budget (principle~{\bf P1}), and constrain stock creation and flow triggering according to the expected structure of benign behavior ({\bf P2}). The \quotacount is not a privacy budget; it simply restricts the number of quota budgets activated under different domains from a single user action. Together, these three quotas constrain consumption of the global budget, forcing the attacker to behave within the bounds of expected workload, and tie this restricted consumption to something tangible, such as intentional user actions across sites ({\bf P3}).
\S\ref{sec:resilience-to-dos-depletion} formalizes the resilience properties afforded by this quota architecture, while our evaluation quantifies them empirically. Finally, our quota enforcement is done separately for each epoch, ensuring that \sysname preserves device-epoch IDP, a property we discuss in \S\ref{sec:big-bird-privacy} ({\bf P4}).

\heading{Example execution.}
\F\ref{fig:big-bird-architecture} shows the workflow.
\step{1}--\step{2} A user sees ads on \code{news.ex} and \code{blog.ex}.
\step{3} After purchasing on \code{shoes.ex}, two \code{measureConversion()} calls are issued: (i) one for \code{shoes.ex} (to measure a creative) and (j) one for \code{adtech.ex} (to optimize its placement model).
\step{4} \sysname processes each call by selecting relevant impressions and attempting to deduct budgets atomically per epoch: querier, global, conversion-site, and matching impression-site quotas. If any deduction fails, that epoch's contribution is dropped. \step{5} Two encrypted reports are returned, one per querier.

Reports (i) and (j) use parameters: \code{epsilon=0.2} (privacy budget requested), \code{value=75} (conversion value), and \code{maxValue=150} (maximum conversion value). Assume the relevant impressions occur in epochs $e1$ and $e2$, while the conversion occurs in $e3$. \T\ref{tab:example-budget-state} shows the remaining budget and quota capacities after releasing reports (i) and (j), assuming all budgets start at the listed capacities.
For report (i), following the IDP rule used in \ppa, each relevant epoch incurs privacy loss $(\text{value}/\text{maxValue})\cdot\epsilon = 0.1$, while other epochs incur none. Only sites with matching impressions (\code{news.ex} in $e1$ and \code{blog.ex} in $e2$) consume impression-site quota. Since all budgets for $e1$ and $e2$ have sufficient capacity, deductions commit atomically in each epoch, and the histogram is generated from these epochs. Report (j) performs the same computation but deducts from \code{adtech.ex}'s querier budget instead of \code{shoes.ex}'s. %

\begin{table}[t]
    \centering
    \scriptsize
    \setlength{\tabcolsep}{3pt}
    \resizebox{\columnwidth}{!}{%
        \begin{tabular}{|c|c|c|c|c|c|c|c|}
            \hline
            \multirow{2}{*}{\textbf{epoch}} &
            \multicolumn{3}{c|}{\makecell{\textbf{querier}                                                                       \\ $\epsQuerier=0.5$}} &
            \multirow{2}{*}{\makecell{\textbf{global}                                                                            \\ $\epsGlobal=4$}} &
            \multicolumn{2}{c|}{\makecell{\textbf{imp-quota}                                                                     \\ $\eqimp=2$}} &
            \makecell{\textbf{conv-quota}                                                                                        \\ $\eqconv=0.75$} \\
            \cline{2-4} \cline{6-7} \cline{8-8}
                                            & {\tiny shoes.ex} & {\tiny adtech.ex} & {\tiny adtech2.ex} &
                                            & {\tiny news.ex}  & {\tiny blog.ex}   & {\tiny shoes.ex}                            \\
            \hline
            e1                              & 0.4              & 0.4               & 0.4                & 3.8 & 1.8 & 2   & 0.55 \\
            \hline
            e2                              & 0.4              & 0.4               & 0.4                & 3.8 & 2   & 1.8 & 0.55 \\
            \hline
            e3                              & 0.5              & 0.5               & 0.5                & 4   & 2   & 2   & 0.75 \\
            \hline
        \end{tabular}%
    }
    \vspace{0.1em}
    \caption{{\bf Quota and budget states after example execution.}}
    \label{tab:example-budget-state}
\end{table}

\heading{\sysname algorithm.}
Alg.~\ref{alg:big-bird} enforces quotas while preserving device-epoch IDP. Each user action creates a context \code{uaCtx}, shared across all API calls causally triggered by
that action. Browsers already have notions of intentional user action, namely the user-activation API~\cite{user-activation-api}, often used to gate access to various APIs.
Within each \code{uaCtx}, \sysname bounds the {\em number of distinct first-party domains}, not API calls. This prevents attackers from spawning many Sybil domains while allowing arbitrary activity within each domain. This matches normal web behavior: a benign user action may involve one or two redirects (e.g., email links or short URLs), but is not expected to traverse many domains. As a result, small \quotacount values (e.g.,~2--3) should comfortably support benign workloads while preventing attackers from spawning many distinct quotas to fuel depletion.

\newif\ifcombineimpconvcap
\combineimpconvcaptrue    %

\algrenewcommand\algorithmicfunction{def}
\algrenewcommand\textproc{}

\ifcombineimpconvcap
  \begin{algorithm}[t]
    \caption{\footnotesize \bf \sysname algorithm}
    \label{alg:big-bird}
    \footnotesize
    \begin{algorithmic}[0] %

      \Function{\textbf{onUserAction}}{uaCtx}
      \State accessedSites[uaCtx][e] $\gets \emptyset$ for each epoch e
      \EndFunction  %
      \Statex

      \vspace{-0.20cm}

      \Function{\textbf{saveImpression}}{uaCtx, impSite, impression}
      \State e = getCurrentEpoch()
      \If{$|\text{accessedSites[uaCtx][e]} \cup \text{\{impSite\}}| > \mathquotacountVar$}
      \State \Return  \graycomment{can't allow new quotas from this user action}
      \EndIf
      \State $\text{accessedSites[uaCtx][e]}\ \cup = \text{\{impSite\}}$
      \State ImpressionStore[e].add(uaCtx, impression) \ \ \graycomment{save uaCtx}
      \State \Return
      \EndFunction  %
      \Statex

      \vspace{-0.20cm}

      \Function{\textbf{measureConversion}}{uaCtx, impSites, querier, convSite, params}
      \State \graycomment{Separately for each epoch, match impressions and do budget}
      \State \graycomment{checks and privacy accounting (Steps (4.1)-(4.3) in \F\ref{fig:big-bird-architecture}).}
      \For{$e \in \text{params.attributionWindow}$}
      \State \graycomment{Check the \quotacountVarName.}
      \If{$|\text{accessedSites[uaCtx][e]} \cup \text{convSite}| > \mathquotacountVar$}
      \State $I_e \gets \emptyset$  \graycomment{drop epoch's impression data from consideration}
      \State \continue
      \EndIf
      \State $\text{accessedSites[uaCtx][e]}\ \cup= \text{\{convSite\}}$

      \State \graycomment{Match impressions, filtering out impressions that querier and convSite}
      \State \graycomment {are not permitted to measure, including impressions from this uaCtx.}
      \State $I_e \gets$ MatchingFn(uaCtx, ImpressionStore[e], params)

      \State \graycomment{Compute IDP loss at two granularities w/ Cookie Monster algo.}
      \State $\epsilon_e \gets$ computeEpochLoss($I_e$, params)
      \For{$i \in$ impSites}
      \State $\epsilon_e^i \gets$ computeEpochImpSiteLoss($I_e[i]$, params)
      \EndFor  %

      \State \graycomment{Execute budget checks \& deductions as a transaction.}
      \State tx = beginTransaction()
      \If{querierBudget[e][querier].tryConsume($\epsilon_e$, tx) \textbf{and }
      \newline\hspace*{1.5em} globalBudget[e].tryConsume($\epsilon_e$, tx) \textbf{and }
      \newline\hspace*{1.5em} convQuota[e][convSite].tryConsume($\epsilon_e$, tx) \textbf{and }
      \newline\hspace*{1.5em} (impQuota[e][$i$].tryConsume($\epsilon_e^i$, tx) $\forall i \in$ impSites)}
      \State tx.commit()  \graycomment{all budgets deducted successfully}
      \Else \ \ \graycomment{a budget deduction failed}
      \State tx.abort() \graycomment{undo any other budget deductions}
      \State $I_e \gets \emptyset$ \graycomment{drop epoch's impression data from consideration}
      \EndIf
      \EndFor  %
      \State \graycomment{Generate attribution histogram from remaining impressions} \\
      \Return AttributionFn$(I_{e_1}, \dots, I_{e_w})$  \ \ \graycomment{$e_i \in \text{params.attributionWindow}$}
      \EndFunction   %

    \end{algorithmic}
  \end{algorithm}

\else  %

  \begin{algorithm}[t]
    \caption{\footnotesize \bf \sysname algorithm}
    \label{alg:big-bird}
    \footnotesize
    \begin{algorithmic}[0] %
      \Function{\textbf{onUserAction}}{ }
      \State uaCtx = createUserContext()
      \State accessedImpSites[uaCtx] $\gets \emptyset$
      \State accessedConvSites[uaCtx][e] $\gets \emptyset$ for each epoch e
      \State \Return uaCtx
      \EndFunction
      \Statex

      \vspace{-0.15cm}

      \Function{\textbf{saveImpression}}{uaCtx, impSite, impression}
      \If{$|\text{accessedImpSites[uaCtx]} \cup \text{\{impSite\}}| > \mathquotacountVar$}
      \State \Return  \graycomment{can't allow new imp-quotas from this user action}
      \EndIf
      \State $\text{accessedImpSites[uaCtx]}\ \cup = \text{\{impSite\}}$
      \State ImpressionStore[current epoch].add(uaCtx, impression) \ \ \graycomment{save uaCtx}
      \State \Return
      \EndFunction
      \Statex

      \vspace{-0.15cm}

      \Function{\textbf{measureConversion}}{uaCtx, impSites, querier, convSite, params}
      \State \graycomment{Separately for each epoch, match impressions and do budget}
      \State \graycomment{checks and privacy accounting (Steps (4.1)-(4.3) in \F\ref{fig:big-bird-architecture}).}
      \For{$e \in \text{params.attributionWindow}$}
      \If{$|\text{accessedConvSites[uaCtx][e]} \cup \text{convSite}| > \mathquotacountVar$}
      \State \graycomment{Can't allow further conv-quota access from this user action.}
      \State $I_e \gets \emptyset$  \graycomment{drop epoch's impression data from consideration}
      \State \continue
      \EndIf
      \State $\text{accessedConvSites[uaCtx][e]}\ \cup= \text{\{convSite\}}$

      \State \graycomment{Match impressions, filtering out impressions that querier and convSite}
      \State \graycomment {are not permitted to measure, including impressions from this uaCtx.}
      \State $I_e \gets$ MatchingFn(uaCtx, ImpressionStore[e], params)
      \State \graycomment{Compute IDP loss at two granularities w/ Cookie Monster algo.}
      \State $\epsilon_e \gets$ computeEpochLoss($I_e$, params)
      \For{$i \in$ impSites}
      \State $\epsilon_e^i \gets$ computeEpochImpSiteLoss($I_e[i]$, params)
      \EndFor  %

      \State \graycomment{Execute budget checks \& deductions as a transaction.}
      \State tx = beginTransaction()
      \If{querierBudget[e][querier].tryConsume($\epsilon_e$, tx) \textbf{and }
      \newline\hspace*{1.5em} globalBudget[e].tryConsume($\epsilon_e$, tx) \textbf{and }
      \newline\hspace*{1.5em} convQuota[e][convSite].tryConsume($\epsilon_e$, tx) \textbf{and }
      \newline\hspace*{1.5em} (impQuota[e][$i$].tryConsume($\epsilon_e^i$, tx) $\forall i \in$ impSites)}
      \State tx.commit()  \graycomment{all budgets deducted successfully}
      \Else \ \ \graycomment{a budget deduction failed}
      \State tx.abort() \graycomment{undo any other budget deductions}
      \State $I_e \gets \emptyset$ \graycomment{drop epoch's impression data from consideration}
      \EndIf
      \EndFor  %
      \State \graycomment{Generate attribution histogram from remaining impressions} \\
      \Return AttributionFn$(I_{e_1}, \dots, I_{e_w})$  \ \ \graycomment{$e_i \in \text{params.attributionWindow}$}
      \EndFunction
      \Statex

      \vspace{-0.15cm}

      \Function{\textbf{onEpochChange}}{uaCtx}
      \State destroyUserContext(uaCtx)
      \EndFunction
    \end{algorithmic}
  \end{algorithm}

\fi  %

Privacy accounting occurs in \code{measureConversion()}.
For each epoch in the attribution window, the function (i) matches impressions,
(ii) computes device-epoch IDP loss for the conversion and the finer-grained
device-epoch--impression-site losses used by impression-site quotas, and
(iii) attempts to deduct all relevant budgets in a single {\em atomic transaction}.
The finer-grained accounting prevents wasting quota on impression sites that
displayed no relevant impressions. If any deduction fails, that epoch's data
contribution is dropped with no budget impact; otherwise, all deductions commit
atomically. This prevents budget waste, preserves the stock-and-flow structure,
and ensures that decisions based on one epoch's data never affect another's
privacy accounting.

A subtle but critical constraint shapes Alg.\ref{alg:big-bird}'s structure.
A natural alternative would be to enforce the \quotacount by computing, inside
\code{measureConversion()}, how many impression-site quotas supply stock and how
many conversion-site quotas trigger flow for the report, and to nullify the
entire report if this exceeds a threshold. \sysname deliberately avoids this.
A single report may aggregate impressions and conversions from multiple epochs,
and any check that reasons about those epochs jointly would couple
their outcomes and thus their privacy losses. Because device-epoch IDP requires each epoch's privacy loss to depend only on that epoch's data, such cross-epoch checks are unsound.

Instead, \sysname enforces the \quotacount using {\em strict per-epoch separation}.
The count of activated quotas is maintained not only per \code{uaCtx} but also {\em per epoch}. If the cap is exceeded in epoch~$e$, \sysname nullifies only epoch~$e$'s data contribution, forcing its privacy loss to zero while leaving all other epochs unaffected. No check aggregates privacy consumption across epochs, and no failure in one epoch propagates to another. This preserves device-epoch IDP while still enforcing meaningful DoS limits tied to user actions.

Finally, we note a subtlety of the atomic check-and-deduct procedure, specifically for impression-site quota deductions within an epoch. As shown in \S\ref{sec:evaluation}, atomicity strengthens resilience by forcing attackers to be selective about which Sybil impressions they declare as relevant: a single impression whose impression-site quota is exhausted suffices to nullify the entire epoch's contribution, preventing any global-budget consumption. This same mechanism can, however, harm benign workloads: if multiple impression sites are relevant in an epoch and one exhausts its quota, the querier cannot obtain attribution information from the remaining sites even if they retain budget. Supporting finer-grained behavior -- e.g., selectively nullifying only the depleted impression site's contribution while preserving device-epoch IDP -- appears feasible, but we leave its design to future work.
Our strongest-attacker experimental results in \S\ref{sec:evaluation} provide an upper bound on resilience if impression-site quotas are removed from the atomic transaction.

\heading{Quota configuration.}
We choose quota capacities via a utility-first approach: specify workload requirements, then select the smallest values that satisfy them.
Taking the querier budget $\epsQuerier$ and intermediary fraction $r$ as base parameters, and using browser-tunable parameters that reflect expected benign workload scale, we define four constraints:
(i) a conversion site must be able to spend its full querier budget, and intermediaries may spend a fraction $r$ of that, giving $\convQuotaEps \ge (1+r)\persiteFilterEps$;
(ii) no single conversion site should drain the global budget, so with $N$ expected conversion sites per epoch, $\globalFilterEps \ge N\convQuotaEps$;
(iii) likewise, the global budget should exceed an impression site's stock, so with $M$ expected impression sites, $\globalFilterEps \ge M\impQuotaEps$; and
(iv) because multiple conversion sites may query the same impression site, a fan-out parameter $n$ requires $\impQuotaEps \ge n\convQuotaEps$.

We then minimize $\epsGlobal$ subject to these constraints, yielding the capacities in \T\ref{tab:filter-configs} and a tight global-privacy guarantee consistent with utility.
Configuring \sysname is thus a three-step procedure:
(1)~choose the base parameters $\epsQuerier$ and $r$;
(2)~conservatively estimate the workload parameters $M, N, n$ from a production sample at a chosen percentile; and
(3)~derive all quota capacities in closed form from
\T\ref{tab:filter-configs}.
The percentile is the only real knob, trading utility against the tightness of $\epsGlobal$.
\S\ref{sec:evaluation:quota-parameters-criteo} instantiates this procedure on Criteo.

\begin{table}[t]
    \centering
    \footnotesize
    \begin{tabular}{|l|l|}
        \hline
        \multicolumn{2}{|p{8cm}|}{
        {\bf ``Normal'' workload parameters:}
        }                                                           \\
        \multicolumn{2}{|p{8cm}|}{
        {\bf M}: {\scriptsize max \# of impression sites contributing nonzero loss in an epoch.}
        }                                                           \\
        \multicolumn{2}{|p{8cm}|}{
        {\bf N}: {\scriptsize max \# of conversion sites requesting nonzero loss from an epoch.}
        }                                                           \\
        \multicolumn{2}{|p{8cm}|}{
        {\bf n}: {\scriptsize max \# of conversion sites requesting nonzero loss from (epoch, impSite).}
        }                                                           \\
        \multicolumn{2}{|p{8cm}|}{
        {\bf r}: {\scriptsize max budget consumed by an intermediary's cross-advertiser queries on a single conversion site, as a fraction of the intermediary's $\persiteFilterEps$.}
        }                                                           \\
        \hline \hline
        \textbf{Cap}            & \textbf{Capacity configuration}   \\
        \hline
        {Per-querier filter}    & $\enc$: configuration parameter   \\
        \hline
        {Global filter}         & $\ec=\max(N, n \cdot M)(1+r)\enc$ \\
        \hline
        {Impression-site quota} & $\eqimp=n(1+r)\enc$               \\
        \hline
        {Conversion-site quota} & $\eqconv=(1+r)\enc$               \\
        \hline
        {\quotacount}           & $\mathquotacountVar = 2$ or $3$   \\
        \hline
    \end{tabular}
    \vspace{0.1cm}
    \caption{{\bf \sysname budget configurations}.}
    \label{tab:filter-configs}
\end{table}
\vspace{-0.2cm}
\subsection{Resilience Guarantee}
\label{sec:resilience-to-dos-depletion}

\sysname restores proportionality between attacker impact and strength along the three dimensions in \S\ref{sec:threat-model}. Its resilience, stated in \Thm~\ref{thm:resilience-three-dimensions}, stems from: (a) per-domain quota limits, (b) user-action-based quota creation, and (c) bounded-sensitivity DP aggregation. Indeed, \ppa serves population-level {\em DP queries} in which each device-epoch contributes only through a report with {\em bounded sensitivity}. As a result, even if an attacker depletes budget on some device-epochs, the resulting error in a benign query scales only with the number of affected device-epochs.

\begin{theorem}[{Resilience to DoS depletion -- proof in \ref{appendix:online-algorithm:dos-proofs}}]
    \label{thm:resilience-three-dimensions}
    Consider an attacker who (1) creates $M^{\textrm{adv}}$ and $N^{\textrm{adv}}$ \impQuotaVar and \convQuotaVar quotas, respectively; (2) collects $U^{\textrm{adv}}$ user actions across the quotas' domains for device $d$; and (3) depletes budget from $k$ device-epochs.

    Denote by $\globalFilterEps^{\textrm{adv}}$ the maximum budget that the attacker can consume from the \globalFilter on any epoch from $d$. Denote by $\|Q(D) - \tilde Q(D)\|$ the error induced by the attacker, for a query $Q$ where reports have sensitivity $\Delta$ and at most $m$ reports belong to any single device.
    Then, we have:

    \begin{enumerate}[(i)]
        \item $\globalFilterEps^{\textrm{adv}} \leq \min(M^{\textrm{adv}} \impQuotaEps, \ N^{\textrm{adv}} \convQuotaEps)$
        \item $\globalFilterEps^{\textrm{adv}} \leq (1+r)\persiteFilterEps \cdot \mathquotacountVar \cdot U^{\textrm{adv}}$
        \item $\|Q(D) - \tilde Q(D)\| \le m \cdot k \cdot \Delta$, and $k \leq U^{\mathsf{adv}} / \lceil \frac{\ec}{n(1+r)\enc \cdot \text{\quotacountVar}} \rceil$.
    \end{enumerate}
\end{theorem}
\vspace{0.1cm}

(i) Even if user actions were free, the global budget that an attacker can drain on a given device-epoch is capped by how many impression-site and conversion-site quotas it has managed to create, scaled by their capacities. This is the stock-and-flow constraint: to move global budget, the attacker must both create stock (impression-site quotas) and trigger flow (conversion-site quotas), and each is individually bounded.
(ii) We connect this to intentional user actions via the \quotacountVarName. Each user action can create at most $\mathquotacountVar$~new quotas of each type, and only within a single epoch. Across $U^{\textrm{adv}}$ actions, the attacker can therefore unlock at most $O(\mathquotacountVar \cdot U^{\textrm{adv}})$ useful quotas, which yields the stated bound on $\globalFilterEps^{\textrm{adv}}$. This enforces proportionality to user actions: substantial depletion now requires many genuine user interactions, not just many cheap domains.
(iii) We lift the guarantee from a single device-epoch to the query level. With bounded report sensitivity $\Delta$,  even if the attacker fully depletes the global budget on $k$ device-epochs, its impact on any aggregate is linear in $k$. $k$ itself is bounded by the number of actions across devices.

\heading{Limitations.}
We identify two limitations of the resilience guarantees provided by \sysname. First, resilience assumes the attacker cannot scale simultaneously along all three strength dimensions. Enforcing this ultimately requires API-governance mechanisms beyond budget management. For instance, browsers could require API invokers to register with trusted authorities to limit Sybil domains, or adopt a conservative definition of \emph{user activation} to ensure that interactions the user did not intend as attribution actions (e.g., clickjacking, deceptive UI, automatic redirects, or invisible frames) do not trigger budget consumption. Our contribution is to {\em expose the necessary control points} within the privacy-budget layer so that, when combined with governance mechanisms, \ppa can be rigorously hardened against attacks.

Second, \Thm\ref{thm:resilience-three-dimensions} bounds depletion of the \emph{global} budget, but quota budgets themselves may still be targeted -- albeit only in ways excluded by our threat model (\S\ref{sec:threat-model}). For example, a malicious intermediary could exhaust its own conversion-site quota to block other intermediaries, or target the impression-site quota of a site that embeds its ads. These attacks are localized: the impact is confined to queriers interacting (directly or transitively) with the affected site, while unrelated sites retain access to the global budget. Moreover, our quota configuration dampens such effects: exhausting an impression-site quota requires fan-out across $n$ conversion sites, while exhausting a conversion-site quota requires $(1+r)$ queriers. Fully addressing quota-targeting attacks will require \ppa to support mechanisms for managing and subdividing quotas (e.g., per-partner allocation) to prevent abuse. We leave this to future work, as it likely requires integration with ecosystem-level constructs such as business relationships and auction pricing.
\vspace{-0.2cm}
\subsection{Privacy Guarantee}
\label{sec:big-bird-privacy}

Our privacy analysis of \sysname aims for a high-fidelity model that provides confidence that the design satisfies global device-epoch-level IDP. We capture: individual DP; on-device report generation with centralized noise addition; all public information exposed by the protocol; query, budget, and data adaptivity; multiple epochs and queriers; the atomic interaction of multiple budgets within a single transaction; and user-action-based limits on how many quotas may be invoked. To our knowledge, this is one of the most faithful system models in the DP literature.
Appendix~\ref{appendix:online-filter-management} gives two versions: a simplified one for educational use (\S\ref{sec:simplified_model}) and the complete version for formal analysis (\S\ref{appendix:online-algorithm:algos}). Using the complete model, we prove that \sysname preserves {\em global device-epoch IDP}:
Here and below, $\Pr[V=v]$ denotes the joint density or probability mass at $v$, as appropriate.

\begin{theorem}[{Global IDP Guarantee -- proof in \S\ref{appendix:online-algorithm:privacy-proofs}}]
    \label{thm:privacy-guarantee}
    Consider $x \in \cX$ with \c capacity $\ec$.
    Then, \sysname satisfies individual device-epoch $\ec$-DP for $x$ under public information $\cC$.
    That is, for every data generation process $\cG$ and all queriers $\cA^1, \dots, \cA_k$, the output $V^{(\chal)}$ of \Alg\ref{alg:e2e_setup} under challenge bit $\chal$ satisfies, for all $v \in \cR^{e_{\max} \times (t_{\max} + 1) \times B}$:
    \begin{align*}
        \left| \ln \left( \frac{\Pr[V^{(0)} = v]}{\Pr[V^{(1)} = v]} \right) \right| \leq \ec.
    \end{align*}
\end{theorem}
\vspace{-0.2cm}
\subsection{Prototype as Reference Implementation}
\label{sec:prototype}

We implement \sysname both as a reference implementation for the emerging W3C standard and as a foundation for broader deployments, including in mobile OS location analytics. The implementation has two components.
(1) {\pdslibbf} is a general-purpose IDP Rust library that subsumes Cookie Monster and \sysname while exposing a generic interface beyond Attribution. Clients (websites or mobile apps) register events (e.g., ad views, location visits), request reports (e.g., attributions, model updates), and receive encrypted, privacy-budgeted responses.
(2) An integration in Firefox's Attribution prototype~\cite{moz-pa} replaces report counting with full IDP and resilient enforcement based on \pdslib.
Mobile recommendation workloads, which share sparsity and structure with advertising, are a promising future instantiation.
\newcommand{\percentiles}{p50_99}
\newcommand{\appendpercentiles}[1]{#1_\percentiles.pdf}

\vspace{-0.2cm}
\section{Evaluation}
\label{sec:evaluation}

We evaluate \sysname's utility and resilience on a real-world Criteo dataset, simulating DoS attackers inspired by \F\ref{fig:example-attack-conversion-sites}. Our evaluation asks two questions: (1) How do quota configurations affect benign-query utility, both without attack and under a fixed-strength attack?
(2) How does attacker strength along different resource dimensions translate into damage to benign queries?

\vspace{-0.1cm}
\subsection{Methodology}
\label{sec:methodology}

\heading{Dataset.}
CriteoPrivateAd~\cite{criteo2025} is a 30-day snapshot of production traffic (104M impressions across 220k publisher sites and 10k conversion sites) released for benchmarking W3C's \ppa API. To our knowledge, it is the only real-world dataset with multiple conversion sites---a requirement for exercising \sysname's
conversion-site quotas. The dataset includes timing, contextual features, conversions, and a daily-reset device ID. Because impressions are subsampled, most devices appear only once. Using Criteo's resampling method, we reconstruct the device-level distribution, yielding {\em 1.4M devices}, {\em 4.6M impressions}, and {\em 5.6M conversions}. The median device has 2 impressions and 4 conversions. While realistic, the dataset reflects a single intermediary and likely underestimates load in multi-intermediary deployments.

\heading{Benign workload.}
We evaluate single-advertiser measurement, where each batch of conversions produces a DP histogram over five contextual buckets. We evaluate utility using RMSRE$_\tau$~\cite{AksuGKKMSV24}, the root mean squared relative error across buckets, clipping counts below $\tau = 5\%$ of batch size. This captures both DP noise (common across baselines) and bias from budget/quota depletion. To ensure low-noise baselines, we (i) restrict to advertisers with $\ge 100$ conversions/day (526 total), (ii) batch $\sim$10 days of conversions (capped at 8,000), and (iii) set $\epsilon$ to target RMSRE$_\tau \approx 5\%$, similar to how we expect an advertiser would tune their measurement.

\heading{Attack workload.}
We inject synthetic adversarial traffic atop real events. The attacker is given impression and conversion streams for the 10 busiest sites, reusing timestamps and device IDs to simulate organic traffic. Each event is treated as a user action enabling budget depletion. The attacker controls a pool of Sybil domains (25 by default) and, per user action, triggers a redirection chain (bounded by $\mathquotacountVar$). Each Sybil registers one fresh impression and requests a report on a previously unqueried impression with $\epsilon=\enc$.

A naive attacker that references all Sybils per action is neutralized by \sysname's atomic deductions: a single over-quota Sybil invalidates the entire request, consuming zero global budget. We confirm this empirically and focus on two stronger strategies. The {\bf Random Attacker} samples a subset of Sybils per conversion (35\% by default, minimizing atomic rejection), modeling limited tracking capabilities. The {\bf Omniscient Attacker} has perfect knowledge of each device's remaining quotas and uses only Sybils with remaining quota, eliminating rejection waste and representing the optimal attack within the \ppa interface. Because these capabilities---perfect repeat-user recognition and full
visibility into every device's quota state---are ones real attackers
will likely lack, we read omniscient-attacker error as a deliberate worst-case
ceiling rather than a typical operating point; real attackers fall
between the two extremes.

\heading{Baselines and defaults.}
We compare \sysname to two baselines spanning the utility--privacy spectrum:
(1) {\em \ppa w/o global budget} (Cookie Monster~\cite{TKM+24}), which enforces per-querier IDP but no shared limit. It is immune to depletion and serves as the {\em utility ceiling}, but lacks global privacy guarantees.
(2) {\em \ppa w/ global budget}, which adds a shared privacy filter but no quota mechanisms, representing the strongest defense without \sysname.

Unless stated otherwise, we set $\enc = 1.0$, $\ec = 8.0$, $\eqimp = 2.0$, and $\eqconv = 1.0$. We use $\mathquotacountVar = 2$ and 25 Sybils at maximum popularity. Epochs are one day, matching Criteo's daily device-ID reset.

\vspace{-0.2cm}
\subsection{Quota Configuration on Criteo}
\label{sec:evaluation:quota-parameters-criteo}

\begin{wraptable}{r}{0.5\columnwidth}
    \vspace{-4pt}
    \caption{{\bf Workload params in Criteo.}}
    \label{tab:filter-config-percentiles}
    \centering
    \footnotesize
    \setlength{\tabcolsep}{2pt}
    \begin{tabular}{r|ccc|cc}
        \toprule
        \textbf{$\%$ile} & $\tilde{N}$ & $\tilde{M}$ & $\tilde{n}$ & $\ec$ & $\eqimp$ \\
        \midrule
        p50              & 2           & 1           & 2           & 2     & 2        \\
        p80              & 4           & 2           & 2           & 4     & 2        \\
        p85              & 4           & 2           & 4           & 8     & 4        \\
        p90              & 4           & 3           & 4           & 12    & 4        \\
        p95              & 6           & 3           & 4           & 12    & 4        \\
        p99              & 8           & 4           & 8           & 32    & 8        \\
        \bottomrule
    \end{tabular}
    \vspace{-10pt}
\end{wraptable}

We instantiate the three-step configuration procedure of
\S\ref{sec:big-bird-overview} (``Quota configuration'' header) on Criteo.
\emph{Step (1):} we use $\enc = 1$ (our default from
\S\ref{sec:methodology}) and, because our evaluation models only
single-advertiser queries, $r = 0$, giving $\eqconv = \enc$; general
workloads would set $r > 0$.
\emph{Step (2):} the distributions of $N, M, n$ across devices could be
computed exactly by running full conversion attribution, but a deployer
can instead use cheap, conservative per-device statistics that
upper-bound them: unique impression sites ($\tilde{M}\!\ge\!M$), unique
conversion sites ($\tilde{N}\!\ge\!N$), and unique conversion sites per
(device, impression-site) pair ($\tilde{n}\!\ge\!n$);
\T\ref{tab:filter-config-percentiles} reports these statistics at
several percentiles.
\emph{Step (3):} the same table shows the resulting capacities; e.g.,
the p85 estimates $(\tilde{N}, \tilde{M}, \tilde{n}) = (4, 2, 4)$ yield
$\ec = \max(\tilde{N}, \tilde{n}\tilde{M})(1+r)\enc = 8$ and
$\eqimp = \tilde{n}(1+r)\enc = 4$.
Only these values are Criteo-specific: the procedure itself is
dataset-driven, and any browser vendor can run it on its own
production sample.

The percentile choice governs the tradeoff between quota size (and thus
utility) and the tightness of the global DP guarantee $\ec$.
Supporting 100\% of devices maximizes utility, since it expects no quota-induced errors, but yields very loose $\ec$. The 85th percentile offers a balanced configuration: tight enough for a single-digit $\ec = 8$ while avoiding quota errors for the vast majority of devices. As in \S\ref{sec:evaluation:quota-configuration}, setting $\eqimp$ below the table value (e.g., $\eqimp = 2$ instead of~4) substantially improves resilience at a negligible utility cost: the resulting slack in $\ec$ ensures enough budget is left for benign queries even under attack.

\subsection{Utility and Resilience with Quota Configuration}
\label{sec:evaluation:quota-configuration}

\begin{figure}[t]
    \centering
    \begin{subfigure}{0.49\linewidth}
        \centering
        \includegraphics[width=\linewidth]{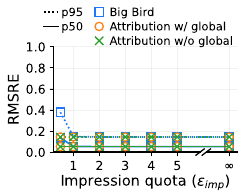}
        \caption{Benign error without attack}
        \label{fig:benign_error}
    \end{subfigure}
    \begin{subfigure}{0.49\linewidth}
        \centering
        \includegraphics[width=\linewidth]{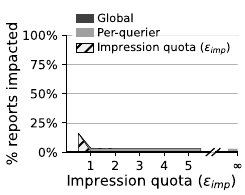}
        \caption{Error causes without attack}
        \label{fig:benign_causes}
    \end{subfigure}\\\vspace{0.2cm}
    \begin{subfigure}{0.49\linewidth}
        \centering
        \includegraphics[width=\linewidth]{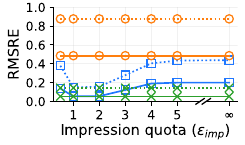}
        \caption{Benign error under attack}
        \label{fig:attack_error}
    \end{subfigure}
    \begin{subfigure}{0.49\linewidth}
        \centering
        \includegraphics[width=\linewidth]{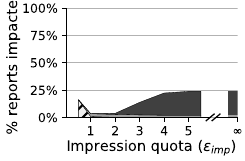}
        \caption{Error causes under attack}
        \label{fig:attack_causes}
    \end{subfigure}
    \vspace{0.2cm}
    \caption{{\bf Utility and resilience with quota configuration.} (a), (b): Benign error and root causes {\em without attack}. (c), (d): Benign error and root causes {\em under the random attacker}.
    }

    \label{fig:online-eval}
\end{figure}

We fix all default parameters and sweep the impression-site quota $\eqimp$ and \quotacountVarName \quotacountVar to determine the range of quota values that preserve benign utility while confining the attacker.

\heading{Benign error without attack (\F\ref{fig:benign_error}, \F\ref{fig:benign_causes}).}
We first measure utility in a purely benign setting, varying the impression-site quota at $\ec = 8$ (derived from the p85 workload in \T\ref{tab:filter-config-percentiles} with $\enc = 1$).
\F\ref{fig:benign_error} plots median and 95th-percentile RMSRE$_\tau$ as $\eqimp$ grows. Both \ppa baselines, which lack impression-site quotas, show constant and identical error across $\eqimp$ (median 5.2\%).
Their agreement confirms that the global budget alone does not reduce utility without an attack, an important sanity check that isolates quota effects from global-budget effects. In contrast, \sysname performs poorly at $\eqimp = 0.5$, where aggressive quota exhaustion produces null reports (median error 14.0\%, a 2.7$\times$ increase). Once $\eqimp$ reaches 2, quotas no longer affect accuracy: \sysname matches the baselines at 5.2\%. This holds through arbitrarily large quota values, including \sysname with the impression-site quota disabled entirely. The p85 quotas from \T\ref{tab:filter-config-percentiles} therefore suffice for normal operation and appear conservative, suggesting that true workload parameters for most devices are well below the upper-bound estimates we use to size $N$, $M$, and $n$.

\F\ref{fig:benign_causes} breaks down the bias component of \sysname's RMSRE$_\tau$
by attributing each blocked report to the first budget in priority order that was exhausted. At $\eqimp = 0.5$, the impression-site quota blocks 16.2\% of reports, explaining the elevated error in \F\ref{fig:benign_error}. At $\eqimp = 1$, this drops to 0.8\%. For $\eqimp \ge 2$, the impression-site quota blocks fewer than 0.002\% of reports; the only residual blocking comes from per-querier budgets (2.8\% of reports). This residual appears in all baselines, and is inherent to IDP systems when device contributions exhaust the querier's budget. The takeaway is that, once quotas are sized to the p85 workload, \sysname introduces no bias beyond what the underlying \ppa accounting already imposes.

\heading{Benign error under attack (\F\ref{fig:attack_error}, \F\ref{fig:attack_causes}).}
We next evaluate resilience when an attacker targets the global budget. \F\ref{fig:attack_error} reports error for benign queries.

\ppa w/o global budget is unaffected (median 5.2\%), as expected: without a global budget, there is nothing to deplete. But this immunity comes at the cost of no global privacy guarantee (\S\ref{sec:contribution1}). \ppa w/ global budget suffers severe degradation: its median RMSRE$_\tau$ reaches 48.5\%, a $9.3\times$ increase, and its p95 reaches 87.6\%. This confirms \sysname's premise: simply adding a global budget to fix per-querier unsoundness is trivially defeated by Sybil attacks.

\begin{figure*}[t!]
    \centering
    \includegraphics[width=0.98\linewidth]{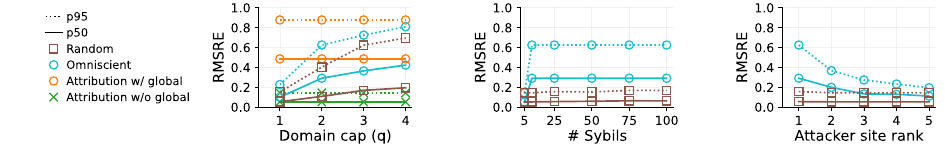}
    \hspace*{0.233\linewidth}%
    \begin{subfigure}{0.25\linewidth}
        \vspace{-1cm}
        \caption{Impact of \quotacountVarName}
        \label{fig:attack-strategies}
    \end{subfigure}%
    \hspace*{0.017\linewidth}%
    \begin{subfigure}{0.25\linewidth}
        \vspace{-1cm}
        \caption{Impact of Sybil domains}
        \label{fig:attacker_sybils}
    \end{subfigure}%
    \begin{subfigure}{0.25\linewidth}
        \vspace{-1cm}
        \caption{Impact of site popularity}
        \label{fig:attacker_popularity}
    \end{subfigure}

        \caption{{\bf (a) Impact of per-action domain cap and (b), (c) attacker strength on resilience.} (b) Number of attacker-controlled Sybil domains capable of accessing the API and (c) popularity of the attacker-controlled sites in terms of number of user actions across all distinct users in the Criteo dataset.
    }
    \label{fig:stuff}
\end{figure*}

In contrast, \sysname achieves near-baseline utility {\em and} the privacy guarantee of a system with a global budget. With a well-sized quota, it nearly matches \ppa w/o global budget at a median of 5.5\% and a p95 of 14.8\%. Moreover, \sysname is robust across a meaningful range of impression-site quota values $\eqimp \in [1, 2]$, for which the budget the attacker can consume remains well below $\ec$, leaving substantial slack in the global budget even after the attack.

We also observe that, as in the benign case, quotas that are too small degrade utility. But under attack, quotas that are {\em too large} also degrade utility, for a fundamentally different reason: they allow the attacker to route enough traffic through Sybils to drain the global budget despite the quota mechanism. At $\eqimp = 3$, median error rises to 12.5\% and by $\eqimp = 5$ it plateaus near 20.1\%. \F\ref{fig:attack_causes} corroborates this transition: for $\eqimp$ up to 2, fewer than 1\% of benign reports are blocked by global-budget depletion; at $\eqimp = 3$ this rises to approximately 11\%, and at $\eqimp = 4$ to approximately 19\%, dominating all other error sources.

\heading{Impact of \quotacountVarName{} and attacker strategy.}
We now fix $\eqimp = 2$ (our default) and sweep \quotacountVar from 1 to 4 to evaluate how the \quotacountVarName{} impacts utility under attack.
Additionally, we compare the omniscient attacker to the random attacker, revealing how adversary sophistication affects resilience.
\F\ref{fig:attack-strategies} shows that across both attackers, $\mathquotacountVar= 1$ keeps median benign RMSRE$_\tau$ at or below 10.0\%, and $\mathquotacountVar= 2$ (our default) at or below 28.6\%. Both remain far below the unprotected \ppa API (48.3--48.4\%).
$\mathquotacountVar = 1$ minimizes worst-case error at the expense of flexibility in the benign case, allowing only one site per user action.
\sysname's quota mechanism provides the structural foundation for resilience, bounding depletion even against the optimal attacker.
Atomic deduction dramatically reduces damage against more realistic adversaries: the random attacker achieves only 5.4\% at $\mathquotacountVar = 2$, versus the omniscient attacker's 28.6\%.

\rigidsubsection{Impact of Attacker Strength}
\label{sec:evaluation:attacker-strength}

The preceding section fixed attacker strength and varied the defense configuration. We now fix the defense ($\mathquotacountVar = 2$, $\eqimp = 2$) and vary attacker resources to test the guarantee from \S\ref{sec:threat-model}: damage should scale with the number of Sybil domains (dimension~(i)) and the volume of genuine user traffic the attacker attracts (dimensions~(ii) and~(iii)).

\heading{Number of Sybil domains (\F\ref{fig:attacker_sybils}).}
We vary the attacker’s Sybil pool from 5 to 100 domains, holding site popularity at maximum. The random attacker rises only from 5.2\% at 5 Sybils to 6.4\% at 100, as the tight $\eqimp = 2$ quota sharply limits per-domain budget consumption regardless of pool size. The omniscient attacker saturates early: RMSRE$_\tau$ reaches 29.4\% at 10 Sybils and remains flat through 100. With perfect tracking, 10 domains already exhaust the available per-device quota headroom; additional Sybils provide no benefit once the global budget becomes the bottleneck. This matches \Thm~\ref{thm:resilience-three-dimensions}(i): depletion is bounded by the product of domain count and quota capacity, so once $M^{\textrm{adv}} \cdot \eqimp$ exceeds the global budget, extra Sybils add nothing.

\heading{Site popularity (\F\ref{fig:attacker_popularity}).}
We vary which of Criteo's busiest sites the attacker controls,
from the top~10 sites to ranks 41--50. Lower-ranked sites have fewer user actions and fewer distinct users, jointly exercising dimensions~(ii) and~(iii) of the threat model.
The random attacker falls from 5.4\% to 5.2\%, remaining near baseline across all ranks --- the tight source quota renders the random attacker essentially harmless regardless of popularity. The omniscient attacker falls from 29.0\% to 10.9\%, a $2.7\times$ reduction. This confirms that damage scales with the volume of genuine user traffic the attacker attracts, as \Thm~\ref{thm:resilience-three-dimensions}(ii--iii) bounds depletion as a function of $U^{\textrm{adv}}$ and the number of affected devices. This dimension is critical: attracting genuine traffic requires costly real-world investment, unlike cheap domain registration.

\vspace{-0.2cm}
\section{Additional Related Work}
\label{sec:related-work}

\S\ref{sec:contribution1} covers DP enforcement granularities. Other work follows.

\heading{Ad-measurement APIs.}
Our primary application is the W3C \ppa API.
Cookie Monster~\cite{TKM+24} introduced device-epoch IDP accounting. Google's ARA line~\cite{AksuGKKMSV24,DGK+23,GHH+25}, IPA~\cite{ipa}, PAM~\cite{pam}, and Hybrid~\cite{hybrid-proposal} explored alternative, albeit now obsolete, privacy architectures. In these proposals, per-querier enforcement is treated as a natural semantic, with collusion recognized as the primary limitation. We show that even \emph{without} collusion, per-querier DP is unsound.
Attribution's measurement-only scope, with targeting unsupported,
is being debated~\cite{hogan2026making}; the debate concerns what the standard should protect, not whether its guarantees hold. We argue that measurement is a sensible place to start, separable from targeting and already treated distinctly by some
U.S. privacy laws~\cite{colorado-privacy-act}.

\heading{Privacy budget management.}
This paper contributes to the broad challenge of {\em privacy budget management}, a crucial but understudied area in DP.
Budget allocation \emph{within} a query is well studied~\cite{matrix_mechanism,Abowd20222020}, but managing budgets \emph{across} queriers, especially untrusted ones, remains relatively unexplored.
\citet{Pujol2021, PSFM22} balance utility across analysts sharing a global budget, but without per-analyst utility assurances.
Relevant systems include those for global budget scheduling~\cite{dpf, dpack, cohere}, %
but none of these works address a Sybil DoS threat. Finally, work on privacy filters~\cite{filters_rogers,aim,adaptive_top_k,lecuyer2021practicalprivacyfiltersodometers,FZ21} provides building blocks on which \sysname implements budgets.

\heading{DoS and quota mechanisms.}
DoS defense is studied extensively in systems and networking (e.g.,~\cite{surgeprotector,algorithmicdos,tcpdos,scout}), and quotas are ubiquitous for conventional resources.
But these approaches do not apply directly to privacy budgets, whose consumption reflects \emph{data contribution}, not request volume.
To our knowledge, no work treats privacy budgets as a \emph{DoS target} and provides Sybil-resilient budget management that preserves DP semantics and utility.
\sysname's novelty does not rest on the notion of quota, but on how the quota \emph{architecture} is shaped by DP semantics and workload structure.

\section{Conclusions}
\label{sec:conclusions}

We study DP enforcement in the W3C Attribution draft standard for privacy-preserving advertising and find that its current reliance on per-querier device-epoch IDP is not a sound foundation: even without collusion, realistic multi-querier behaviors such as cross-querier adaptivity and shared limits can violate its guarantees. We therefore advocate global device-epoch IDP and present \sysname, a budget manager that makes it deployable by tying privacy-budget consumption to the stock-and-flow structure of benign workloads via IDP-aligned quotas and per-action domain caps. Our results show that \sysname achieves rigorous global privacy, resilience to depletion attacks, and strong utility under adversarial conditions. Both contributions are now part of the draft standard and have broader implications for DP systems in open, untrusted environments.

\begin{acks}
We thank the anonymous reviewers and the PATCG/PATWG participants for their constructive feedback.
Nikos Goutzoulias contributed to the implementation of an early prototype.
This work was supported by NSF grant CNS-2143868, NSF Cooperative Agreement No. EEC-2133516, and Google, Meta, Microsoft, and Sloan Faculty Fellowships.
Co-author L\'ecuyer acknowledges the Canada CIFAR AI Chair program and Amii.
Co-author Geambasu was partially employed by Meta during this project, and subsequently by Google; the work was done exclusively in the context of Geambasu's Columbia University research.
\end{acks}

\bibliographystyle{ACM-Reference-Format}
\bibliography{refs/all,refs/thesis}

\clearpage
\newpage

\renewenvironment{theorem}
{\origtheorem}
{\endorigtheorem}

\appendix
\noindent {\em Note: This appendix has not been peer-reviewed.}
\section{Per-Querier DP Analysis (\S\ref{sec:contribution1} Proofs)}
\label{sec:appendix:analysis-per-querier}

\subsection{Data Adaptivity}
\label{sec:appendix:analysis-per-querier-data-adaptivity}

We provide more details on the model introduced in \S\ref{sec:analysis-per-querier-data-adaptivity}.
Our formalism with a data generation process follows existing work \cite{sage}.
Other adaptive stream definitions have been proposed \cite{GS13, JRSS23}.
Adaptive data generation definitions capture stronger adversaries than non-adaptive adjacency definitions for streams \cite{CSS11}.
Note that when there is only one querier, a single adversary can be both the data generation process and the querier.

Without loss of generality, the opt-out record and the time at which it is inserted are known upfront. We could let $\cG$ adaptively decide when and what to insert, while enforcing that this happens at most once, but this decision would not actually depend on the challenge bit and would therefore follow a distribution known by $\cG$ upfront.

\Thm\ref{thm:data-adaptivity} shows that data adaptivity results in excess per-querier privacy loss growing linearly in the number of queriers.
The proof exhibits queriers and a data generation process, where results from queriers $\cQ^1, \dots, \cQ^{n}$ are first encoded into the dataset itself. 
$\cQ^{n+1}$ later decodes this information to gain additional confidence about the value of $\chal$ beyond what its own queries should allow according to the per-querier filter.

\begin{numberedthm}[\ref{thm:data-adaptivity}]
  Consider \Alg\ref{alg:adaptivity_model_simplified}, where $\cF^k$ are instantiated with pure DP filters \cite{filters_rogers}, each with capacity $\epsilon$.
  Denote by $V^{(\chal)}$ the output of the algorithm on challenge bit $\chal$.

  When $K=1$, \Alg\ref{alg:adaptivity_model_simplified} reduces to single-querier fully adaptive composition on adaptively generated streams.
  We have:

  \begin{align*}
    \forall x_0 \in \cX, \forall t \in [\tmax], \forall \cG, \forall \cQ^1, \\
    \forall S \subseteq \cR^{t_{\max}}, \left| \ln \left( \frac{\Pr[V^{1,(0)} \in S]}{\Pr[V^{1, (1)} \in S]} \right) \right| \leq \epsilon.
  \end{align*}

  However, for $K = n + 1 > 1$, there exists $\epsilon' > 0$ such that:
  \begin{align*}
    \exists x_0 \in \cX, \exists t_0 \in [\tmax], \exists \cG, \exists (\cQ^1, \dots, \cQ^{n+1}), \\
    \exists S \subseteq \cR^{t_{\max}}, \left| \ln \left( \frac{\Pr[V^{n+1, (0)} \in S]}{\Pr[V^{n+1,(1)} \in S]} \right) \right| = \epsilon + n\epsilon' > \epsilon.
  \end{align*}

\end{numberedthm}
\begin{proof}
  When $K=1$, we are exactly in the setting of \citet{sage}.
  Briefly, when data can change adaptively based on past results, inserting a single record can have cascading effects on future data and future queries.
  Still, after conditioning on a prefix $V_1^1, \dots, V_t^t$, the databases in both worlds differ only by at most one record.

  Now, consider $K = n + 1$ with $n > 0$.
  Take $x_0 = 1 $ and $t_0 = 1$.
  Without loss of generality, take $\tmax = 2$ and $\cX = \{1\}$.
  We use histogram notation for the database, but since $\cX$ is a singleton, we write $D \in \N$ instead of $D \in \N^\cX$.
  In multiset notation, we could take $\sum_{x \in D} x$ to get an integer representation of the database.
  Define $\cG$ and $\cQ^1, \dots, \cQ^n, \cQ^{n+1}$ as follows:
  \begin{enumerate}
    \item At $t = 1$:
      \begin{enumerate}
        \item The data generation process does not return any data: $\cG(\emptyset, \dots, \emptyset) = \emptyset$. Thus $D_1 = 1$ if $\chal = 1$, and $D_1 = 0$ otherwise, that is:

          \begin{align}
            \label{eq:d1}
            D_1 = \chal
          \end{align}

        \item For $k \in \{1..n\}$, $\cQ^k$ submits $\cM_1^k: D \mapsto D + \lap(1/\epsilon)$, which consumes $\epsilon$ from the per-querier filter $\cF^k$.
        \item $\cQ^K$ submits a query that consumes no budget from the per-querier filter: $\cQ^K(\emptyset) = \cM_1^K : D \mapsto 0$. \label{item:v1}

      \end{enumerate}
    \item At $t = 2$:
      \begin{enumerate}
        \item The data generation process inserts up to $n$ new records, conditionally on outputs $V_1^1, \dots, V_1^n$ from $\cQ^1, \dots, \cQ^n$:
          $\cG(V_1^1, \dots, V_1^K) = |\{k \in \{1..n\}: V_1^k > 1/2\}|$.
          That is:
          \begin{align}
            \label{eq:d2}
            D_2 = \sum_{k=1}^n \mathds{1}[D_1 + Z_k > 1/2]
          \end{align}
          where $Z_k \sim \lap(1/\epsilon)$. Since $V_1^k$ is more likely to be greater than $1/2$ if $\chal = 1$, $D_2$ is more likely to be high if $\chal = 1$.
          In other words, $\cG$ encodes what $\cQ^1, \dots, \cQ^n$ have learned into the dataset itself.
        \item For $k \in \{1..n\}$, $\cQ^k$ submits a query that consumes no budget from the per-querier filter: $\cQ^k(V_1^k) = \cM_2^k : D \mapsto 0$.
        \item $\cQ^K$ submits a query that consumes $\epsilon$ from the per-querier filter $\cF^K$: $\cQ^K(V_1^K) = \cM_2^K: D \mapsto D + \lap(1/\epsilon)$, and therefore:

          \begin{align}
            \label{eq:v2}
            V_2^K = D_1 + D_2 + Z
          \end{align}

          where $Z \sim \lap(1/\epsilon)$.
          We will show that $V^K_2$ is disproportionally more likely to be high if $\chal = 1$.
          In other words, $\cQ^K$ retrieves the information encoded into the dataset, that originally came from $\cQ^1, \dots, \cQ^n$.
      \end{enumerate}
  \end{enumerate}

  Recall that $\Pr[Z_k > 1/2] = \frac{1}{2}e^{-\epsilon / 2}$.
  Also, $\Pr[1 + Z_k > 1/2] = \Pr[-Z_k < 1/2] = 1 - \frac{1}{2}e^{-\epsilon / 2}$, since $-Z_k \sim Z_k$.
  Set $p := \frac{1}{2}e^{-\epsilon / 2}$.
  $D_2$ follows a binomial distribution:
  $D_2^{(0)} \sim \cB(n, p)$ and $D_2^{(1)} \sim \cB(n, 1-p) \sim n - \cB(n, p)$.
  Hence, \Eq\ref{eq:v2} gives:
  \begin{align}
    \begin{cases}
      V_2^{K,(0)} \sim \cB(n, p) + \lap(1/\epsilon) \\
      V_2^{K,(1)} \sim 1 + n - \cB(n, p) + \lap(1/\epsilon)
    \end{cases}
  \end{align}

  We already have $\E[V_2^{K,(0)}] = np < n/2$ and $\E[V_2^{K,(1)}] = 1 + n(1-p) > 1 + n/2$.
  We now exhibit an event for which $V^{K, (0)}_2 \ll V^{K, (1)}_2$.

  Take $B \sim \cB(n, p)$.
  Take $N \ge n +1$.
  Recall that since $Z \sim \lap(\epsilon)$, for all $z \ge 0$ and $t \in \R$ such that $z+t\ge0$, we have:
  \begin{align}
    \label{eq:lap-shift-lemma}
    \Pr[Z \ge z] = e^{\epsilon t} \Pr[Z \ge z + t]
  \end{align}

  On one hand, we have $\Pr[V^0 \ge N] = \Pr[B + Z \ge N] = \Pr[Z \ge N - B]$.
  On the other hand, $\Pr[V^1 \ge N] = \Pr[1 + n - B + Z \ge N] = \Pr[Z \ge N - 1 - n + B]$.
  Since $N - 1 - n + B \ge 0$, we have: $\Pr[Z \ge N - 1 - n + B] = e^\epsilon \Pr[Z \ge N - n + B]$.
  Hence,
  \begin{align}
    \frac{\Pr[V^1 \ge N]}{\Pr[V^0 \ge N]} = e^\epsilon \frac{\Pr[Z \ge N - n + B]}{\Pr[Z \ge N - B]}
    \label{eq:binomial_plus_laplace}
  \end{align}

  For all $k \in [0,n]$ we have $N - n + k \ge 0$ and $N - k \ge 0$, so we can compute the exact probabilities:
  {\small
  \begin{align*}
    \Pr & [Z \ge N - n + B] = \sum_{k=0}^n  {n\choose k} p^k(1-p)^{n-k} \Pr[Z \ge N - n + k]        \\
    & = \sum_{k=0}^n  {n\choose k} p^k(1-p)^{n-k} \frac{1}{2} e^{-(N-n+k)\epsilon}              \\
    & = \frac{1}{2} e^{-N\epsilon} \sum_{k=0}^n  {n\choose k} p^k(1-p)^{n-k} (e^\epsilon)^{n-k} \\
    & = (p + (1-p)e^\epsilon)^n
  \end{align*}
  }

  Similarly,
  {\small
  \begin{align*}
    \Pr & [Z \ge N - B] = \sum_{k=0}^n  {n\choose k} p^k(1-p)^{n-k}\Pr[Z \ge N - k]            \\
    & =\frac{1}{2} e^{-N\epsilon} \sum_{k=0}^n  {n\choose k} p^k(1-p)^{n-k} (e^\epsilon)^k \\
    & = (pe^\epsilon + 1-p)^n
  \end{align*}
  }

  Hence,
  \begin{align*}
    \frac{\Pr[V^1 \ge N]}{\Pr[V^0 \ge N]} = e^\epsilon \left( \frac{p + (1-p)e^\epsilon}{pe^\epsilon + 1-p} \right)^n
  \end{align*}

  Moreover, we have:
  \begin{align*}
    p + (1-p)e^\epsilon - (pe^\epsilon + 1-p) & = (1-2p)e^\epsilon - (1-2p) \\
    & = (1-2p)(e^\epsilon - 1)    \\
    & > 0
  \end{align*}
  since $p < 1/2$ and $\epsilon > 0$.
  Define $\epsilon' := \ln \left( \frac{p + (1-p)e^\epsilon}{pe^\epsilon + 1-p} \right) > 0$. Then:
  \begin{align*}
    \ln \left( \frac{\Pr[V^1 \ge N]}{\Pr[V^0 \ge N]} \right) = \epsilon + n\epsilon'
  \end{align*}

\end{proof}

\subsection{Shared Limits}
\label{sec:appendix:analysis-per-querier-shared-limits}

Since we know that data adaptivity already threatens per-querier budget semantics, we consider a simpler setting where data is known upfront. \Alg\ref{alg:shared_limits_model} presents this variant of \Alg\ref{alg:adaptivity_model_simplified}.
\Thm\ref{thm:shared-limit} then shows that adding a shared filter results in excess per-querier privacy loss, also growing linearly in the number of queriers.
The proof exhibits $n$ queriers $\cQ_1, \dots, \cQ_n$ that first ask queries. These queriers then adaptively encode what they have learned into the privacy parameters of their next queries. The shared privacy filter tracks privacy loss across queriers, so the state of that filter now depends on all queriers' past queries. Finally, a querier $\cQ_{n+1}$ can decode the information stored into the shared filter by observing whether its own queries get rejected by the shared filter. This gives $\cQ_{n+1}$ additional information about the dataset, beyond what its own per-querier filter allows. This happens without different queriers sharing their DP outputs, which is what people have referred to as collusion before. In fact, even if $\cQ_1, \dots, \cQ_n$ do not follow a particular strategy, and $\cQ_{n+1}$ doesn't know about them, $\cQ_{n+1}$'s view can still be impacted beyond what its per-querier filter allows.

\begin{algorithm}[!ht]
  \caption{Formal model for shared limits}
  \label{alg:shared_limits_model}
  \begin{algorithmic}[1] %
    \Input
    \State Database $D$
    \State Queriers $\cQ^1, \dots, \cQ^K$
    \State Per-querier filters $\cF^1, \dots, \cF^K$
    \State Shared limit $\cL$
    \EndInput

    \Output
    \State View $ V = (V^1, \dots, V^K) \in \cR^{t_{\max} \times K}$ of all $\cQ^k$
    \EndOutput

    \For{$t \in [\tmax]$}
    \For{$k \in [K]$}
    \Statex \ \ \ \ \ \graycomment{Query based on $\cQ^k$'s {\em own} previous results}
    \State $\cM_t^k \gets \cQ^k(V^k_{<t})$
    \Statex \ \ \ \ \ \graycomment{Execute if (1) $\cQ^k$ has enough budget and (2) shared limit accepts the query.}
    \If{$\cF^k(\cM_1^k, \dots, \cM_t^k) \wedge \cL(\cM_1^1, \dots, \cM_1^K, \dots \cM_t^k)$}
    \State $V^k_{t} \gets \cM_t^k(D)$
    \Else
    \State $V^k_t \gets \bot$
    \EndIf
    \EndFor
    \EndFor
  \end{algorithmic}
\end{algorithm}

\begin{numberedthm}[\ref{thm:shared-limit}]
  Take $K = n + 1 > 1$, per-querier filters with capacity $\epsilon$ and shared filter with capacity $\epsilon_g < (n + 1)\epsilon$ (allowing executions that trigger the shared limit).

  For $n$ large enough that $\lceil \frac{n}{2} e^{-\epsilon/2}\rceil < n /2$,
  there exists neighboring databases $D,D'$ such that:
  \begin{align*}
    & \exists (\cQ^1, \dots, \cQ^{n+1}), \exists S \subseteq \cR^{t_{\max}},                              \\
    & \left| \ln \left( \frac{\Pr[V^{n+1, (0)} \in S]}{\Pr[V^{n+1,(1)} \in S]} \right) \right|> \epsilon.
  \end{align*}

  Moreover, there exists $\epsilon' >0$
  and neighboring databases $D,D'$ such that
  for all $n \ge \frac{4\ln2}{(1-e^{-\epsilon/2})^2}$,
  \begin{align*}
    & \exists (\cQ^1, \dots, \cQ^{n+1}), \exists S \subseteq \cR^{t_{\max}},                                             \\
    & \left| \ln \left( \frac{\Pr[V^{n+1, (0)} \in S]}{\Pr[V^{n+1,(1)} \in S]} \right) \right| \ge \epsilon + n\epsilon'.
  \end{align*}

\end{numberedthm}

\begin{proof}
  Consider $D = 0$, $D' = 1$ from $\cN^\cX = \N$ with $\cX = \{1\}$.
  Define $\cQ^1, \dots, \cQ^n$ as follows.
  Take $\epsilon_0, \epsilon_1, \epsilon_2 >0$ that we will calibrate later.
  \begin{enumerate}
    \item At $t = 1$:
      \begin{enumerate}
        \item For $k \in \{1..n\}$, $\cQ^k$ submits $\cM_1^k: D \mapsto D + \lap(1/\epsilon_1)$, which consumes $\epsilon_1$.
        \item Then, $\cQ^{n+1}$ submits a query that consumes all but $\epsilon_0$ from its budget: $\cQ^{n+1}(\emptyset) = \cM_1^{n+1} : D \mapsto D + \lap(1/(\epsilon - \epsilon_0))$.

      \end{enumerate}
    \item At $t = 2$:
      \begin{enumerate}
        \item For $k \in \{1..n\}$, $\cQ^k$ submits a query that consumes no budget if $V_1^k$ was small, or $\epsilon_2$ if $V_1^k$ was large: $\cQ^k(V_1^k) = \cM_2^k : D \mapsto \mathds{1}[V_1^k > 1/2] (D + \lap(1/\epsilon_2))$.
          In other words, $\cQ^k$ encodes the information it learned at $t=1$ into the shared limit itself, through the adaptively chosen budget associated with its request at time $t=2$.
        \item $\cQ^{n+1}$ submits a query that attempts to consume $\epsilon_0$: $\cQ^{n+1}(V_1^{n+1}) = \cM_2^{n+1}: D \mapsto D + \lap(1/\epsilon_0)$, and therefore $V_2^{n+1} \sim
          D + \lap(1/\epsilon_0)$ if the shared limit and the per-querier filter both accept the query, and $V_2^{n+1} = \bot$ otherwise.
          In other words, $\cQ^{n+1}$ retrieves the information encoded into the shared limit that originally came from $\cQ^1, \dots, \cQ^n$'s own requests.
      \end{enumerate}
  \end{enumerate}

  Now, we configure $\epsilon_1$ and $\epsilon_2$ to ensure that no query prior to $V_2^{n+1}$ gets filtered.
  That is, $\epsilon-\epsilon_0 + n(\epsilon_1 + \epsilon_2) \le \epsilon_g = n_0 \epsilon$ and $\epsilon_1 + \epsilon_2 \le \epsilon$.
  To saturate the global filter when the first $n$ queriers all ask for $\epsilon_2$, we want:

  \begin{align}
    \label{eq:saturation}
    \epsilon-\epsilon_0 + n(\epsilon_1 + \epsilon_2) = \epsilon_g
  \end{align}

  This means that the state $f_2^{n, (\chal)}$ of the shared filter just before query $t=2, k=n+1$, on input $D^{(\chal)}$, is:
  \begin{align}
    f_2^n = n\epsilon_1 + \epsilon-\epsilon_0 + \sum_{k=1}^n \Pr[V_1^k > 1/2] \epsilon_0
  \end{align}

  Set $p = \frac{1}{2}e^{-\epsilon/2}$. We have:
  \begin{align}
    \begin{cases}
      f_2^{n,(0)} \sim n\epsilon_1 + \epsilon-\epsilon_0 + \cB(n,p) \epsilon_2 \\
      f_2^{n,(1)} \sim n\epsilon_1 + \epsilon-\epsilon_0 + (n - \cB(n,p))\epsilon_2
    \end{cases}
  \end{align}

  In particular, $\E[f_2^{n,(0)}] = \epsilon-\epsilon_0 + n(\epsilon_1 + p \epsilon_2)$ and $\E[f_2^{n,(1)}] = \epsilon-\epsilon_0 + n(\epsilon_1 + (1- p) \epsilon_2)$.
  We take $\epsilon_2$ so that $\cQ^{n+1}$ is more likely to trigger the global filter on $D^{(1)}$ and more likely to succeed on $D^{(0)}$:
  \begin{align}
    \label{eq:overflow}
    n\epsilon_1 + \epsilon-\epsilon_0 + n\epsilon_2/2 + \epsilon_0 = \epsilon_g
  \end{align}

  Subtracting \Eq\ref{eq:saturation} from \Eq\ref{eq:overflow}, we get $\epsilon_0 - n\epsilon_2/2 = 0$,
  \ie $\epsilon_2 = 2\epsilon_0/n$.

  Define $B^{(\chal)} \sim \chal(n-B) + (1-\chal)B$, the number of queriers that ask for $\epsilon_2$ at $t=2$,
  with $B \sim \cB(n,p)$ as in the proof of \Thm\ref{thm:data-adaptivity}.
  By construction, $\epsilon_0$ passes $\cQ^{n+1}$'s per-querier filter, so we can only return $\bot$ if $\epsilon_0$ hits the shared filter's capacity:

  \begin{align}
    & V_2^{n+1, (\chal)} = \bot \iff f_2^{n,(0)} + \epsilon_0 > \epsilon_g                                                           \\
    & \iff n\epsilon_1 + \epsilon-\epsilon_0 +B^{(\chal)} \epsilon_2 + \epsilon_0 > \epsilon-\epsilon_0 + n(\epsilon_1 + \epsilon_2) \\
    & \iff B^{(\chal)}\epsilon_2 + \epsilon_0> n\epsilon_2                                                                           \\
    & \iff 2 B^{(\chal)} /n +1 > 2                                                                                                   \\
    & \iff B^{(\chal)} > n/2 \label{eq:v2_bot}
  \end{align}

  We now upper-bound $\Pr[B^{(0)} > n/2]$ and lower-bound $\Pr[B^{(1)} > n/2]$.
  By Hoeffding's inequality, we have:
  \begin{align}
    \Pr[B^{(0)} > n/2] & = \Pr[B > n/2]               \\
    & \le \Pr[B \ge n/2]           \\
    & = \Pr[B - np \ge n(1/2 - p)] \\
    & \le \exp{(-2 (1/2 - p)^2 n)} \\
  \end{align}

  Similarly:
  \begin{align}
    \Pr[B^{(1)} > n/2] & = \Pr[n-B > n/2]                 \\
    & = 1 - \Pr[B \ge n/2]             \\
    & \ge 1 - \exp{(-2 (1/2 - p)^2 n)}
  \end{align}

  Finally, set $S = \{ (v, \bot), v > 1\}$.
  Since $\Pr[V_1^{n+1, (1)} > 1 ] = e^{\epsilon - \epsilon_0} \Pr[V_1^{n+1, (0)} > 1 ]$ and $V_1^{n+1, (\chal)}$ and $V_2^{n+1, (\chal)}$ are independent, we get:

  \begin{align}
    & \frac{\Pr[V^{n+1, (1)} \in S]}{\Pr[V^{n+1,(0)} \in S]} = e^{\epsilon - \epsilon_0} \frac{ \Pr[V_2^{n+1, (1)} = \bot]}{\Pr[V_2^{n+1, (0)} = \bot]} \\
    & = e^{\epsilon - \epsilon_0} \frac{\Pr[B^{(1)} > n/2]}{\Pr[B^{(0)} > n/2]}                                                                         \\
    & \ge e^{\epsilon - \epsilon_0} \left(\exp{(2 (1/2 - p)^2 n)} - 1\right)
  \end{align}

  Now, since the median of $B$ is $\lfloor np \rfloor$ or $\lceil np \rceil$, we already have $\Pr[B^{(0)} > n/2] < 1/2 < \Pr[B^{(1)} > n/2]$ when $\lceil np \rceil < n/2$, which happens for $n$ large enough or $p$ small enough (\ie $\epsilon$ large enough).

  We can get a more informative bound asymptotically.
  Set $\alpha_n := \exp{((1/2 - p)^2 n)}$.
  We have $\frac{\Pr[V^{n+1, (1)} \in S]}{\Pr[V^{n+1,(0)} \in S]} \ge \alpha_n^2 - 1$.
  For $\alpha_n$ large enough, for instance any $\alpha_n \ge 2$, we have $\alpha_n^2 - 1 \ge \alpha_n$.
  We have $\alpha_n \ge 2 \iff n \ge \frac{\ln 2}{(1/2 - p)^2} = \frac{4\ln2}{(1-e^{-\epsilon/2})^2}$, so for such $n$ we get:
  \begin{align}
    & \frac{\Pr[V^{n+1, (1)} \in S]}{\Pr[V^{n+1,(0)} \in S]} \\
    & \ge e^{\epsilon - \epsilon_0} \alpha_n                 \\
  & = e^{\epsilon - \epsilon_0} (\exp{(1/2 - p)^2)})^n
\end{align}

That is:

\begin{align}
  \left| \ln \left( \frac{\Pr[V^{n+1, (0)} \in S]}{\Pr[V^{n+1,(1)} \in S]} \right) \right| \ge \epsilon + n\epsilon' > \epsilon
\end{align}

with $\epsilon' := (1/2 - p)^2 > 0$.

\end{proof}

\heading{Remark about atomicity.}
We clarify notation for \Alg\ref{alg:adaptivity_model_simplified}, \Line\ref{line:shared-limit}, where a query is executed only if both the per-querier filter and the shared limit permit it.
The joint condition
\[
  \cF^k(\cM_1^k, \dots, \cM_t^k)
  \wedge
  \cL(\cM_1^1, \dots, \cM_1^K, \dots, \cM_t^k)
\]
means that each filter takes its decision independently.
In particular, if $\cF^k$ accepts a query $q_t^k$ while $\cL$ rejects it, $q_t^k$ will not be executed, but $\cF^k$ will not know that fact when it is time to make a decision about $q_{t+1}^k$.
That means $\cF^k$ can be overly conservative, by deducting budget for queries that are in fact not executed.
While this preserves privacy, it can hurt utility.
We propose a solution to this conundrum in \S\ref{sec:contribution2}, thanks to a new form of atomic filter that is designed to avoid spending per-querier budget needlessly.
This can be modeled by a new object $(\cF^k \circ \cL)(\cM_1^1, \dots, \cM_1^K, \dots \cM_t^k)$ that has access to all the previous queries. The shared state in this object means that per-querier budgets act as heuristics (as opposed to formal per-querier DP guarantees) unless a dedicated siloing assumption holds.

\subsection{Extension to Individual DP filters}
\label{sec:appendix:extension_idp}

The model from \Alg\ref{alg:adaptivity_model_simplified} considers standard filters \cite{filters_rogers}, where queriers can directly observe whether a query gets rejected.
However, systems such as Cookie Monster \cite{TKM+24} or ARA \cite{GHH+25} operate under individual DP filters \cite{FZ21}, which behave differently.
Because an IDP budget depends on the data itself, IDP filters cannot return $\bot$ explicitly when they are out of budget. Instead, they silently drop the out-of-budget individuals from the dataset and return a noisy answer.

This raises the question: since filter outputs are not visible under IDP, do shared limits still leak extra information?
We give a positive answer.
First, we adapt the formalism to IDP filters in \Alg\ref{alg:shared_limits_model_idp}, with changes from \Alg\ref{alg:shared_limits_model} in \textcolor{blue}{blue}.
Next, \Thm\ref{thm:shared_limits_idp} shows a variant of \Thm\ref{thm:shared-limit} for IDP filters.
Finally, \Thm\ref{thm:data-adaptivity} immediately applies since the counterexample from the proof does not trigger any filter, and therefore works identically under IDP filters.

\begin{algorithm}[!ht]
\caption{Formal model for shared limits \textcolor{blue}{with IDP filters}}
\label{alg:shared_limits_model_idp}
\begin{algorithmic}[1] %
  \Input
  \State Database $D$
  \State Queriers $\cQ^1, \dots, \cQ^K$
  \State Per-querier \textcolor{blue}{individual filters $(\cF^1_x, \dots, \cF^K_x)_{x \in D}$}
  \State Shared limit \textcolor{blue}{$(\cL_x)_{x \in D}$}
  \EndInput

  \Output
  \State View $ V = (V^1, \dots, V^K) \in \cR^{t_{\max} \times K}$ of all $\cQ^k$
  \EndOutput

  \For{$t \in [\tmax]$}
  \For{$k \in [K]$}
  \Statex \ \ \ \ \ \graycomment{Query based on $\cQ^k$'s {\em own} previous results}
  \State $\cM_t^k \gets \cQ^k(V^k_{<t})$
  \Statex \ \ \ \ \ \graycomment{Execute if (1) $\cQ^k$ has enough budget and (2) shared limit accepts the query.}
  \textcolor{blue}{
    \State $S_t^k \gets D$
    \For{$x \in D$}
    \If{$\cF^k_x(\cM_1^k, \dots, \cM_t^k) \wedge \cL_x(\cM_1^1, \dots, \cM_1^K, \dots \cM_t^k)$}
    \Statex \ \ \ \ \ \ \ \ \graycomment{Drop out-of-budget records.}
    \State $S_t^k \gets S_t^k \setminus\{x\}$ \label{line:idp_drop}
    \EndIf
    \EndFor
    \State $V^k_{t} \gets \cM_t^k(S_t^k)$
  }
  \EndFor
  \EndFor
\end{algorithmic}
\end{algorithm}

\begin{theorem}
\label{thm:shared_limits_idp}
Take $K = n + 1 > 1$, per-querier individual filters with capacity $\epsilon$ and shared individual filter with capacity $\epsilon_g < (n + 1)\epsilon$ (allowing executions that trigger the shared limit).

There exists $\epsilon' >0$
and neighboring databases $D,D'$ such that
\begin{align*}
  & \exists (\cQ^1, \dots, \cQ^{n+1}), \exists S \subseteq \cR^{t_{\max}},                                             \\
  & \left| \ln \left( \frac{\Pr[V^{n+1, (0)} \in S]}{\Pr[V^{n+1,(1)} \in S]} \right) \right| \ge \epsilon + n\epsilon'.
\end{align*}

\end{theorem}

\begin{proof}
We modify the execution from the proof for \Thm\ref{thm:shared-limit} to match IDP semantics, where records get dropped conditionally on past results.
This emulates a form of data adaptivity, except that datasets $S_2^{n+1}$ now differ because of shared filters for various records, instead of through a data generation process.
We then reuse the calculation from the proof of \Thm\ref{thm:shared-limit}.

The changes in execution are as follows:
\begin{itemize}
  \item $D'$ contains $n+1$ distinct records $\{x_0, x_1, \dots, x_{n}\}$ and $D$ only contains $\{x_1, \dots, x_{n}\}$. We no longer have $D \in \N$, instead $D \in \N^\cX$ with $\cX = \{x_0,\dots, x_{n}\}$.
  \item At $t=1$, queriers submit $q_1^k: D \mapsto |\{x \in D: x = x_0\}| + \lap(1/\epsilon)$. This consumes $x_0$'s full individual budget.
  \item At $t=2$:
    \begin{itemize}
      \item For $k\in[n]$, querier $k$ requests $x_k$ with full budget, conditionally on past results. That is:

      $\cQ^k(V_1^k) = \cM_2^k : D \mapsto \mathds{1}[V_1^k \le 1/2] \cdot (|\{x \in D: x = x_k\}| + \lap(1/\epsilon))$

    \item Finally, querier $n+1$ requests all the records $x_1, \dots, x_n$:

      $\cQ^{n+1}(V_1^{n+1}) = \cM_2^{n+1}: D \mapsto |\{x \in D: x \in \{x_1, \dots, x_{n}\}\}| + \lap(1/\epsilon)$
  \end{itemize}
\end{itemize}

Since $\{x_1, \dots, x_{n}\}$ are present in both worlds, $V^{{n+1},(1)}$ and $V^{{n+1},(0)}$ differ only by the points that get dropped at \Line\ref{line:idp_drop}.
Point $x_k$ gets dropped only if $V_1^k \le 1/2$, which is more likely to happen in world 0 (on input $D$).
More precisely, we have:
\begin{align}
V^{{n+1},(0)} \sim B + Z \text{ and } V^{{n+1},(1)} \sim n - B + Z
\end{align}

with $B \sim \cB(n,p)$ with $p = \frac{1}{2}\exp(-\epsilon/2)$ and $Z \sim \lap(1/\epsilon)$.

This is exactly the same situation as \Eq\ref{eq:binomial_plus_laplace}, where the $e^\epsilon$ factor comes from the privacy loss at $t=1$.
We conclude the same way.

\end{proof}

\subsection{Querier-Siloing Assumption}
\label{sec:appendix:siloing}

\begin{lemma}
Consider \Alg\ref{alg:adaptivity_model_simplified} with $K \in \N$, per-querier filters with capacity $\epsilon > 0$ and shared limit $\cL$.
Suppose that the following assumption holds:
\begin{align}
\label{eq:siloing-assumption}
&\forall k \in [K], \forall u, w \in \cR^{t}:  u^k = w^k,\\
&\cL\left(\cM_1^1, \dots, \cM_1^K, \dots, \cM_t^1(u^1_{<t}), \dots \cM_t^k(u^k_{<t})\right) \cdot \cM_t^k\left(D_{\le t}(u_{<t})\right) \nonumber \\
&=\cL\left(\cM_1^1, \dots, \cM_1^K, \dots, \cM_t^1(w^1_{<t}), \dots \cM_t^k(w^k_{<t})\right) \cdot \cM_t^k\left(D_{\le t}(w_{<t})\right) \nonumber
\end{align}

Then, for all $k \in [K]$:
\begin{align*}
\forall x_0 \in \cX, \forall t_0 \in [\tmax], \forall \cG, \forall (\cQ^1, \dots, \cQ^{n+1}), \\
\forall S \subseteq \cR^{t_{\max}}, \left| \ln \left( \frac{\Pr[V^{k,(0)} \in S]}{\Pr[V^{k,(1)} \in S]} \right) \right| \le \epsilon.
\end{align*}

In other words, if querier $k$'s results don't depend on other queriers, either through data or shared limits, then \Alg\ref{alg:adaptivity_model_simplified} is per-querier $\epsilon$-DP for querier $k$.

\end{lemma}
\begin{proof}
Take $x \in \cX$, $\cG$, $\cQ^1, \dots, \cQ^K$, $k \in [K]$, and $v^k \in \cR^\tmax$.
We have:

\begin{align*}
\Pr[V^{k} = v^k] &= \prod_{t=1}^{\tmax} \Pr[V_t^k = v_t^k | V_1^k = v_1^k, \dots, V_{t-1}^k = v_{t-1}^k]
\end{align*}

Take $t \in [\tmax]$.
We now express $V_t^k$ in terms of queries and filter decisions, since the filter properties are necessary for the privacy proof.
Without loss of generality, we treat filters as functions that return 0 for reject and 1 for accept at \Line\ref{line:shared-limit}, and we output $0$ instead of $\bot$ at \Line\ref{line:v_bot}.
Hence, we can express $V_t^k$ as:

\begin{align}
\label{eq:v_with_filters}
V_t^k = \cF^k(\cM_1^k, \dots, \cM_t^k) \cdot \cL(\cM_1^1, \dots, \cM_1^K, \dots \cM_t^k) \cdot \cM_t^k(D_{\le t})
\end{align}

The terms in this expression depend on various variables:
\begin{itemize}
\item     Once we condition on $v_1^k, \dots, v_{t-1}^k$, we know the queries asked by $\cQ^k$:

  \begin{align}
    q_1^k = \cQ^k(), q_2^k = \cQ^k(v_1^k), \dots, q_t^k = \cQ^k(v_1^k, \dots, v_{t-1}^k).
  \end{align}

\item As a consequence, $\cF^k(\cM_1^k, \dots, \cM_t^k)$ is also fully determined by $v_1^k, \dots, v_{t-1}^k$.
\item However, $\cL(\cM_1^1, \dots, \cM_1^K, \dots \cM_t^k)$ can depend on other queriers' past results $V^i_t$ for $i\neq k$, since these outputs are used to select $\cM_t^i$.
\item Similarly, $D_{\le t}$ can also depend on other queriers' past results.
\end{itemize}

Pose $V^{\neq k} := (V_i^j)_{i \in [t], j \in [K] \setminus \{k\}}$.
By the law of total probability, we have:

\begin{align}
\Pr [V_t^k = v_t^k \mid {} & V^k_{<t} = v^k_{<t}] \nonumber \\
= \int_{v^{\neq k} \in \cR^{t \times (K - 1)}} &\Pr[V_t^k = v_t^k \mid V^k_{<t} = v^k_{<t}, V^{\neq k} = v^{\neq k}] \nonumber \\
&\cdot \Pr[V^{\neq k} = v^{\neq k}] dv^{\neq k} \label{eq:vk_integral}
\end{align}

Take any $v^{\neq k} \in \cR^{t \times (K - 1)}$.
Define $u \in \cR^{t \times K}$ by concatenating $v^k$ and $v^{\neq k}$ such that $u^k = v^k$ and $u^{\neq k} = v^{\neq k}$.
Thanks to \Eq\ref{eq:v_with_filters} with explicit conditioning on all past results, we have:

\begin{align}
\Pr&[V_t^k = v_t^k | V_1^k = v_1^k, \dots, V_{t-1}^k = v_{t-1}^k, V^{\neq k} = v^{\neq k}]  \nonumber \\
= & ~\cF^k(\cM_1^k, \dots, \cM_t^k(u^k_{<t})) \nonumber \\
&\cdot \cL\left(\cM_1^1, \dots, \cM_1^K, \dots, \cM_t^1(u^1_{<t}), \dots \cM_t^k(u^k_{<t})\right) \nonumber \\
&\cdot \cM_t^k\left(D_{\le t}(u_{<t})\right) \label{eq:vk}
\end{align}

Define $\tilde v \in \cR^{t \times K}$ by concatenating $v^k$ and $(\bot, \dots, \bot) \in \cR^{t \times (K - 1)}$.
We have $\tilde v^k = u^k$, so by \Eq\ref{eq:siloing-assumption}, \Eq\ref{eq:vk} becomes:

\begin{align}
\Pr&[V_t^k = v_t^k | V_1^k = v_1^k, \dots, V_{t-1}^k = v_{t-1}^k, V^{\neq k} = v^{\neq k}] \\
= & ~\cF^k(\cM_1^k, \dots, \cM_t^k(\tilde v^k_{<t})) \\
&\cdot \cL\left(\cM_1^1, \dots, \cM_1^K, \dots, \cM_t^1(\tilde v^1_{<t}), \dots \cM_t^k(\tilde v^k_{<t})\right) \\
&\cdot \cM_t^k\left(D_{\le t}(\tilde v_{<t})\right)
\end{align}

This term is independent of $v^{\neq k}$, so we can express it as a function of $v^k_{<t}$ and take it out of the integral at \Eq\ref{eq:vk_integral}, which gives:

\begin{align}
\Pr &[V_t^k = v_t^k | V_1^k = v_1^k, \dots, V_{t-1}^k = v_{t-1}^k] \\
&= \left(\cF^k(\cM_1^k, \dots, \cM_t^k) \cdot \cL(\cM_1^1, \dots, \cM_t^k)  \cdot \cM_t^k(D_{\le t})\right) \left(v^k_{<t}\right)
\end{align}

Now we can bound the per-querier privacy loss:
\begin{align}
&\left|\ln\left( \frac{\Pr[V^{k} = v^k]}{\Pr[V^{k} = v^k]}\right)\right| \nonumber \\
&=\left|\ln\left(\prod_{t=1}^{\tmax} \frac{\Pr[V_t^{k, (0)} = v_t^k | V_1^{k, (0)} = v_1^k, \dots, V_{t-1}^{k, (0)} = v_{t-1}^k]}{\Pr[V_t^{k, (1)} = v_t^k | V_1^{k, (1)} = v_1^k, \dots, V_{t-1}^{k, (1)} = v_{t-1}^k]}\right)\right| \nonumber \\
&\le \sum_{t=1}^{\tmax} \left|\ln \left(
  \frac{\Pr\left[\substack{\cF^{k}(\cM_1^k, \dots, \cM_t^k) \cdot \cL(\cM_1^1, \dots, \cM_t^k) \\
  {}\cdot \cM_t^k(D_{\le t}^{(0)}) \{v^k_{<t}\} = v_t^k}\right]}
  {\Pr\left[\substack{\cF^{k}(\cM_1^k, \dots, \cM_t^k) \cdot \cL(\cM_1^1, \dots, \cM_t^k) \\
  {}\cdot \cM_t^k(D_{\le t}^{(1)}) \{v^k_{<t}\} = v_t^k}\right]}
\right)\right|\nonumber \\
& \le \sum_{t=1}^{\tmax} \left|\ln \left(
  \frac{\Pr[\cF^{k}(\cM_1^k, \dots, \cM_t^k)  \cdot \cM_t^k(D_{\le t}^{(0)}) \{v^k_{<t}\} = v_t^k]}{\Pr[\cF^{k}(\cM_1^k, \dots, \cM_t^k) \cdot \cM_t^k(D_{\le t}^{(1)}) \{v^k_{<t}\} = v_t^k]}
\right)\right| \label{eq:total_privacy_loss}
\end{align}

The last inequality comes from what follows.
$\cL(\cM_1^1, \dots, \cM_t^k)$ is entirely determined by $v^k_{<t}$ and thus identical in the numerator and denominator.
For $t$ where $\cL(\cM_1^1, \dots, \cM_t^k) = 1$, the $\cL$ factor is one and can be omitted from the product.
For $t$ where $\cL(\cM_1^1, \dots, \cM_t^k) = 0$, we have:
\begin{align*}
&\left|\ln \left(
  \frac{\Pr\left[\substack{\cF^{k}(\cM_1^k, \dots, \cM_t^k) \cdot \cL(\cM_1^1, \dots, \cM_t^k) \\
  {}\cdot \cM_t^k(D_{\le t}^{(0)}) \{v^k_{<t}\} = v_t^k}\right]}
  {\Pr\left[\substack{\cF^{k}(\cM_1^k, \dots, \cM_t^k) \cdot \cL(\cM_1^1, \dots, \cM_t^k) \\
  {}\cdot \cM_t^k(D_{\le t}^{(1)}) \{v^k_{<t}\} = v_t^k}\right]}
\right)\right|\\
&= 0 \le
\left|\ln \left( \frac{\Pr[\cF^{k}(\cM_1^k, \dots, \cM_t^k)  \cdot \cM_t^k(D_{\le t}^{(0)}) \{v^k_{<t}\} = v_t^k]}{\Pr[\cF^{k}(\cM_1^k, \dots, \cM_t^k) \cdot \cM_t^k(D_{\le t}^{(1)}) \{v^k_{<t}\} = v_t^k]}\right)\right|
\end{align*}

The rest of the proof is identical to standard filter guarantees on adaptively generated data.
Suppose the filter uses an upper bound $\epsilon_i$ on the privacy loss for each query at step $i \in [\tmax]$:

\begin{align}
\left|\ln \left( \frac{\Pr[\cM_i^k(D_{\le i}^{(0)}) \{v^k_{<i}\} = v_i^k]}{\Pr[\cM_i^k(D_{\le i}^{(1)}) \{v^k_{<i}\} = v_i^k]}  \right)\right|\le \epsilon_i(v^k_{<i})
\end{align}

Then, by definition of pure DP filters, we get the following recursive property for $t \in [\tmax]$:
$\cF^{k}(\cM_1^k, \dots, \cM_t^k) = 1$ iff $\sum_{i < t} \cF^{k}(\cM_1^k, \dots, \cM_t^k) \cdot \epsilon_i(v^k_{<i}) + \epsilon_t(v^k_{<t})  \le \epsilon$.

We conclude by applying this property at $t = \tmax$ in \Eq\ref{eq:total_privacy_loss}:

\begin{align*}
\left|\ln\left( \frac{\Pr[V^{k} = v^k]}{\Pr[V^{k} = v^k]}\right)\right| \le \epsilon
\end{align*}
\end{proof}

\begin{theorem}\label{thm:siloing-assumption}
Suppose that the following two assumptions hold:

\begin{enumerate}
\item Querier $k$'s queries operate on a restricted data stream that only depends on its own past results. That is, $D_t \gets \cG(V^1_{<t}, \dots V^K_{<t})$ is of the following form:

  $$ D_t^1 \gets \cG(V^1_{<t}), \dots, D_t^K \gets \cG(V^K_{<t})$$

  with $q_t^k(D_{\le t}) = q_t^k(D_{\le t}^k)$.
\item Shared limits do not filter $k$'s reports depending on other queriers' results:

  \begin{align*}
    &\forall k \in [K], \forall u, w \in \cR^{t}:  u^k = w^k,\\
    &\cL\left(\cM_1^1, \dots, \cM_1^K, \dots, \cM_t^1(u^1_{<t}), \dots \cM_t^k(u^k_{<t})\right) \\
    &= \cL\left(\cM_1^1, \dots, \cM_1^K, \dots, \cM_t^1(w^1_{<t}), \dots \cM_t^k(w^k_{<t})\right)
  \end{align*}

  Sufficient conditions for this property include queriers being non-adaptive, never reaching the shared limit, or operating on disjoint sets of devices (for on-device accounting with IDP filters).
\end{enumerate}

Then condition \Eq\ref{eq:siloing-assumption} holds, and thus per-querier guarantees hold.
\end{theorem}
\begin{proof}
The first point implies:

$$q_t^k(v_{<t}^k)(D_{\le t}(v_{<t})) = q_t^k(v_{<t}^k)(D_{\le t}(v_{<t}^k))$$

Combining this with the second point, we directly get the assumption:
\begin{align}
&\forall k \in [K], \forall u, w \in \cR^{t}:  u^k = w^k,\\
&\cL\left(\cM_1^1, \dots, \cM_1^K, \dots, \cM_t^1(u^1_{<t}), \dots \cM_t^k(u^k_{<t})\right) \cdot \cM_t^k\left(D_{\le t}(u_{<t})\right) \nonumber \\
&=\cL\left(\cM_1^1, \dots, \cM_1^K, \dots, \cM_t^1(w^1_{<t}), \dots \cM_t^k(w^k_{<t})\right) \cdot \cM_t^k\left(D_{\le t}(w_{<t})\right) \nonumber
\end{align}

\end{proof}

\section{Global IDP with \sysname (\S\ref{sec:contribution2} Proofs)}
\label{appendix:online-filter-management}

\subsection{Data and query model for \sysname}
\label{appendix:epoch-site-sensitivity}
\label{appendix:data-model}\label{appendix:query-model}

This section sets up the formalism that underpins the main theorems, proved in subsequent sections and evoked in the body of the paper.
We begin by defining \sysname's data and query model, broadly adapted from Cookie Monster's formalism.
We then provide useful properties that will be needed in subsequent sections. Indeed, while Cookie Monster analyzes global and individual sensitivities of queries at the device-epoch level, \sysname additionally needs such analyses at the device-epoch-site level.

\begin{definition}[User actions, sites and events]
    \label{def:user-action-context}
    Let $\cU$ be the set of user-action identifiers, $\cS$ the set of site identifiers, $\cI$ the set of impression payloads and $\cC$ the set of conversion payloads.

    An {\em event} is either an impression event $(u,i,\text{imp})$ or a conversion event $(u,c,\bfb, \text{conv})$:
    \begin{itemize}
        \item $u \in \cU$ is the user action that caused the impression or conversion. The same user action can cause multiple events, and is denoted $\mathsf{uaCtx}$ in the body of the paper. On a fixed device, all accepted impression events carrying the same user-action identifier occur in a single epoch.
        \item  $i \in \cS$ (resp. $c \in \cS$) is the impression  (resp. conversion) site where the event occurred.
        \item Conversions contain the set of queriers $\bfb \subset \cS$ that will receive the conversion report.
              \ifdoubleblind
              \else  %
                  ($\bfb$ is a singleton unless we use the optimization from \S\ref{appendix:cross-report-optimization}.)
              \fi
        \item The payload $\text{imp} \in \cI$ or $\text{conv} \in \cC$ can be used to define concrete attribution functions.
    \end{itemize}

    We write sets of all impression events and all conversion events as follows:
    \begin{align*}
        \Ev^{\mathsf{imp}}  & := \cU\times\cS\times\cI,               &
        \Ev^{\mathsf{conv}} & := \cU\times\cS\times\cP(\cS)\times\cC,
    \end{align*}
    and we write $\Ev := \Ev^{\mathsf{imp}}\cup\Ev^{\mathsf{conv}}$ for the set
    of all events.

\end{definition}

\begin{definition}[Database of device-epoch records]
    \label{def:db-per-site}
    Let $\cD$ be the set of device identifiers and $\cE$ the set of epoch identifiers.
    A database $D$ is a set of device-epoch records where each record $x = (d,e,F) \in \cX = \cD \times \cE \times \cP(\Ev)$ contains a device $d$, an epoch $e$ and a set of impression and conversion events $F \subset \Ev$.
    We write $\D$ for the set of all databases where each $(d,e)$ pair appears at most once.
\end{definition}

\begin{definition}[Conversion-site and impression-site event sets from events] \label{def:site-sets}
    For any event set $F$, define:
    \begin{align*}
        \mathsf{ImpSites}(F) := \{  & i \in \cS : \exists u\in \cU, \text{imp} \in \cI, (u,i,\text{imp}) \in F\}, \\
        \mathsf{ConvSites}(F) := \{ & c \in \cS: \exists u\in \cU, \bfb \subset \cS, \text{conv} \in \cC,         \\
                                    & (u,c, \bfb, \text{conv}) \in F \}.
    \end{align*}
    Furthermore, for any user action $u \in \cU$, define the user-action-scoped variants of these sets:
    \begin{align*}
        \mathsf{ImpSites}_u(F)
        := \{ & i \in \cS : \exists \text{imp} \in \cI,(u,i,\text{imp}) \in F\}, \\
        \mathsf{ConvSites}_u(F)
        := \{ & c \in \cS: \exists \bfb \subset \cS, \text{conv} \in \cC,        \\
              & (u,c, \bfb, \text{conv}) \in F \},
    \end{align*}
    and let $\mathsf{AccessedSites}_u(F) := \mathsf{ImpSites}_u(F) \cup \mathsf{ConvSites}_u(F)$.
    We also write $\mathsf{ImpActions}(F) := \{u \in \cU : \exists i \in \cS, \text{imp} \in \cI, (u,i,\text{imp}) \in F\}$ for the user actions that saved the impressions of $F$.
\end{definition}

We now define queries and reports, with a slight adaptation of Cookie Monster's definitions to our per-site semantics.
While Cookie Monster comes with an arbitrary set of public events $P$ for each querier $b$, here for simplicity we assume that all the conversions for a querier are public.
Using Cookie Monster's terminology, that means we set $P := \{(u,c,\bfb,\text{conv}) \in \Ev^{\mathsf{conv}} : b \in \bfb\}$.
Also, while Cookie Monster defines a set of {\em relevant events}, potentially including conversions, in \Def\ref{def:attribution-function} we only consider {\em relevant impressions} for simplicity. In particular, this hardcodes "Case 1" from \cite[\Thm 1]{TKM+24}.

\begin{definition}[Neighboring databases]\label{def:neighboring-databases}
    $D \sim_x D'$ iff database $D'$ is obtained from $D$ by removing all impression events of a single device-epoch $x$. All conversion events are identical for $D$ and $D'$.
\end{definition}

\begin{definition}[Attribution function, adapted from Cookie Monster]
    \label{def:attribution-function}
    Fix a set of {\em relevant impressions} $F_A \subset  \Ev^{\mathsf{imp}} $ such that there exists $G_A \subset \cS \times \cI$ with $F_A = \cU \times G_A$ (\ie user actions play no role in defining relevant impressions).
    Fix $k,m \in \N^*$ where $k$ is a number of epochs.
    An {\em attribution function} is a function $A : \cP(\Ev)^k \to \R^m$ that takes $k$ event sets $F_1, \dots, F_k$ from $k$ epochs and outputs an $m$-dimensional vector $A(F_1, \dots, F_k)$, such that only {\em relevant events} contribute to $A$. That is, for all $(F_1, \dots, F_k) \in \cP(\Ev)^k$, we have:
    \begin{align}
        A(F_1, \dots, F_k) = A(F_1 \cap F_A, \dots, F_k \cap F_A).
        \label{eq:relevance-identity}
    \end{align}
    In particular, conversion events never contribute to $A$.

\end{definition}

\begin{definition}[Report identifier and attribution report, same as Cookie Monster]
    \label{def:attribution-report}
    Fix a domain of report identifiers $\Z$.
    Consider a mapping $d(\cdot): \Z \to \cD$ from report identifiers to devices that gives the device $d_r$ that generated a report $r$.

    Given an attribution function $A$, a set of epochs $E$ and a report identifier $r \in \Z$,
    the {\em attribution report} $\rho_{r, A, E}$, or $\rho_r$ for short, is a function over the whole database $D$ defined by:

    \begin{align}
        \rho_r: D\in \D \mapsto A(D^{ E}_{d_r}),
        \label{eq:attribution-report}
    \end{align}
    When several reports are in play, we write $A_r$, $E_r$ and $F_{A_r}$ for the attribution function, the set of epochs and the set of relevant impressions attached to report $r$, so that $\rho_r: D \mapsto A_r(D^{E_r}_{d_r})$.
    The mechanism also stores, with each identifier, the \convsite $c_r$ and the opaque action context $u_r$ of the \code{measureConversion} request that created $r$ (\Alg\ref{alg:e2e_setup}). Neither changes the function $\rho_r$: $u_r$ is only used for the per-action domain cap.

\end{definition}

\begin{definition}[Query, same as Cookie Monster]
    \label{def:query}
    Consider a set of report identifiers $R \subset \Z$, and a set of attribution reports $(\rho_r)_{r\in R}$ each with output in $\R^m$.
    The {\em query} for $(\rho_r)_{r\in R}$ is the function $Q: \D \to \R^m$ defined as $Q(D) := \sum_{r \in R} \rho_r(D)$ for $D \in \D$.
\end{definition}

\begin{definition}[Device-epoch sensitivity, same as Cookie Monster]
    \label{def:device-epoch-global-sensitivity}
    Fix a report $\rho: \D \to \R^m$ for some $m$. We define the {\em individual $L_1$ sensitivity of $\rho$ for a device-epoch $x \in \cX$} as follows:
    \begin{align} \Delta_x(\rho) := \max_{D, D'\in \mathbb{D}: D' \sim_{x} D} \| \rho(D) - \rho(D') \|_1.
    \end{align}

    We also define the  {\em device-epoch global $L_1$ sensitivity of $\rho$} as follows:

    \begin{align} \Delta(\rho) := \max_{x\in \mathcal{X}} \Delta_{x}(\rho)
    \end{align}

\end{definition}

\begin{definition}[Device-epoch-impression-site sensitivity]
    \label{def:device-epoch-site-global-sensitivity}
    Fix a report $\rho: \D \to \R^m$ for some $m$, and an impression site $i\in\cS$.
    We define the {\em individual $L_1$ sensitivity of $\rho$ for a device-epoch-site $x\in\cX, i \in \cS$} as follows:

    \begin{align} \Delta_{x,i}(\rho) := \max_{D, D'\in \mathbb{D}: D' \sim_{x,i} D} \| \rho(D) - \rho(D') \|_1
    \end{align}

    where $D \sim_{x,i} D'$ iff database $D'$ is obtained from $D$ by removing all impression events on a single site $i$ of a single device-epoch $x$. Equivalently, there exists $D_0 \in \D$, a record $x \in \cX$ such that $\{D,D'\} = \{D_0 + x, D_0 + x^{i \to \emptyset}\}$, where $x^{i \to \emptyset} := (d,e,F^{i \to \emptyset})$ is the record obtained by removing all the impressions on $i$ from $x$.

    We also define the  {\em device-epoch-site global $L_1$ sensitivity of $\rho$}  as follows:

    \begin{align} \Delta_i(\rho) := \max_{x\in \mathcal{X}} \Delta_{x,i}(\rho)
    \end{align}
\end{definition}

To simplify subsequent results, we define some notation:
\begin{definition}[Zeroing-out]
    Fix a vector of impression sets $\bfF = (F_1, \dots, F_k) \in \cP( \Ev^{\mathsf{imp}} )^k$ for any $k > 0$.
    For $i \in \cS$ and $j \in [k]$ we define:
    \begin{itemize}
        \item $\mathbf{F}^{j\rightarrow \emptyset} := (F_1, \dots, F_{j-1}, \emptyset, F_{j+1}, \dots, F_k)$, \ie we zero-out the $j$th epoch.
        \item $\mathbf{F}^{j,i\rightarrow \emptyset} := (F_1, \dots, F_{j-1}, F_j \setminus \cI_i, F_{j+1}, \dots, F_k)$, \ie we zero-out all the impressions $\cI_i := \{(u, i, \text{imp}): u\in\cU, \text{imp} \in \cI\}$ belonging to site $i$ from the $j$th epoch.
    \end{itemize}

\end{definition}

Plugging \Def\ref{def:attribution-function} into \Def\ref{def:device-epoch-global-sensitivity} and \Def\ref{def:device-epoch-site-global-sensitivity} immediately gives:
\begin{lemma}[Global sensitivity of reports]
    \label{lem:sensitivity_exact_formula}
    Fix a device $d$, a set of $k$ epochs $E$, an attribution function $A$ and the corresponding report $\rho: D \mapsto A(D^E_d)$. We have:

    \begin{align}
        \Delta(\rho) = \max_{\substack{\bfF \in \cP( \Ev^{\mathsf{imp}} )^k, j \in [k]}} \|A(\bfF) - A(\bfF^{j \to \emptyset})\|_1. \label{eqn:global_sensitivity_guarantee}
    \end{align}
    Moreover, for any impression site $i\in\cS$ we have:
    \begin{align}
        \label{eq:site-level-global-sensitivity}
        \Delta_i(\rho) =  \max_{\substack{\bfF \in \cP( \Ev^{\mathsf{imp}} )^k, j \in [k]}} \|A(\bfF) - A(\bfF^{j,i \to \emptyset})\|_1.
    \end{align}
\end{lemma}

\begin{theorem}[Individual sensitivity of reports per epoch-site]\label{thm:individual-sensitivity-of-reports}
    Fix a report identifier $r$, with device $d_r$, epochs $E_r = \{e_1^{(r)}, \ldots, e_k^{(r)}\}$, attribution function $A_r$ and relevant impressions $F_{A_r}$, and let $\rho_r$ be the corresponding report from \Eq\ref{eq:attribution-report}.
    Fix an impression site $i \in \cS$ and a device-epoch record $x = (d,e,F) \in \cX$, and write $F_i := F \cap \cI_i$ for the impressions of $x$ on site $i$.
    Finally, let
    \begin{align}
        I_r(x) := \mathsf{ImpSites}(F \cap F_{A_r})
    \end{align}
    be the impression sites of $x$ that can contribute to $\rho_r$. We can upper bound the individual sensitivity of $\rho_r$ per epoch-site as follows:
    \begin{align}
        \Delta_{x,i}(\rho_r) \leq
        \begin{cases}
            0
             & \text{if $d\neq d_r$, $e\notin E_r$,}       \\
             & \text{or $i\notin I_r(x)$,}                 \\
            \|A_r(F_i) - A_r(\emptyset)\|_1
             & \text{if $d = d_r$, $E_r=\{e\}$}            \\
             & \text{and $I_r(x)=\{i\}$,}                  \\
            \Delta_i(\rho_r)
             & \text{if $d = d_r$, $e\in E_r$,}            \\
             & \text{$i\in I_r(x)$, and}                   \\
             & \text{$|E_r| \geq 2$ or $|I_r(x)| \geq 2$.}
        \end{cases}\label{eqn:single-epoch-ind-sensitivity}
    \end{align}
\end{theorem}

\begin{proof}
    By \Def\ref{def:device-epoch-site-global-sensitivity}, each pair $(D,D')$ in the maximum defining $\Delta_{x,i}(\rho_r)$ satisfies $\{D,D'\} = \{D_0 + x, D_0 + x^{i \to \emptyset}\}$ for some $D_0 \in \D$, where $x^{i \to \emptyset} = (d,e,F\setminus \cI_i)$. Since $D_0 + x \in \D$, the database $D_0$ has no record at $(d,e)$. W.l.o.g we assume $D = D_0 + x$ and $D' = D_0 + x^{i \to \emptyset}$. We now take the three cases in turn.

    \begin{itemize}
        \item {\em Case 1: $d\neq d_r$, or $e\notin E_r$, or $i\notin I_r(x)$.} If $d \neq d_r$, then $D$ and $D'$ only differ in a record of device $d$, so $D^{E_r}_{d_r} = (D_0)^{E_r}_{d_r} = (D')^{E_r}_{d_r}$. If $e \notin E_r$, then $D$ and $D'$ only differ in an epoch that $\rho_r$ does not read, and again $D^{E_r}_{d_r} = (D')^{E_r}_{d_r}$. Finally, suppose $i \notin I_r(x)$, and assume $d = d_r$ and $e \in E_r$. The records at $(d,e)$ do differ, but not in any relevant impression: $F \cap F_{A_r}$ contains no impression on site $i$, so
              \begin{align}
                  (F\setminus \cI_i) \cap F_{A_r} = (F \cap F_{A_r}) \setminus \cI_i = F \cap F_{A_r}.
              \end{align}
              Then \Eq\ref{eq:relevance-identity} gives $A_r(D^{E_r}_{d_r}) = A_r((D')^{E_r}_{d_r})$.

        \item {\em Case 2: $d = d_r$, $E_r=\{e\}$ and $I_r(x)=\{i\}$.} Here $k=1$ and $\rho_r(D) = A_r(D^{e}_{d_r})$. As $D_0$ has no record at $(d_r,e)$, we get $D^{e}_{d_r} = F$ and $(D')^{e}_{d_r} = F \setminus \cI_i$. Since site $i$ has all the relevant impressions of $x$, we have $F \cap F_{A_r} \subseteq \cI_i$, and therefore
              \begin{align}
                  F \cap F_{A_r}                = F_i \cap F_{A_r}, &  &
                  (F\setminus \cI_i)\cap F_{A_r} = \emptyset.
              \end{align}
              \Eq\ref{eq:relevance-identity} then gives $A_r(F) = A_r(F_i)$ and $A_r(F \setminus \cI_i) = A_r(\emptyset)$, so every pair $(D,D')$ yields the same value $\|A_r(F_i) - A_r(\emptyset)\|_1$.

        \item {\em Case 3: $d = d_r$, $e\in E_r$, $i\in I_r(x)$, and $|E_r| \geq 2$ or $|I_r(x)| \geq 2$.} Let $j \in [k]$ be the index of $e$ in $E_r$ and $\bfF := D^{E_r}_{d_r}$, so that $(D')^{E_r}_{d_r} = \bfF^{j,i\to\emptyset}$ and
              \begin{align}
                  \|\rho_r(D) - \rho_r(D')\|_1 = \|A_r(\bfF) - A_r(\bfF^{j,i\to\emptyset})\|_1 \leq \Delta_i(\rho_r),
              \end{align}
              by \Eq\ref{eq:site-level-global-sensitivity}. Unlike in the first two cases, this quantity depends on $D_0$ (through the events that the other epochs of $E_r$ contribute to $\bfF$). Taking the maximum over $D_0$ can only give back the global bound $\Delta_i(\rho_r)$ in the worst case.
    \end{itemize}
\end{proof}

Next, we show that $\Delta_i(\rho) \leq 2\Delta(\rho)$ for any report.
For reports using certain attribution functions, we can have a tighter $\Delta_i(\rho) = \Delta(\rho)$ bound, but it does not hold in general.

\begin{lemma}[Relation between Device-Epoch-Site and Device-Epoch Sensitivities]
    \label{lem:relationship-des-and-de-sensitivities}
    For any report $\rho$ with attribution function $A$, let $F$ denote the full dataset, $F_j$ denote all data in $F$ pertaining to epoch $j$, and $F_{j,i}$ denote all data in $F_j$ pertaining to site $i$. The following inequality holds:
    \begin{align}
        \Delta_i(\rho) \leq 2\Delta(\rho)
    \end{align}
\end{lemma}

\begin{proof}
    Recall the definition of device-epoch-site global sensitivity:
    \begin{align}
        \Delta_i(\rho) = \max_{\substack{\bfF \in \cP( \Ev^{\mathsf{imp}} )^k, j \in [k]}} \|A(\bfF^{j,i \to \emptyset}) - A(\mathbf{F})\|_1.
    \end{align}

    We can decompose this expression using the triangle inequality:
    \begin{align}
         & \|A(\bfF^{j,i \to \emptyset}) - A(\mathbf{F})\|_1                                                                                                                  \\
        =
         & \|A(\bfF^{j,i \to \emptyset}) - A(\bfF^{j \to \emptyset}) + A(\bfF^{j \to \emptyset}) - A(\mathbf{F})\|_1                                                          \\
        \leq
         & \|A(\bfF^{j,i \to \emptyset}) - A(\bfF^{j \to \emptyset})\|_1 + \|A(\bfF^{j \to \emptyset}) - A(\mathbf{F})\|_1\label{eqn:trig-ineq-upper-bound-for-site-glob-sen}
    \end{align}

    For the first term, note that $A(\bfF^{j,i \to \emptyset})$ uses events $F_{j,\neg i}$ from epoch $j$ where $F_{j,\neg i} = F_j \setminus F_{j,i}$, while $A(\bfF^{j \to \emptyset})$ uses no events from epoch $j$. Since $F_{j,\neg i} \subseteq F_j$ and all events from epoch $j$ for a given device come from a single device-epoch record, we can view this as the change from adding a single record containing events $F_{j,\neg i}$. This is bounded by the definition of $\Delta(\rho)$:
    \begin{align}
        \|A(\bfF^{j,i \to \emptyset}) - A(\bfF^{j \to \emptyset})\|_1 \leq \Delta(\rho).
    \end{align}

    The second term represents the sensitivity to removing events from the entire epoch $j$. In the individual DP setting with device-epoch records, all events from epoch $j$ for a given device come from a single device-epoch record. Removing this entire record corresponds exactly to one of the cases considered in the definition of $\Delta(\rho)$. Therefore:
    \begin{align}
        \|A(\bfF^{j \to \emptyset}) - A(\mathbf{F})\|_1 \leq \Delta(\rho).
    \end{align}

    Therefore, substituting the two upper bounds into \Eq(\ref{eqn:trig-ineq-upper-bound-for-site-glob-sen}):
    \begin{align}
        \Delta_i(\rho) \leq \Delta(\rho) + \Delta(\rho) = 2\Delta(\rho).
    \end{align}

\end{proof}

\subsection{Comprehensive \sysname Model}
\label{appendix:online-algorithm:algos}

\heading{Overview.}
\sysname manages per-site and global privacy filters using the quota
mechanisms described in \S\ref{sec:big-bird-overview}.
\Alg\ref{alg:e2e_functional_view} depicts the functionality triggered
on receiving a report request (i.e., \code{measureConversion()} and
\code{getReport()}). \sysname checks and consumes budget from the
relevant filters and prunes the resulting report based on filter status.
First, \sysname ensures all filters and quotas are initialized.
Second, it computes
the privacy losses incurred both at the epoch level \cite[\S
    C]{TKM+24} and at the epoch-impression-site level
(\Def\ref{defn:EpochImpSiteBudget}). Third, it checks whether all
filters have sufficient budget and attempts to consume it. To avoid wasting budget from some filters (or quotas) when other filters are out of budget, \sysname uses a transactional protocol to deduct privacy losses from multiple
filters atomically. \Alg\ref{alg:atomic_filter_check} formalizes this: if any
filter cannot afford its share of the privacy loss, the report is
zeroed out, and no budget is consumed from any filter for that report.

\heading{Enforcing the  \quotacountVarName.}
In addition to the filter-based budget checks, \sysname enforces a $\quotacountValue$ cap  on the number of distinct sites that can access the Attribution API within a single user action context (\Def\ref{def:user-action-context}). This mechanism limits adversarial creation of quota filters. As formalized in \Alg\ref{alg:e2e_functional_view}, the \quotacountVarName check occurs \emph{before} the atomic filter check: if the check fails, the epoch's data is zeroed out ($F_e \gets \emptyset$, using the zeroing-out notation from Appendix~\ref{appendix:data-model}), and no filters are accessed for that epoch.

\heading{Subroutines.} \Def\ref{def:epoch-budget}, \ref{defn:EpochImpSiteBudget},  \ref{defn:filters-canconsum-and-tryconsume} and \Alg\ref{alg:atomic_filter_check} define subroutines used in \Alg\ref{alg:e2e_functional_view}. The first two definitions rely on the sensitivity bounds from \Thm\ref{thm:individual-sensitivity-of-reports} (for the impression-site quota) and \cite{TKM+24} (for the filters and the other quotas).

\begin{definition}[EpochImpSiteBudget]\label{defn:EpochImpSiteBudget}
    Let $x=(d,e,F)\in \cX$ be a device-epoch record, $i\in \cS$ an impression site, $\rho$ an attribution report, and $\lambda>0$ the Laplace noise scale applied to $\rho$.
    Given the upper bound $\tilde \Delta_{x,i}(\rho) \ge \Delta_{x,i}(\rho)$ on the per-epoch-site individual sensitivity given by \Thm\ref{thm:individual-sensitivity-of-reports}, we can upper bound the epoch-site privacy loss consumed by $\rho$ at $(x,i)$ by
    \begin{align}
        \text{EpochImpSiteBudget}(x,i,\rho,\lambda) :=
        \frac{\tilde \Delta_{x,i}(\rho)}{\lambda}.
    \end{align}
\end{definition}

\begin{definition}[EpochBudget, from \cite{TKM+24}]
    \label{def:epoch-budget}
    Fix a device-epoch record $x \in \cX$,
    $\rho$ an attribution report, and $\lambda>0$ the Laplace noise scale applied to $\rho$.
    Given the upper bound $\tilde \Delta_{x}(\rho) \ge \Delta_{x}(\rho)$ on the per-epoch individual sensitivity given by \cite{TKM+24}, the individual privacy loss for device-epoch record $x$ is:
    \begin{align}
        \operatorname{EpochBudget}(x,\rho,\lambda) := \frac{\tilde \Delta_x(\rho)}{\lambda}
    \end{align}
\end{definition}

\begin{definition}[Privacy filter $\mathcal{F}$, adapted from \cite{filters_rogers}]
    \label{defn:filters-canconsum-and-tryconsume}
    Privacy filters track privacy loss of adaptively chosen mechanisms against a predetermined budget. We adapt the standard definition for pure DP privacy filters and define a filter $\mathcal{F}$ as a stateful object storing its remaining capacity, initialized to $\epsilon_\text{initial}$ (the filter capacity).

    We then recursively define $\cF.\operatorname{canConsume}$,
    $\cF.\operatorname{tryConsume}$ and $\pass_{\mathcal{F},j}$, the indicator that the $j$th budget was granted, for a sequence of adaptively chosen privacy budgets $\epsilon^1, \epsilon^2, \dots$, as follows:
    \begin{itemize}
        \item $\cF.\operatorname{canConsume}(\epsilon^j)$: Returns \true if the filter can accommodate additional privacy loss $\epsilon^j$, \ie $\epsilon^j \leq \epsilon_\text{initial} - \sum_{j' < j} \epsilon^{j'} \cdot \pass_{\mathcal{F},j'}$
              where $\epsilon_\text{initial}$ is the filter capacity.
        \item $\cF.\operatorname{tryConsume}(\epsilon^j)$: Calls $\operatorname{canConsume}(\epsilon^j)$. If successful, sets $\pass_{\mathcal{F},j} = 1$ to deduct $\epsilon^j$ from the filter's remaining capacity; otherwise, sets $\pass_{\mathcal{F},j} = 0$.
    \end{itemize}

    By definition of $\operatorname{canConsume}$ we obtain the following {\em capacity invariant}:
    \begin{align}
        \sum_j \epsilon^j\pass_{\cF,j} \le \epsilon_{\text{initial}}.
        \label{eq:capacity-invariant}
    \end{align}
\end{definition}

\begin{definition}[Per-action domain cap $\cK$]
    \label{def:domain-cap}

    We define a {\em domain cap} as a stateful object $\cK \subseteq \cS$, initialized to $\emptyset$ with a cap $ \mathquotacountVar$ and two methods, for a domain $s \in \cS$:

    \begin{itemize}
        \item $\cK.\operatorname{canAdmit}(s) \iff |\cK \cup \{s\}| \le \mathquotacountVar$,
        \item $\cK.\operatorname{tryAdmit}(s)$: if
              $\cK.\operatorname{canAdmit}(s)$, set
              $\cK\gets\cK\cup\{s\}$; otherwise leave $\cK$ unchanged.
    \end{itemize}

    Note that only the first admission for a given domain can be rejected, and the invariant $|\cK| \le \mathquotacountVar$ bounds the number of distinct domains.
\end{definition}

\begin{definition}[Per-device-epoch state $\cF_{(d,e)}$]
    \label{def:quota-count-state}
    For each device-epoch record $x = (d, e, F)$, we maintain four families of pure DP filters (global, per querier $b$, per conversion site $c$, per impression site $i$) and domain caps for each user action $u$.
    We denote the resulting state $\cF_{d,e}$:
    \begin{align}
        \cF_{d,e} := \Bigl(
         & \cF^{\globalFilterVarName}_{d,e},\
        \bigl(\cF^{\persiteFilterVarName[b]}_{d,e}\bigr)_{b\in\cS},\
        \bigl(\cF^{\convQuotaVarName[c]}_{d,e}\bigr)_{c\in\cS}, \notag \\
         & \bigl(\cF^{\impQuotaVarName[i]}_{d,e}\bigr)_{i\in\cS},\
        \bigl(\cK_{d,e}[u]\bigr)_{u\in\cU}
        \Bigr),
        \label{eq:state-entry}
    \end{align}

    We lazily initialize these filters/cap instantiated with respective capacities $\ec, \enc, \eqimp$, $\eqconv$, $\mathquotacountVar$.

\end{definition}

We make two remarks. First, the state $\cF_{d,e}$ can change depending on the adversary's queries, even when the underlying data is fixed (\eg when the data is generated in chunks epoch by epoch). Second, since the pair $(d,e)$ appears only once in valid databases, we can index by $x = (d,e,F)$ to obtain the same $\cF_x$ even as $F$ changes, \eg when new events are added to the same device.

\begin{algorithm}[t]
    \caption{\sysname algorithm on a single device}
    \label{alg:e2e_functional_view}
    \footnotesize
    \begin{algorithmic}[1]
        \Input
        \State Device $d$ with growing on-device database $D_d \gets \emptyset$
        \State State $(\cF_{d,e})_{e \in \cE}$ lazily initialized with caps~$\ec,\enc,\eqimp,\eqconv,\mathquotacountVar$

        \EndInput
        \Statex

        \Function{$\operatorname{SaveImpression}$}{$i, \mathsf{imp}, u$}
        \State $e \gets \operatorname{CurrentEpoch}()$\label{line:save-impression-epoch}
        \If{$\cF_{d,e}.\cK[u].\operatorname{canAdmit}(i) = \false$}
        \State \Return \graycomment{Silently reject impression if cap exceeded}
        \EndIf
        \State $\cF_{d,e}.\cK[u].\operatorname{tryAdmit}(i)$
        \State \graycomment{Store impression event}
        \State $D_d^e \gets D_d^e \cup \{(u,i,\mathsf{imp})\}$
        \State \Return
        \EndFunction
        \Statex

        \Statex \graycomment{Generate report and update on-device budget}
        \Function{$\operatorname{GenerateReport}$}{$D, \rho, \lambda, u, b$}
        \State Read $\rho$ to get device $d$, \convsite $c$,
        target epochs $E$,
        attribution function $A$ with relevant impressions $F_A$.
        \For{$e \in E$}
        \State $x \gets (d,e,D_d^e)$

        \If{$\cF_{d,e}.\cK[u].\operatorname{canAdmit}(c) = \false$}  \graycomment{\Def\ref{def:quota-count-state}}
        \State $F_e \gets \emptyset$\label{line:cap-zero-out} \graycomment{Zero out epoch if \quotacountVarName exceeded}
        \State \textbf{continue} \graycomment{Skip to next epoch}
        \EndIf
        \State $\cF_{d,e}.\cK[u].\operatorname{tryAdmit}(c)$
        \State $F_e \gets D_d^e \cap F_A$ \label{line:report-relevant-impressions} \graycomment{Effective relevant impressions}
        \State $\mathbf{i}_e \gets \mathsf{ImpSites}(F_e)$ \graycomment{Effective impression sites}
        \State $\epsilon_x \gets \operatorname{EpochBudget}(x,\rho, \lambda)$\label{line:epoch_budget} \graycomment{\Def\ref{def:epoch-budget}}
        \State $\epsilon_\mathbf{x}^{\mathbf{i}} \gets \{\}$
        \For{$i \in \mathbf{i}_e$}
        \State $\epsilon_x^i \gets \operatorname{EpochImpSiteBudget}(x,i,\rho,\lambda)$ \label{line:epoch-site-privacy-loss} \graycomment{\Def\ref{defn:EpochImpSiteBudget}}
        \State $\epsilon_\mathbf{x}^{\mathbf{i}}[i] \gets \epsilon_x^i$
        \EndFor
        \Statex \ \ \ \ \graycomment{Check and update filters/quotas atomically (\Alg\ref{alg:atomic_filter_check})}
        \If{$\operatorname{AtomicFilterCheckAndConsume}(\cF_x, b, c, \mathbf{i}_e, \epsilon_x, \epsilon_\mathbf{x}^{\mathbf{i}}) = \false$}
        \label{line:2pc-return}
        \State $F_e \gets \emptyset$ \graycomment{Empty the epoch if any check fails}
        \label{line:zero-out}
        \EndIf
        \EndFor

        \State $\hat\rho \gets A((F_e)_{e \in E})$ \graycomment{Clipped attribution report}
        \State\Return $\hat\rho$
        \EndFunction
        \State
    \end{algorithmic}
\end{algorithm}

\begin{algorithm}[t]
    \caption{AtomicFilterCheckAndConsume}
    \label{alg:atomic_filter_check}
    \begin{algorithmic}[1]
        \Input
        \State Budgets $\cF_x$, querier $b$, conversion site $c$, impression sites $\mathbf{i}$
        \State $\epsilon_x$: epoch-level privacy loss
        \State $\epsilon_\mathbf{x}^{\mathbf{i}}$: epoch-site-level privacy loss
        \EndInput

        \State \graycomment{Phase 1: check if all filters can consume (\Def \ref{defn:filters-canconsum-and-tryconsume})}
        \If{$\cF_x^{\persiteFilterVarName[b]}.\operatorname{canConsume}(\epsilon_x) = \false$} \label{line:per-site-can-consume}
        \State \Return $\false$
        \EndIf

        \If{$\cF^\globalFilterVarName_x.\operatorname{canConsume}(\epsilon_x) = \false$}
        \State \Return $\false$
        \EndIf

        \If{$\mathcal{F}^{\convQuotaVarName[c]}_x.\operatorname{canConsume}(\epsilon_x) = \false$}
        \State \Return $\false$
        \EndIf

        \For{$i \in \mathbf{i}$}
        \If{$\mathcal{F}^{\impQuotaVarName[i]}_x.\operatorname{canConsume}(\epsilon_\mathbf{x}^{\mathbf{i}}[i]) = \false$}
        \State \Return $\false$
        \EndIf
        \EndFor

        \State \graycomment{Phase 2: consume from all filters (\Def \ref{defn:filters-canconsum-and-tryconsume})}
        \State $\cF_x^{\persiteFilterVarName[b]}.\operatorname{tryConsume}(\epsilon_x)$ \label{line:per-site-try-consume}
        \State $\cF^\globalFilterVarName_x.\operatorname{tryConsume}(\epsilon_x)$
        \State $\mathcal{F}^{\convQuotaVarName[c]}_x.\operatorname{tryConsume}(\epsilon_x)$
        \For{$i \in \mathbf{i}$}
        \State $\mathcal{F}^{\impQuotaVarName[i]}_x.\operatorname{tryConsume}(\epsilon_\mathbf{x}^{\mathbf{i}}[i])$
        \EndFor

        \State \Return $\true$
    \end{algorithmic}
\end{algorithm}

\subsection{Global Privacy Guarantees}
\label{appendix:online-algorithm:privacy-proofs}

We now state and prove the privacy guarantees, stated for a formal privacy game modeling \Alg\ref{alg:e2e_functional_view} execution over the whole system.

\heading{Execution model.}
\Alg\ref{alg:e2e_setup} presents an abstract model of \sysname's
operation, capturing how it answers queries sequentially for all queriers.
At each time step $t$, querier $k$ can formulate a query $Q_t^k$, which is a batch of reports as defined in \Def\ref{def:query}. When a querier has nothing to ask, it can send an empty query with zero sensitivity.
\code{AnswerQuery} then processes each individual report $\rho_r$ (for $r\in R$, where $R$ is the set of report identifiers in $Q_t^k$).
For each report, \code{GenerateReport} creates an individual clipped attribution report.
Finally, reports are summed and noised before returning the query result.

\begin{algorithm}[t]
    \caption{Formalism for DP analysis}
    \label{alg:e2e_setup}
    \footnotesize
    \begin{algorithmic}[1] %

        \Input
        \State Challenge bit $\chal \in \{0,1\}$
        \State In-or-out device-epoch $(d_0, e_0)$
        \State Data generation process $\cG$
        \State Queriers $\cQ^1, \dots, \cQ^K$
        \EndInput

        \Output
        \State View $ V = (V^1, \dots, V^K) \in \cR^{e_{\max} \times (t_{\max} + 1) \times K}$ of all $\cQ^k$
        \EndOutput

        \For{$e \in [e_{\max}]$}
        \Statex \graycomment{Devices and events for epoch $e$, generated from {\em all previous results}}
        \State $G \gets \cG(V^1_{<e}, \dots V^K_{<e})$

        \Statex \graycomment{View of each querier, initialized empty for conversions}
        \State $(V^k_{e,0})_{k \in [K]} = (\emptyset)_{k \in [K]}$
        \For{$(d,f) \in G$}
        \If{$f = (u,i,\mathrm{imp}) \in \Ev^{\mathsf{imp}}$}
        \If{$\chal = 0$ and $d = d_0, e = e_0$} \label{line:opt-out}
        \State \textbf{skip}  \graycomment{Keep impression events out in World 0}
        \Else
        \State $\operatorname{SaveImpression}(i, \mathsf{imp}, u)$ on $d$
        \EndIf
        \Else
        \Statex \graycomment{Each querier receives its conversions and corresponding report identifiers}
        \State $D^e_d \gets D^e_d + f$, with $f = (u,c,\bfb,\text{conv})$
        \State Generate report identifier $r
            \overset{{\scriptscriptstyle\$}}{\leftarrow} U(\Z)$
        \Statex \ \ \ \ \ \ \graycomment{Save the device and action context that generated $r$}
        \State  $(d_r,u_r) \gets (d,u)$
        \For{$k \in \bfb$}
        \State $V^k_{e,0} \gets V^k_{e,0} \cup \{(r,f)\}$  \label{line:public-info-output} \graycomment{Output conversion and report identifier}
        \EndFor
        \EndIf
        \EndFor

        \State \graycomment{Queriers ask queries, round robin}
        \For{$t \in [t_{\max}]$} \label{line:query_step_t}
        \For{$k \in [K]$}
        \Statex \ \ \ \ \ \ \graycomment{Querier $k$ chooses based on {\em its own previous results}}
        \State $Q_t^k, \lambda_t^k \gets \cQ^k(V^k_{\le e, <t})$
        \State $V^k_{e,t} \gets \operatorname{AnswerQuery}(D^{\le e}, Q_t^k,
            \lambda_t^k, k)$ \label{line:query-output}

        \EndFor
        \EndFor
        \EndFor

        \State \graycomment{Collect, aggregate and noise reports to answer $Q$}
        \Function{$\operatorname{AnswerQuery}$}{$D, Q, \lambda, b$}
        \State $(\rho_r)_{r \in R} \gets Q$ \graycomment{Get report
            identifiers from $Q$}
        \For{$r \in R$}            %
        \State $\hat \rho_r \gets \operatorname{GenerateReport}(D, \rho_r, \lambda, u_r, b)$
        \EndFor
        \State Sample $X \sim \lap(\lambda)$
        \State \Return $\sum_{r \in R} \hat \rho_r + X$
        \EndFunction
    \end{algorithmic}
\end{algorithm}

At \Line\ref{line:opt-out}, world $\chal = 0$ skips all the impressions on
$(d_0,e_0)$, following the neighboring relation from \Def\ref{def:neighboring-databases}, and every other call
is executed identically in both worlds.

\begin{numberedthm}[\ref{thm:privacy-guarantee}]
    Consider $x \in \cX$ with global filter capacity $\ec$.
    Then, \sysname satisfies individual device-epoch $\ec$-DP for $x$ with respect to the neighboring relation in Definition~\ref{def:neighboring-databases}. That is, for all data generation processes $\cG$, for all queriers $\cQ^1, \dots, \cQ^K$, the output $V^{(\chal)}$ of \Alg\ref{alg:e2e_setup} under challenge bit $\chal$ verifies:

    \begin{align}
        \label{eq:privacy-want-to-show-appendix}
        \forall v \in \cR^{e_{\max} \times (t_{\max} + 1) \times K}, \left| \ln \left( \frac{\Pr[V^{(0)} = v]}{\Pr[V^{(1)} = v]} \right) \right| \leq \ec.
    \end{align}
\end{numberedthm}

\begin{proof}

    Take the in-or-out device-epoch $x = (d_0,e_0,F) \in \mathcal{X}$.
    Denote by $x_{\cC} = (d_0,e_0,F\cap \Ev^{\mathsf{conv}})$ the device-epoch obtained by keeping only public events from $x$, where public events are the set of all conversions.
    Take $v \in \cR^{e_{\max} \times (t_{\max} + 1) \times K}$ and $\chal \in \{0,1\}$.
    $v$ is the vector of values returned at different points in \Alg\ref{alg:e2e_setup}. We split it into $v = (v_{1,0}, v_{1,1}, \ldots, v_{1,\tmax}, \dots, v_{e,0}, \dots v_{e,t})$ where $v_{e,0} = (v_{e,0}^1, \dots, v_{e,0}^K)$ is a value for the output from \Line\ref{line:public-info-output} of \Alg\ref{alg:e2e_setup} (public events from epoch $e$ for each querier $k \in [K]$), and $v_{e,t} = (v_{e,t}^1, \dots, v_{e,t}^K)$, where $v_{e,t}^k$ is the output for the query $Q^k_t$ for querier $k$ at step $t$ from epoch $e$ (\Line\ref{line:query-output}).
    By conditioning over past outputs at each time step $(e,t,k) \in [\emax] \times \{0, ..., \tmax\} \times [K]$ we get:
    \begin{align}
        \label{eq:bayes_pub}
         & \Pr[V^{(\chal)} = v]                                                                                              \\
        =
         & \prod_{e=1}^{\emax} \prod_{t=0}^{\tmax} \prod_{k=1}^K \Pr[V^{k, (\chal)}_{e,t} = v_{e,t}^k | v_{\le e, <t}^{<k}].
    \end{align}

    Take $e \in [\emax]$, $t \in [\tmax]$ and $k \in [K]$.
    By Algorithm \ref{alg:e2e_setup}, we have:
    \begin{align}
          & \Pr[V^{k, (\chal)}_{e,t} = v_{e,t}^k | v_{\le e, <t}^{<k}]                                                  \\
        = & \Pr[\operatorname{AnswerQuery}(D^{(\chal)}, Q^k_t, \lambda_t^k, k ; \cF^{k, (\chal)}_{e,t}) =  v_{e,t}^k ],
    \end{align}
    where the database $D^{(\chal)}$, the query $Q^k_{t}$ with associated noise $\lambda_t^k$ and the state $\cF^{k, (\chal)}_{e,t}$ are functions of past results $v_{\le e, <t}^{<k}$.
    The generated data and the queries are chosen by $\cG$ and $\cQ^k$ conditionally on past results but independently of $\chal$.
    The database can depend on $\chal$, but only at $(d_0,e_0)$, since world $0$ skips the impression saves on that device-epoch at \Line\ref{line:opt-out}.
    The state can also depend on $\chal$, albeit in a limited fashion, since we have:

    \begin{align}
        \label{eq:filters-identical}
        \forall x' \in D^{(0)} \cap D^{(1)}, (\cF^{k, (0)}_{e,t})_{x'} = (\cF^{k, (1)}_{e,t})_{x'}
    \end{align}

    Indeed, \Alg\ref{alg:e2e_functional_view} shows that the budget spent for $x'$ at each step only depends on past queries, which are identical in both worlds once we condition on $v_{\le e, <t}^{<k}$, and on the record $x'$ itself, which is identical in both worlds. This is thanks to device-epoch level accounting in \Def\ref{def:epoch-budget}, \ref{defn:EpochImpSiteBudget}. The same holds for domain caps $\cK_{d',e'}[u]$: even though $\operatorname{GenerateReport}$ iterates over the whole window $E$, each iteration reads and writes only $\cK_{d',e'}[u]$, decides from that entry, and skips only that epoch if it hits the cap.

    Let $\rho_r(D; \cF) = \operatorname{GenerateReport}(D, \rho_r, \lambda ; \cF)$ denote the filtered report returned by \Alg\ref{alg:e2e_functional_view}, which maintains the state $\cF$.
    Then:

    \begin{align}
         & \Pr[V^{k, (\chal)}_{e,t} = v_{e,t}^k | v_{\le e, <t}^{<k}] \nonumber                                                 \\
         & \qquad = \Pr\left[\sum_{r \in R_{e,t}^k} \rho_r(D^{(\chal)}; \cF_{e, t, r}^{(\chal)}) + X_{e,t} = v_{e,t}^k \right],
    \end{align}
    where $X_{e,t}$ is the Laplace noise added at time $e,t$.

    We omit the $e,t,k$ indices when they are clear from context.
    We denote by $D$ the database generated conditionally on $v_{\le e, <t}^{<k}$ before adding the in-or-out record, giving: $D^{(0)} = D + x_\cC$ and $D^{(1)} = D + x$ for $e \ge e_0$ (and $D^{(0)} = D^{(1)}$ for $e < e_0$).
    By \Def\ref{def:attribution-function}, since $F_A \subseteq \Ev^{\mathsf{imp}}$ and $F \cap \Ev^{\mathsf{conv}} \cap \Ev^{\mathsf{imp}} = \emptyset$, we have:

    \begin{align*}
        \rho_r(D^{(0)} ; \cF_r^{(0)} )
         & = \rho_r(D + (d_0,e_0,F\cap \Ev^{\mathsf{conv}} \cap F_A)) \\
         & = \rho_r(D; \cF_r^{(0)} )
    \end{align*}

    By \Alg\ref{alg:e2e_functional_view}, remark that the state $(\cF_r^{(0)})_{x'}$ for $x' \in \{x, x_\cC\}$ has no impact on the output: replacing the corresponding data by $\emptyset$ at \Line\ref{line:cap-zero-out} (when hitting a per-action domain cap) or \Line\ref{line:zero-out} (when running out of budget or quota) is a no-op.
    Combining this observation with \Eq\ref{eq:filters-identical} gives:

    \begin{align}
        \rho_r(D^{(0)} ; \cF_r^{(0)} ) = \rho_r(D; \cF_r^{(1)} )
    \end{align}

    We can now bound the privacy loss at step $e,t,k$:

    \begin{align}
         & \left| \ln \left( \frac{\Pr[V^{k, (0)}_{e,t} = v_{e,t}^k | v_{\le e, <t}^{<k}]}{\Pr[V^{k, (1)}_{e,t} = v_{e,t}^k | v_{\le e, <t}^{<k}]} \right) \right|                                                                     \\
         & \le \left| \ln \left( \frac{\Pr\left[\sum_{r \in R_{e,t}^k} \rho_r(D; \cF_r^{(1)}) + X_{e,t} = v_{e,t}^k \right]}{\Pr\left[\sum_{r \in R_{e,t}^k} \rho_r(D + x; \cF_r^{(1)}) + X_{e,t} = v_{e,t}^k \right]} \right) \right|
        \label{eq:loss_one_query_base}
    \end{align}

    We now omit $\cF_r^{(1)}$ since it is present on both sides.
    Take $r \in R_t$.
    We define the boolean $\pass_r$ by $\pass_r =0$ if $F_e$ is zeroed out (either by the domain cap at \Line \ref{line:cap-zero-out} or by the atomic check at \Line\ref{line:zero-out}), in the world where $x$ is present.
    If $\pass_r = 0$, we have $\rho_r(D+x) = \rho_r(D)$ because of \Alg\ref{alg:e2e_functional_view}, \Line\ref{line:zero-out}.
    Otherwise, if $\pass_r = 1$, we have $\left\|\rho_r(D+x) - \rho_r(D)\right\|_1 \leq \Delta_x(\rho_r)$.

    Hence,
    \begin{align}
        \label{eq:loss_one_report}
        \|  \rho_r(D) -  \rho_r(D + x)\|  \le \Delta_x(\rho_r) \pass_r
    \end{align}

    Thus by triangle inequality followed by \Def\ref{def:epoch-budget} we have:
    \begin{align}
        \label{eq:loss_one_query}
        \| \sum_{r \in R_t} \rho_r(D) - \sum_{r \in R_t} \rho_r(D + x)\| & \le \sum_{r \in R_t} \Delta_x(\rho_r) \pass_r             \\
                                                                         & \le \sum_{r \in R_t} \lambda_t^k \cdot \epsilon_r \pass_r
    \end{align}

    And since  $X_t \sim \lap(\lambda_t^k)$, by property of the Laplace distribution \Eq\ref{eq:loss_one_query_base} becomes:
    \begin{align}
        \left| \ln \left( \frac{\Pr[V^{k, (0)}_{e,t} = v_{e,t}^k | v_{\le e, <t}^{<k}]}{\Pr[V^{k, (1)}_{e,t} = v_{e,t}^k | v_{\le e, <t}^{<k}]} \right) \right|
        \leq \sum_{r \in R_t} \epsilon_r \pass_r\label{ineq:sensitivity-guarantee-on-global-filter}
    \end{align}

    We can now bound the total privacy loss.
    First, we use  \Eq\ref{eq:bayes_pub} and the
    fact that the outputs $v_{e,0}^k$ at \Line\ref{line:public-info-output} are identical across both worlds by definition of $x_{\cC}$ to drop indices where $t=0$:
    \begin{align}
         & \left| \ln \left( \frac{\Pr[V^{(0)} = v]}{\Pr[V^{(1)} = v]} \right) \right| \\
         & = \left| \ln \left(
        \prod_{e=1}^{\emax} \prod_{t=0}^{\tmax} \prod_{k=1}^K
        \frac{
            \Pr[V^{k, (0)}_{e,t} = v_{e,t}^k | v_{\le e, <t}^{<k}]}
        {
            \Pr[V^{k, (1)}_{e,t} = v_{e,t}^k | v_{\le e, <t}^{<k}]
        } \right) \right|                                                              \\
         & = \left| \sum_{e=1}^{\emax} \sum_{t=1}^{\tmax} \sum_{k=1}^K
        \ln \left(
        \frac{
            \Pr[V^{k, (0)}_{e,t} = v_{e,t}^k | v_{\le e, <t}^{<k}]}
        {
            \Pr[V^{k, (1)}_{e,t} = v_{e,t}^k | v_{\le e, <t}^{<k}]
        }
        \right)
        \right|
    \end{align}

    By triangle inequality and \Eq\ref{ineq:sensitivity-guarantee-on-global-filter}, we get:
    \begin{align}
         & \left| \ln \left( \frac{\Pr[V^{(0)} = v]}{\Pr[V^{(1)} = v]} \right) \right|                \\
         & \le \sum_{e=1}^{\emax} \sum_{t=1}^{\tmax} \sum_{k=1}^K \sum_{r \in R_t} \epsilon_r \pass_r
    \end{align}

    Finally, each $\epsilon_r \pass_r$ is an amount granted by $\cF^{\globalFilterVarName}_{d_0,e_0}$, so its capacity invariant \Eq\ref{eq:capacity-invariant} gives:

    \begin{align}
        \sum_{e=1}^{\emax} \sum_{t=1}^{\tmax} \sum_{k=1}^K \sum_{r \in R_t} \epsilon_r \pass_r \le \ec
    \end{align}

    Hence we have shown \Eq\ref{eq:privacy-want-to-show-appendix}.
\end{proof}

\subsection{DoS Resilience}
\label{appendix:online-algorithm:dos-proofs}

This section proves our main resilience result for \sysname's quota-based online algorithm: \Thm\ref{thm:resilience-three-dimensions}. First, \Lem~\ref{lemma:2pc_guarantees} shows that the 2-PC check (\Alg~\ref{alg:atomic_filter_check}) ensures (1) atomic consumption across all filters relevant to a query, and (2) when all filters have sufficient budget, each consumes an amount proportional to its level-specific sensitivity either at epoch or at epoch-site level. Next, \Lem~\ref{lem:first-upper-bounds-on-global-filter} and \ref{lem:second-upper-bounds-on-global-filter} bound the total privacy budget the adversary can consume from the \c at any device-epoch, using the atomic consumption guarantees of \Lem\ref{lemma:2pc_guarantees}. These first two lemmas give \Thm\ref{thm:resilience-three-dimensions}(i).
\Lem\ref{lem:quotas-per-user-action-restated} then connects user actions to obtain \Thm\ref{thm:resilience-three-dimensions}(ii), and
\Lem\ref{thm:graceful-degradation-population} lifts the bound to the query level
for \Thm\ref{thm:resilience-three-dimensions}(iii).

\subsubsection{Preliminaries}

\mbox{}\\
\heading{Notation.}
Everything in this section (\S\ref{appendix:online-algorithm:dos-proofs}) is stated inside one fixed execution of \Alg\ref{alg:e2e_setup}.
Let $s_{\max}:=e_{\max}t_{\max}K$. We index the queries from
\Alg\ref{alg:e2e_setup} by $s\in[s_{\max}]$ instead of triples $(e,t,k)$, so
each execution of $\operatorname{AnswerQuery}$ is associated with a unique
$s$, with corresponding report identifiers $R_s$ and querier $b_s$.

For a device-epoch $x=(d,e,F)$, a query $s$ and $r \in R_s$ we write $c_r$ for the conversion site of $r$,  $\mathbf i_{r,e}=\mathsf{ImpSites}(F\cap F_{A_r})$ for its effective
impression sites (\ie $I_r(x)$ from \Thm\ref{thm:individual-sensitivity-of-reports}), $\epsilon_{r,s}$ and $\epsilon^i_{r,s}$ (for
$i\in\mathbf i_{r,e}$) for the losses computed at
\Line\ref{line:epoch_budget} and \Line\ref{line:epoch-site-privacy-loss},
and $\pass_{r,s} \in \{0,1\}$ for the indicator that $\operatorname{GenerateReport}$ zeroes
$F_e$ at neither \Line\ref{line:cap-zero-out} nor \Line\ref{line:zero-out}.
Each of $\epsilon_{r,s}$ and $\epsilon^i_{r,s}$ is $0$ when
\Alg\ref{alg:e2e_functional_view} does not compute it at $x$.
When the report index is clear from context, we drop $r$ and write $c_s$, $\mathbf i_s$, $\epsilon^i_s$ and $\pass_s$.

\begin{lemma}[Atomic filter check guarantees]
    \label{lemma:2pc_guarantees}
    Fix a device-epoch record $x=(d,e,F)$, a query $s$ and $r \in R_s$, with querier $b=b_s$.
    Using the definition of $\pass_{r,s}$ and \Alg\ref{alg:atomic_filter_check}, we have:

    \begin{enumerate}
        \item[(a)] $\cF^{\persiteFilterVarName[b]}_{d,e}$,
              $\cF^{\globalFilterVarName}_{d,e}$ and
              $\cF^{\convQuotaVarName[c_r]}_{d,e}$ each grant exactly
              $\epsilon_{r,s}\pass_{r,s}$;

        \item[(b)] for each $i\in\mathbf i_{r,e}$,
              $\cF^{\impQuotaVarName[i]}_{d,e}$ grants exactly
              $\epsilon^i_{r,s}\pass_{r,s}$.
    \end{enumerate}
\end{lemma}

\begin{proof}
    We prove both properties together. Fix $r\in R_s$ with
    $\pass_{r,s}=1$. By \Alg\ref{alg:atomic_filter_check}, the atomic check
    returns $\true$ exactly when all Phase~1 checks pass and all Phase~2
    operations execute. For this report, with conversion site $c_r$, querier
    $b_s$, and impression sites $\mathbf i_{r,e}$, Phase~2 calls:
    \begin{itemize}
        \item $\cF_x^{\persiteFilterVarName[b_s]}.\operatorname{tryConsume}(\epsilon_{r,s})$
        \item $\cF_x^{\globalFilterVarName}.\operatorname{tryConsume}(\epsilon_{r,s})$
        \item $\cF_x^{\convQuotaVarName[c_r]}.\operatorname{tryConsume}(\epsilon_{r,s})$
        \item For each $i \in \mathbf{i}_{r,e}$: $\cF_x^{\impQuotaVarName[i]}.\operatorname{tryConsume}(\epsilon^i_{r,s})$
    \end{itemize}
    Note that $\epsilon_{r,s}$ is the device-epoch-level individual privacy loss computed for this specific report $r$ (\eg, via $\operatorname{EpochBudget}$ in \Line\ref{line:epoch_budget} of \Alg\ref{alg:e2e_functional_view}). Similarly, each $\epsilon^i_{r,s}$ is the device-epoch-site-level individual privacy loss for impression site $i$ relevant to this report $r$ (via $\operatorname{EpochImpSiteBudget}$ in \Alg\ref{alg:e2e_functional_view}). Therefore, when $\pass_{r,s} = 1$, the conversion-site quota filter (for $c_r$), the per-site filter (for $b_s$), and the \c all consume exactly the same amount $\epsilon_{r,s}$. Concurrently, each relevant impression-site quota filter (for $i \in \mathbf{i}_{r,e}$) consumes its specific amount $\epsilon^i_{r,s}$, which is proportional to its sensitivity at the epoch-impression-site level for this report $r$.
\end{proof}

\subsubsection{Number of sites}

With these atomic guarantees we can bound how much of the \c can be depleted at any device-epoch $x \in \cX$.
The main isolation result is in \Thm\ref{thm:resilience-three-dimensions}, which uses \Lem\ref{lem:first-upper-bounds-on-global-filter},~\ref{lem:imp_quota_allocation_consistency} and \ref{lem:second-upper-bounds-on-global-filter}.

\Needspace{12\baselineskip}
\begin{definition}[Budget consumed and quotas charged]
    \label{def:charged-reports}
    Fix a device-epoch $x=(d,e,F)$, and {\em any} set of reports $\cR_s\subseteq R_s$ for each query
    $s\in[s_{\max}]$.

    Let
    $\cR^+:=\{(s,r):s\in[s_{\max}],\ r\in\cR_s,\ \pass_{r,s}=1\}$ the reports that pass the filters. Then, we define:
    \begin{align}
        \globalFilterEps(\cR)
               & :=\sum_{(s,r)\in\cR^+}\epsilon_{r,s},
        \label{eq:charged-consumption}                           \\
        C(\cR) & :=\bigcup_{(s,r)\in\cR^+}\{c_r\},\notag         \\
        N(\cR) & :=|C(\cR)|,\notag                               \\
        I(\cR) & :=\bigcup_{(s,r)\in\cR^+}\mathbf i_{r,e},\notag \\
        M(\cR) & :=|I(\cR)|.
        \label{eq:charged-quota-sets}
    \end{align}
\end{definition}

\Def\ref{def:charged-reports} defines what a group of report consumes. When reports are associated with an attacker, this defines the impact of the attacker, but the definition itself does not involved any notion of attacker.
By \Lem\ref{lemma:2pc_guarantees}, reports $\cR$ consume
$\globalFilterEps(\cR)$ from $\cF^{\globalFilterVarName}_{d,e}$, and they charge the domains $C(\cR)$ and
$I(\cR)$.
In the lemmas below, we take $\cR$ to be the reports an adversary elicits and note:
$\cR_{\mathsf{adv}}=(R^{\mathsf{adv}}_s)_{s\in[s_{\max}]}$ with
$R^{\mathsf{adv}}_s\subseteq R_s$, writing
$\globalFilterEps^{\mathsf{adv}}:=\globalFilterEps(\cR_{\mathsf{adv}})$ and
likewise $\cR^+_{\mathsf{adv}}$, $C_{\mathsf{adv}}$, $N^{\mathsf{adv}}$, $I_{\mathsf{adv}}$,
$M^{\mathsf{adv}}$.
Note that $I_{\mathsf{adv}}$ contains every impression site charged by an attacker-elicited report, even honest impression sites that happen to allow malicious conversion sites to query their impressions.
This corresponds to the threat model from \S\ref{sec:threat-model}, where $I_{\mathsf{adv}} \cup C_{\mathsf{adv}}$ is the set of domains directly controlled by, or transitively partnering with the attacker.

\begin{lemma}[Conversion-site quotas bound]
    \label{lem:first-upper-bounds-on-global-filter}
    Fix a device-epoch $x=(d,e,F)$ and an adversary that elicits
    the reports $\cR_{\mathsf{adv}}$.  By \Def\ref{def:charged-reports},
    these charge $N^{\mathsf{adv}}$ distinct \convQuotaVarName filters and
    consume $\globalFilterEps^{\mathsf{adv}}$ from
    $\cF^{\globalFilterVarName}_{d,e}$.

    Then, we have:
    \begin{align}
        \globalFilterEps^{\mathsf{adv}} \le N^{\mathsf{adv}}\convQuotaEps.
    \end{align}
\end{lemma}

\begin{proof}
    By definition of $\globalFilterEps^{\mathsf{adv}}$ we have:

    \begin{align*}
        \globalFilterEps^{\mathsf{adv}} & = \sum_{s\in [s_{\max}]}\sum_{r \in \cR_s : \pass_{r,s} = 1} \epsilon_{r,s} \\
                                        & = \sum_{s\in [s_{\max}]}\sum_{r \in \cR_s} \epsilon_{r,s} \pass_{r,s}
    \end{align*}

    Then we use the definition of $C_{\mathsf{adv}}$ to sum over all reports $R_s$ instead of $\cR_s \subset R_s$:

    \begin{align*}
        \globalFilterEps^{\mathsf{adv}} & \le \sum_{c\in C_{\mathsf{adv}}}\ \
        \sum_{s\in[s_{\max}]}\sum_{\substack{r\in R_s                                   \\ c_r=c}}
        \epsilon_{r,s}\pass_{r,s}                                                       \\
                                        & \le \sum_{c\in C_{\mathsf{adv}}}\convQuotaEps
    \end{align*}

    where the last step comes from \Lem\ref{lemma:2pc_guarantees}(a): each $r$ grants
    $\epsilon_{r,s}\pass_{r,s}$ to $\cF^{\globalFilterVarName}_{d,e}$ and the
    same amount to $\cF^{\convQuotaVarName[c_r]}_{d,e}$, followed by the capacity invariant \Eq\ref{eq:capacity-invariant}.

    The definition of $N^{\mathsf{adv}}$ finally gives  $\globalFilterEps^{\mathsf{adv}} \le N^{\mathsf{adv}}\convQuotaEps$.
\end{proof}

\begin{lemma}[Impression-site quota allocation consistency]
    \label{lem:imp_quota_allocation_consistency}
    Fix a device-epoch $x=(d,e,F)$, a query $s$ and a report $r \in R_s$, and take $\tilde \Delta_{x,i}(\rho_r)$ as in \Eq\ref{eqn:single-epoch-ind-sensitivity} with $2\Delta(\rho_r)$ as an upper bound for $\Delta_i(\rho_r)$ (\Lem\ref{lem:relationship-des-and-de-sensitivities}).
    We have:
    \begin{align}
        \epsilon_{r,s} \le \sum_{i \in \mathbf i_{r,e}} \epsilon^i_{r,s}
    \end{align}
\end{lemma}

\begin{proof}
    Write $\lambda$ for the noise scale of query $s$, so that $\epsilon_{r,s} = \tilde \Delta_x(\rho_r)/\lambda$ and $\epsilon^i_{r,s} = \tilde \Delta_{x,i}(\rho_r)/\lambda$ (\Def\ref{def:epoch-budget}, \ref{defn:EpochImpSiteBudget}). We go through the different cases for the upper bound on $\Delta_x(\rho_r)$:
    \begin{itemize}
        \item If $d \neq d_r$, $e \notin E_r$, or $F \cap F_{A_r} = \emptyset$, then $\epsilon_{r,s} = 0$. In that case, by \Thm\ref{thm:individual-sensitivity-of-reports} we also have $\epsilon^i_{r,s} = 0$ for all $i$. This covers in particular $\mathbf i_{r,e} = \emptyset$, since $\mathbf i_{r,e} = \mathsf{ImpSites}(F \cap F_{A_r})$.
        \item Otherwise $\mathbf i_{r,e} \neq \emptyset$, and we fix any $i \in \mathbf i_{r,e}$:
              \begin{itemize}
                  \item If $d = d_r$ and $E_r = \{e\}$:
                        \begin{itemize}
                            \item If $\mathbf i_{r,e} = \{i\}$, we have $\epsilon_{r,s} = \epsilon^i_{r,s} = \|A_r(F) - A_r(\emptyset)\|_1/\lambda$.
                            \item Else, we have $\epsilon^i_{r,s} = 2\Delta(\rho_r)/\lambda \ge \|A_r(F) - A_r(\emptyset)\|_1/\lambda = \epsilon_{r,s}$.
                        \end{itemize}
                  \item Else, we have $\epsilon^i_{r,s} = 2\Delta(\rho_r)/\lambda \ge \Delta(\rho_r)/\lambda \ge \epsilon_{r,s}$.
              \end{itemize}
              In each case $\epsilon_{r,s} \le \epsilon^i_{r,s}$, and the remaining terms of the sum are non-negative.
    \end{itemize}
\end{proof}

Note that the upper bound in \Lem\ref{lem:imp_quota_allocation_consistency} can be quite loose.
Since the \qimp has no privacy meaning and is only used through \Lem\ref{lem:imp_quota_allocation_consistency} to obtain isolation guarantees, instead of using per-epoch-site privacy loss we could use any heuristic that also satisfies \Lem\ref{lem:imp_quota_allocation_consistency}'s $\epsilon_{r,s} \le \sum_{i \in \mathbf i_{r,e}} \epsilon^i_{r,s}$. For instance, we could divide $\epsilon_{r,s}$ uniformly across $\mathbf i_{r,e}$.

\begin{lemma}[Impression-site quotas bound]
    \label{lem:second-upper-bounds-on-global-filter}
    Fix a device-epoch $x=(d,e,F)$ and an adversary that elicits
    the reports $\cR_{\mathsf{adv}}$.  By \Def\ref{def:charged-reports},
    these charge $M^{\mathsf{adv}}$ distinct \impQuotaVarName filters and
    consume $\globalFilterEps^{\mathsf{adv}}$ from
    $\cF^{\globalFilterVarName}_{d,e}$.

    Then, we have:
    \begin{align}
        \globalFilterEps^{\mathsf{adv}} \le M^{\mathsf{adv}}\impQuotaEps.
    \end{align}
\end{lemma}

\begin{proof}
    By definition of $\globalFilterEps^{\mathsf{adv}}$, and then by \Lem\ref{lem:imp_quota_allocation_consistency} applied to each report, we have:

    \begin{align*}
        \globalFilterEps^{\mathsf{adv}} & = \sum_{s\in [s_{\max}]}\sum_{r \in \cR_s} \epsilon_{r,s} \pass_{r,s}                                 \\
                                        & \le \sum_{s\in [s_{\max}]}\sum_{r \in \cR_s} \sum_{i\in \mathbf i_{r,e}} \epsilon^i_{r,s} \pass_{r,s}
    \end{align*}

    Then we use the definition of $I_{\mathsf{adv}}$, which contains $\mathbf i_{r,e}$ for every report with $\pass_{r,s}=1$, to sum over all reports $R_s$ instead of $\cR_s \subset R_s$:

    \begin{align*}
        \globalFilterEps^{\mathsf{adv}} & \le \sum_{i\in I_{\mathsf{adv}}}\ \
        \sum_{s\in[s_{\max}]}\sum_{\substack{r\in R_s                                  \\ i\in \mathbf i_{r,e}}}
        \epsilon^i_{r,s}\pass_{r,s}                                                    \\
                                        & \le \sum_{i\in I_{\mathsf{adv}}}\impQuotaEps
    \end{align*}

    where the last step comes from \Lem\ref{lemma:2pc_guarantees}(b): each $r$ grants
    $\epsilon^i_{r,s}\pass_{r,s}$ to $\cF^{\impQuotaVarName[i]}_{d,e}$ for each
    $i\in \mathbf i_{r,e}$, followed by the capacity invariant \Eq\ref{eq:capacity-invariant}.

    The definition of $M^{\mathsf{adv}}$ finally gives  $\globalFilterEps^{\mathsf{adv}} \le M^{\mathsf{adv}}\impQuotaEps$.
\end{proof}

We can now combine \Lem\ref{lem:first-upper-bounds-on-global-filter} and \ref{lem:second-upper-bounds-on-global-filter} to obtain our main isolation theorem:

\begin{numberedthm}[\ref{thm:resilience-three-dimensions}(i) (Resilience to DoS depletion)]
    Fix a device-epoch $x=(d,e,F)$ and an adversary that elicits the reports
    $\cR_{\mathsf{adv}}$.  By \Def\ref{def:charged-reports}, these charge
    $M^{\textrm{adv}}$ distinct \impQuotaVarName filters and
    $N^{\textrm{adv}}$ distinct \convQuotaVarName filters.

    Then, the adversary's consumption $\globalFilterEps^{\textrm{adv}}$ from
    $\cF^{\globalFilterVarName}_{d,e}$ satisfies:
    \begin{align}
        \globalFilterEps^{\textrm{adv}} \leq \min(M^{\textrm{adv}} \impQuotaEps, \ N^{\textrm{adv}} \convQuotaEps).
    \end{align}
\end{numberedthm}

\begin{proof}
    The conversion-side bound is \Lem\ref{lem:first-upper-bounds-on-global-filter}
    and the impression-side bound is
    \Lem\ref{lem:second-upper-bounds-on-global-filter}, and we take their minimum.
\end{proof}

\subsubsection{Number of user actions}
\label{appendix:user-action-quota-bounds}

We now bound the number of sites by the user actions that resulted in quota deductions.

\begin{lemma}[Quotas per user action]
    \label{lem:quotas-per-user-action-restated}

    Fix a device-epoch $x=(d,e,F)$ and an adversary that elicits
    the reports $\cR_{\mathsf{adv}}$. Separate the actions that carry their
    conversions from those that saved the impressions they read at $x$:
    \begin{align*}
        \cU^{\mathsf{adv}}_{\mathsf{conv}}
         & := \{u_r:(s,r)\in\cR^+_{\mathsf{adv}}\},  \\
        \cU^{\mathsf{adv}}_{\mathsf{imp}}
         & := \bigcup_{(s,r)\in\cR^+_{\mathsf{adv}}}
        \mathsf{ImpActions}(F\cap F_{A_r}).
    \end{align*}

    Write $U^{\mathsf{adv}}_{\mathsf{conv}} := |\cU^{\mathsf{adv}}_{\mathsf{conv}}|$,
    $U^{\mathsf{adv}}_{\mathsf{imp}} := |\cU^{\mathsf{adv}}_{\mathsf{imp}}|$, and $U^{\mathsf{adv}} := |\cU^{\mathsf{adv}}_{\mathsf{conv}} \cup \cU^{\mathsf{adv}}_{\mathsf{imp}}|$.
    Then,
    \begin{align}
        N^{\mathsf{adv}}
         & \le \mathquotacountVar U^{\mathsf{adv}}_{\mathsf{conv}},
        \label{eq:conv-quota-per-action}                            \\
        M^{\mathsf{adv}}
         & \le \mathquotacountVar U^{\mathsf{adv}}_{\mathsf{imp}},
        \label{eq:imp-quota-per-action}                             \\
        N^{\mathsf{adv}},M^{\mathsf{adv}}
         & \le \mathquotacountVar U^{\mathsf{adv}}.
        \label{eq:combined-quota-bound}
    \end{align}
\end{lemma}
\begin{proof}
    For any $c\in C_{\mathsf{adv}}$, some $(s,r)\in\cR^+_{\mathsf{adv}}$
    has $c_r=c$, so  the \code{GenerateReport} path for that
    \code{measureConversion} request passed
    $\cF_{d,e}.\cK[u_r].\operatorname{canAdmit}(c)$. Hence  $c\in\cK_{d,e}[u_r]$ with
    $u_r\in\cU^{\mathsf{adv}}_{\mathsf{conv}}$.

    Similarly, for any  $i\in I_{\mathsf{adv}}$, there is $r$ such that
    $i\in\mathbf i_{r,e}=\mathsf{ImpSites}(F\cap F_{A_r})$.
    Hence $F\cap F_{A_r}$ contains an impression $(u,i,\mathrm{imp})$ with
    $u\in\mathsf{ImpActions}(F\cap F_{A_r})
        \subseteq\cU^{\mathsf{adv}}_{\mathsf{imp}}$.  That
    event was stored by an accepted \code{SaveImpression}, which passed
    $\cF_{d,e}.\cK[u].\operatorname{canAdmit}(i)$, so $i\in\cK_{d,e}[u]$.

    Finally, each cap contains at most $\mathquotacountVar$ domains (\Def\ref{def:domain-cap}). Hence the first bound (\Eq\ref{eq:conv-quota-per-action}) comes from $C_{\mathsf{adv}}$ with the caps indexed by $\cU^{\mathsf{adv}}_{\mathsf{conv}}$, and the second bound (\Eq\ref{eq:imp-quota-per-action}) comes from $I_{\mathsf{adv}}$ with the caps indexed by $\cU^{\mathsf{adv}}_{\mathsf{imp}}$.

    The third bound (\Eq\ref{eq:combined-quota-bound}) takes the union.
\end{proof}

We now prove the main graceful degradation theorem, which bounds the adversary's budget consumption as a function of the number of user actions they can elicit.

\begin{numberedthm}[\ref{thm:resilience-three-dimensions}(ii) (Graceful degradation)]
    Consider an attacker collecting $U^{\mathsf{adv}}$ user actions across their domains for device $d$. Under the configuration of Table~\ref{tab:filter-configs}, the budget $\globalFilterEps^{\mathsf{adv}}$ that this attacker can consume from the \globalFilter is upper-bounded by:
    \begin{align}
        \globalFilterEps^{\mathsf{adv}} \leq (1+r)\enc \times \text{\quotacountVar} \times U^{\mathsf{adv}}.
    \end{align}
\end{numberedthm}

\begin{proof}
    By \Lem\ref{lem:quotas-per-user-action-restated}, and using the configuration from Table~\ref{tab:filter-configs}, where $\impQuotaEps = n(1+r)\enc$ and $\convQuotaEps = (1+r)\enc$,
    \begin{align*}
        M^{\mathsf{adv}}\impQuotaEps \leq n(1+r)\enc \times \text{\quotacountVar} \times U^{\mathsf{adv}} \\
        N^{\mathsf{adv}}\convQuotaEps \leq (1+r)\enc \times \text{\quotacountVar} \times U^{\mathsf{adv}}.
    \end{align*}

    The limiting factor is $N^{\mathsf{adv}}\convQuotaEps$, and applying part (i) of \Thm\ref{thm:resilience-three-dimensions} yields the result.
\end{proof}

Note that under the additional assumption that no domain both saves an impression and measures a conversion under the same user action, each charged quota filter would occupy a distinct domain cap slot, giving
$M^{\mathsf{adv}}+N^{\mathsf{adv}}
    \le \mathquotacountVar U^{\mathsf{adv}}$.  Part (i) of \Thm\ref{thm:resilience-three-dimensions} would then give
$\globalFilterEps^{\mathsf{adv}}
    \le (1+r)\enc \min(nM^{\mathsf{adv}},N^{\mathsf{adv}})
    \le \frac{n}{n+1}(1+r)\enc \mathquotacountVar U^{\mathsf{adv}}$.
We do not make this assumption, and \Thm\ref{thm:resilience-three-dimensions}(ii) uses the weaker bound above.

\subsubsection{Number of devices}

\begin{lemma}[Query error from depleted device-epochs]
    \label{thm:graceful-degradation-population}
    Take a database $D$ with filter state $\cF$, and let $\tilde \cF$ be obtained from $\cF$ by depleting $\globalFilterEps$ at $k$ distinct device-epochs.
    Let $Q = \sum_{r \in R} \rho_r$ be a query in which $\Delta(\rho_r) \le \Delta$ for every $r$, and at most $m$ reports belong to any one device.
    Then

    \[
        \|Q(D; \cF) - Q(D; \tilde\cF)\|_1 \le m \cdot k \cdot \Delta .
    \]
\end{lemma}
\begin{proof}
    Take a query $Q: D \mapsto \sum_{r \in R} \rho_r(D)$ as defined in \Def\ref{def:query}.
    Take a database $D$ and filter states $\cF, \tilde \cF$ where $\tilde \cF$ is obtained from $\cF$ by depleting $\globalFilterEps$ for $k$ distinct device-epochs.
    For a report $\rho_r$, we denote by $\rho_r(D; \cF)$ the filtered report returned by GenerateReport in \Alg\ref{alg:e2e_functional_view}.
    Similarly, we denote by $Q(D; \cF) := \sum_{r \in R} \rho_r(D; \cF)$ the filtered query before adding noise.

    Consider a single report identifier $r \in R$, with attribution function $A_r$ over $n_r := |E_r|$ epochs.
    We have $\rho_r(D; \cF) = A_r(F_1, \dots, F_{n_r})$ where $F_1, \dots, F_{n_r}$ are the relevant events from $D$ after passing the filter check based on $\cF$ in \Alg\ref{alg:e2e_functional_view}.
    Now, without loss of generality, suppose that the first $k_r$ epochs for that report are depleted in $\tilde \cF$. This causes the filter check to fail for these epochs, returning:

    $$
        \rho_r(D; \tilde \cF) = A_r(\emptyset, \dots, \emptyset, F_{k_r+1}, \dots, F_{n_r}).
    $$

    For $j \in \{0, \dots, k_r\}$, write $\bfF^{(j)} := (\emptyset, \dots, \emptyset, F_{j+1}, \dots, F_{n_r})$ for the input with the first $j$ epochs zeroed out, so that $\rho_r(D; \cF) = A_r(\bfF^{(0)})$ and $\rho_r(D; \tilde \cF) = A_r(\bfF^{(k_r)})$.
    Consecutive inputs $\bfF^{(j-1)}$ and $\bfF^{(j)}$ differ in a single epoch, so by definition of global sensitivity (\Def\ref{def:device-epoch-global-sensitivity}) each step is bounded by $\Delta(\rho_r) \le \Delta$.
    After telescoping over the $k_r$ depleted epochs and applying the triangle inequality we get:

    \begin{align*}
        \|\rho_r(D; \cF)  - \rho_r(D; \tilde \cF) \|_1
         & = \Big\|\sum_{j=1}^{k_r} A_r(\bfF^{(j-1)}) - A_r(\bfF^{(j)})\Big\|_1 \\
         & \le \sum_{j=1}^{k_r} \|A_r(\bfF^{(j-1)}) - A_r(\bfF^{(j)})\|_1       \\
         & \le k_r \cdot \Delta
    \end{align*}

    Putting this together, we get:

    \begin{align*}
        \|Q(D; \cF) - Q(D; \tilde\cF)\|_1 & = \Big\| \sum_{r \in R} \rho_r(D; \cF)  - \rho_r(D; \tilde \cF) \Big\|_1 \\
                                          & \le \sum_{r \in R} \|\rho_r(D; \cF)  - \rho_r(D; \tilde \cF) \|_1        \\
                                          & \le \sum_{r \in R} k_r \cdot \Delta                                      \\
                                          & \le m \cdot k \cdot \Delta
    \end{align*}

    where $\sum_{r \in R} k_r \le m \cdot k$: each report reads depleted epochs only on its own device, so each of the $k$ depleted device-epochs is counted by at most the $m$ reports belonging to that device.
\end{proof}

\begin{numberedthm}[\ref{thm:resilience-three-dimensions}(iii) (Bounded query impact)]
    Consider the setting of Lemma \ref{thm:graceful-degradation-population} with an attacker collecting $U^{\mathsf{adv}}$ user actions across their domains for device $d$, whose own reports deplete the \globalFilter at each of the $k$ device-epochs. Under the configuration of Table~\ref{tab:filter-configs}, the error induced by the attacker is bounded by:

    \[
        \|Q(D; \cF) - Q(D; \tilde\cF)\|_1 \le m \cdot k \cdot \Delta ,
    \]
    where
    \[
        k \leq U^{\mathsf{adv}} / \lceil \frac{\ec}{n(1+r)\enc \cdot \text{\quotacountVar}} \rceil.
    \]
\end{numberedthm}

\begin{proof}
    Lemma \ref{thm:graceful-degradation-population} covers the first statement.

    Denote by $X_{\mathsf{dep}}$ the $k$ depleted device-epochs, and take $x \in X_{\mathsf{dep}}$, writing $M_x^{\mathsf{adv}}$ and $U^{\mathsf{adv}}_{\mathsf{imp}}(x)$ for the corresponding quantities at $x$.
    The attacker's own reports consume the whole capacity of $\cF^{\globalFilterVarName}_x$, so $\globalFilterEps^{\mathsf{adv}} = \ec$ at $x$. By \Lem\ref{lem:second-upper-bounds-on-global-filter} and \Eq\ref{eq:imp-quota-per-action}, and using the configuration from Table~\ref{tab:filter-configs} where $\impQuotaEps = n(1+r)\enc$:
    \begin{align*}
        \ec \le M_x^{\mathsf{adv}}\impQuotaEps \le n(1+r)\enc \times \text{\quotacountVar} \times U^{\mathsf{adv}}_{\mathsf{imp}}(x).
    \end{align*}

    Rearranging gives $U^{\mathsf{adv}}_{\mathsf{imp}}(x) \ge \frac{\ec}{n(1+r)\enc \cdot \text{\quotacountVar}}$, and $U^{\mathsf{adv}}_{\mathsf{imp}}(x)$ is an integer, so we can round the right-hand side up. By \Def\ref{def:user-action-context}, all the impressions of one action on $d$ belong to a single device-epoch. The device-epochs of $X_{\mathsf{dep}}$ therefore draw disjoint sets of actions from the $U^{\mathsf{adv}}$ actions collected on $d$:
    \begin{align*}
        U^{\mathsf{adv}} \ge \sum_{x \in X_{\mathsf{dep}}} U^{\mathsf{adv}}_{\mathsf{imp}}(x) \ge k \left\lceil \frac{\ec}{n(1+r)\enc \cdot \text{\quotacountVar}} \right\rceil,
    \end{align*}
    which gives the bound on $k$.
\end{proof}

\subsection{Simplified model (for educational purposes)}
\label{sec:simplified_model}

\begin{algorithm}[t]
    \caption{Simplified version of \sysname formal model. }
    \label{alg:big_bird_model_simplified}
    \begin{algorithmic}[1] %
        \Input
        \State Challenge bit $\chal \in \{0,1\}$
        \State In-or-out record $x_0$ to be inserted at time $t_0$
        \State Public information $\cC$
        \State Data generation process $\cG$
        \State Queriers $\cQ^1, \dots, \cQ^K$
        \State IDP budgets for each $x \in \cX$: $\cF_x^{\globalFilterVarName}$, $(\cF_x^{\persiteFilterVarName}[k])_{k \in [K]}$, $(\cF_x^{\convQuotaVarName}[c])_{c \in [\cS]}$, $(\cF_x^{\impQuotaVarName}[i])_{i \in [\cS]}$ \label{line:idp_budgets}
        \EndInput

        \For{$t \in [t_{\max}]$}
        \Statex \ \ \graycomment{Data generation model:}
        \State $D_t \gets \cG(V^1_{<t}, \dots V^K_{<t})$ \ \graycomment{\footnotesize new data depends on {\bf all} past results}
        \If{$t = t_0$}
        \Statex \ \ \ \ \graycomment{\footnotesize On challenge, insert in-or-out record...}
        \If{$\chal = 0$} \label{line:opt-out-simplified}
        \State $D_t \gets D_t + x_0 \cap \cC$  \graycomment{\footnotesize ...with public events only}
        \Else
        \State $D_t \gets D_t + x_0$ \graycomment{\footnotesize ...with public and private events}
        \EndIf
        \EndIf

        \Statex \ \ \graycomment{Query execution model:}
        \For{$k \in [K]$}
        \State $\cM_t^k \gets \cQ^k(V^k_{<t})$ \graycomment{ \footnotesize Abstract query over whole database}
        \State $S_t^k \gets D$
        \For{$x \in D$}
        \Statex \ \ \ \ \ \graycomment{\footnotesize Compute device-epoch loss and device-epoch-impSite loss (for \impQuotaVarName)}
        \State $\epsilon_x \gets \operatorname{EpochBudget}(x, q_t^k)$
        \For{$i \in \mathbf{i}$}
        \State $\epsilon_x^i \gets \operatorname{EpochImpSiteBudget}(x,i,q_t^k)$ \label{line:two_granularities}
        \EndFor

        \Statex \ \ \ \ \ \graycomment{\footnotesize Atomic transaction across all budgets}
        \If{$(\cF_x^{\globalFilterVarName} \circ \cF_x^{\persiteFilterVarName}[k] \circ \cF_x^{\convQuotaVarName}[c] \circ \cF_x^{\impQuotaVarName}[\mathbf{i}])(\epsilon_x, \epsilon_x^\mathbf{i})$} \label{line:transaction}
        \State $S_t^k \gets S_t^k \setminus\{x\}$ \graycomment{\footnotesize Drop out-of-budget record}
        \EndIf
        \EndFor
        \State $V^k_{t} \gets \cM_t^k(S_t^k)$
        \EndFor
        \EndFor
        \State \Return $V^1, \dots, V^K$

    \end{algorithmic}
\end{algorithm}

\Alg\ref{alg:big_bird_model_simplified} shows a simplified version of the full \sysname system model that we analyze formally, but one in line with the generic model from \S\ref{sec:contribution1:formal-model}. The purpose of including this simplified version is educational: it provides easy-to-read understanding of how one can represent key aspects of system behavior in modeling.

First, the budgets' IDP semantics are visible in the $x \in \cX$ subscript at \Line\ref{line:idp_budgets}, where $x$ corresponds to a device-epoch. This indicates that the filters are each maintained on each device and for each epoch.
Second, privacy loss is computed at a finer granularity (\Line\ref{line:two_granularities}) for \impQuotaVar than for the other budgets.
Third, \sysname uses a shared limit ($\cL$ in \S\ref{sec:contribution1}) consisting of a collection of privacy budgets and quota budgets, denoted $\cF$.
At \Line\ref{line:transaction}, $(\cF_x^{\globalFilterVarName} \circ \dots \circ \cF_x^{\impQuotaVarName}[i])(\epsilon_x, \epsilon_x^i)$ denotes the atomic check and consumption of budgets performed for each record (device-epoch). For comparison, $\cF_x^{\globalFilterVarName}(\epsilon_x) \wedge \dots \wedge \cF_x^{\impQuotaVarName}[i](\epsilon_x^i)$ would check budgets independently, potentially spending from one but not the other, which harms utility. We show that this atomicity, never previously studied, preserves global IDP.

\section{Examples of Cap Interactions with Device-Epoch IDP}
\label{appendix:cap-interactions-with-idp}

\algrenewcommand\algorithmicfunction{def}
\algrenewcommand\textproc{}

\begin{algorithm}[t]
  \caption{Non-IDP capping of budget consumption per user action}
  \label{alg:cap-budget-per-user-action-faulty}
  \footnotesize
  \begin{algorithmic}[0] %
    \Function{\textbf{onUserAction}}{ }
    \State uaCtx = createUserContext()
    \State consumedPrivacy[uaCtx] $\gets 0$
    \State \Return uaCtx
    \EndFunction
    \Statex

    \Function{\textbf{measureConversion}}{uaCtx, impSites, querier, convSite, params}
    \For{$e \in \text{params.attributionWindow}$}
    \State $I_e \gets$ MatchingFn(ImpressionStore[e], params)
    \State \graycomment{Compute IDP loss w/ Cookie Monster algo.}
    \State $\epsilon_e \gets$ computeEpochLoss($I_e$, params)

    \State \graycomment{Ensure that total privacy consumed in this user-action context across}
    \State \graycomment{all epochs stays within the cap. If it exceeds, stop execution and return}
    \State \graycomment{a null report.}
    \If{$\text{consumedPrivacy[uaCtx]} + \epsilon_e > \epsilon_{action}$}
    \State \Return AttributionFn$(\emptyset, \dots, \emptyset)$  \ \ \graycomment{cap exceeded, return null report}
    \EndIf
    \State $\text{consumedPrivacy[uaCtx]}\ += \epsilon_e$

    \State \graycomment{Then deduct from per-epoch global budgets and per-querier budgets.}
    \State tx = beginTransaction()
    \If{querierBudget[e][querier].tryConsume($\epsilon_e$, tx) \textbf{and }
      \newline\hspace*{1.5em} globalBudget[e].tryConsume($\epsilon_e$, tx)}
    \State tx.commit()
    \Else
    \State tx.abort()
    \State $I_e \gets \emptyset$
    \EndIf
    \EndFor  %
    \State \graycomment{Generate attribution histogram from remaining impressions} \\
    \Return AttributionFn$(I_{e_1}, \dots, I_{e_w})$  \ \ \graycomment{$e_i \in \text{params.attributionWindow}$}
    \EndFunction
  \end{algorithmic}
\end{algorithm}

\begin{algorithm}[t]
  \caption{\footnotesize \bf Correct, IDP-respecting capping of budget consumption per user action}
  \label{alg:cap-budget-per-user-action-correct}
  \footnotesize
  \begin{algorithmic}[0] %
    \Function{\textbf{onUserAction}}{ }
    \State uaCtx = createUserContext()
    \State consumedPrivacy[uaCtx][$e$] $\gets 0$ for each epoch $e$
    \State \Return uaCtx
    \EndFunction
    \Statex

    \Function{\textbf{measureConversion}}{uaCtx, impSites, querier, convSite, params}
    \For{$e \in \text{params.attributionWindow}$}
    \State $I_e \gets$ MatchingFn(ImpressionStore[e], params)
    \State \graycomment{Compute IDP loss w/ Cookie Monster algo.}
    \State $\epsilon_e \gets$ computeEpochLoss($I_e$, params)

    \State \graycomment{Ensure that privacy consumed per epoch in this user-action context}
    \State \graycomment{stays within the cap. If it exceeds, eliminate the epoch's data from}
    \State \graycomment{consideration for this report.}
    \If{$\text{consumedPrivacy[uaCtx][e]} + \epsilon_e > \epsilon_{action}$}
    \State $I_e \gets \emptyset$  \ \ \graycomment{nullify contribution of this epoch to the report}
    \EndIf
    \State $\text{consumedPrivacy[uaCtx][e]}\ += \epsilon_e$

    \State \graycomment{Then deduct from per-epoch global budgets and per-querier budgets.}
    \State tx = beginTransaction()
    \If{querierBudget[e][querier].tryConsume($\epsilon_e$, tx) \textbf{and }
      \newline\hspace*{1.5em} globalBudget[e].tryConsume($\epsilon_e$, tx)}
    \State tx.commit()
    \Else
    \State tx.abort()
    \State $I_e \gets \emptyset$
    \EndIf
    \EndFor  %
    \State \graycomment{Generate attribution histogram from remaining impressions} \\
    \Return AttributionFn$(I_{e_1}, \dots, I_{e_w})$  \ \ \graycomment{$e_i \in \text{params.attributionWindow}$}
    \EndFunction
    \Statex

    \Function{\textbf{onEpochChange}}{uaCtx}
    \State destroyUserContext(uaCtx)
    \EndFunction
  \end{algorithmic}
\end{algorithm}

To illustrate how easy it is for seemingly reasonable DoS caps to violate device-epoch IDP, we present a natural but incorrect instantiation of ``capping global-budget consumption per user action,'' followed by its corrected form.

\paragraph{Faulty instantiation.}
Alg.~\ref{alg:cap-budget-per-user-action-faulty} tries to cap the total privacy consumption generated by all reports within a single user-action context. A natural approach would be to cap the \emph{cumulative} privacy loss \emph{across all epochs} accessed by reports following that user action. The algorithm does exactly this by maintaining a single accumulator, \texttt{consumedPrivacy[uaCtx]}, that aggregates privacy loss over every epoch touched by any report in that \texttt{uaCtx}. If, during the execution of a report, this cumulative loss would exceed the cap $\epsilon_{\text{action}}$, the algorithm halts and returns a null report, AttributionFn$(\emptyset,\dots,\emptyset)$.

While intuitive, this design inadvertently couples epochs: whether data from epoch $e$ is included in the report now depends on privacy consumption in \emph{other} epochs accessed earlier in the same \texttt{uaCtx}. This violates device-epoch IDP, which requires that privacy loss in epoch $e$ depend only on that epoch's data. In effect, a querier's interactions in epoch $e_1$ can suppress contributions from epoch $e_2$, which means that privacy loss accounted for $e_2$ now depends on data in $e_1$, violating IDP.

\paragraph{Correct instantiation.}
Alg.~\ref{alg:cap-budget-per-user-action-correct} restores IDP soundness for this candidate approach by isolating caps \emph{per epoch}. Each user-action context maintains a separate tracker for every epoch:
\[
  \texttt{consumedPrivacy[uaCtx][e]}.
\]
When epoch $e$'s loss in this uaCtx exceeds $\epsilon_{\text{action}}$, only that epoch's contribution is nullified; all other epochs proceed independently. This preserves the required per-epoch separability: the decision to include or exclude epoch $e$ depends solely on that epoch's data and budgets, not on privacy loss incurred elsewhere. This same principle underlies \sysname's \quotacountVarName design (Alg.~\ref{alg:big-bird}), where checks are carefully partitioned per epoch and user-action contexts never span epoch boundaries. Together, these mechanisms ensure that resilience mechanisms do not inadvertently break device-epoch IDP.

\ifdoubleblind
\else  %
  \section{Cross-report Privacy Loss Optimization}
\label{sec:cross-report-optimization-adaptive}
\label{appendix:cross-report-optimization}

In \S\ref{appendix:online-filter-management}, we presented a general algorithm where queriers pay for each report separately.
It is common for conversion sites to embed multiple intermediaries and to perform attribution once across impressions displayed by all intermediaries. When multiple reports for the same conversion involve disjoint impression sets (\eg across intermediaries), we prove that one need not double-count against shared budgets (global, conversion-site, and impression-site quotas).
Indeed, the combination of reports from different intermediaries reveals no more than a single attribution across intermediaries.
We present the intuition first with an example and then with detailed formalism and proofs.

\subsection{Example}
\label{appendix:cross-report-optimization:example}

We modify the example in \F\ref{fig:big-bird-architecture} and \S\ref{sec:big-bird-overview} as follows. Instead of working with a single intermediary \code{adtech.ex}, \code{shoes.ex} works with both \code{adtech.ex} and \code{adtech2.ex}.
\code{adtech.ex} displays impressions on \code{news.ex}, and \code{adtech2.ex} on \code{blog.ex}.
Upon conversion, attribution is still computed across all relevant impressions, spanning both intermediaries.
For \code{shoes.ex}'s own measurement queries, accounting is unchanged since intermediaries operate on \code{shoes.ex}'s behalf and are not queriers themselves.
As before, \code{adtech.ex} can request an additional report (j) measuring its own impressions for placement optimization purposes. Here, \code{adtech.ex} receives a report containing a histogram with attributed value 30 (for the impression on \code{news.ex}).
Additionally, \code{adtech2.ex} can measure its own impressions and receive a report (k) containing a histogram with attributed value 30.
Naively, one would expect a cumulative deduction of 0.9 from the global budget (three reports).
Yet, one can observe that the two intermediary reports are shards of a single report with attributed value 60, because \code{adtech.ex} and \code{adtech2.ex} receive disjoint bins of the histogram.
Leveraging this observation, we show that it is possible to deduct only 0.6 from the global budget. The same optimization applies to shared quotas.

\subsection{Problem Statement}

The example in \S\ref{appendix:cross-report-optimization:example} admits a
cross-report optimization, which we formalize here. We define variations of
\Alg\ref{alg:e2e_setup} and \Alg\ref{alg:e2e_functional_view} in
\S\ref{sec:cross-report-alg}.
We focus on histogram reports, defined in \S\ref{sec:histogram-reports}, and show that paying only once for a sequence of properly correlated histogram reports still satisfies global IDP guarantees in \S\ref{sec:cross-report-proofs}.
We leave the generalization to other queries for future work.

\subsection{Algorithm}
\label{sec:cross-report-alg}

\heading{Overview.}
\Alg\ref{alg:cross-report-setup} updates the formalism from \Alg\ref{alg:e2e_setup}.
Instead of generating a report immediately upon conversion, the first querier to request a report attached to a report identifier $r$ calls MeasureConversion to create a stateful attribution object $\alpha_r$, which pays upfront for any valid sequence of histogram reports over non-overlapping impressions.

MeasureConversion performs attribution using the pre-specified attribution function $a$ and the noise scale $\lambda$, which give an upper bound on the total leakage of any sequence of future reports applying $a$ on disjoint sets of impressions. All shared privacy loss is charged upfront and atomically at MeasureConversion time, before any querier-specific interaction occurs.
We run a modified two-phase commit protocol to deduct budget from all shared filters (global and quota filters), but not from querier-local filters, which are charged separately.
\footnote{The querier budget is left out of the 2PC because we don't know ahead of time which queriers will request a report, and we don't want to block some queriers if other queriers are out of budget.}
Finally, $\alpha_r$ stores the attribution function $a_\mathbf{F}$ with corresponding impressions $\bfF$, the privacy parameters $\lambda$, and the support $S$ of impressions requested so far.

GetReport allows a querier to receive a report from $r$, once the attribution object $\alpha_r$ has been created. $\alpha_r$ is only created once, and is reused every time a querier requests a report from $r$.
GetReport checks that privacy parameters match what was committed in the attribution object, and that impressions are not queried twice.
If these checks pass, GetReport spends querier budget and computes the report using the predefined attribution function $a$ stored in $\alpha_r$. We update the support of impressions $S \gets S\sqcup S_\rho$ in $\alpha_r$ each time a new report $\rho$ requests impressions in $S_\rho \subset \cI$.

\heading{Subroutine.}
We define AtomicFilterCheckAndConsume2 as in \Alg\ref{alg:atomic_filter_check}, except that querier filters are not part of the atomic commit.
That is, we do not take $b$ as an input, and we delete \Line\ref{line:per-site-can-consume} and \Line\ref{line:per-site-try-consume} from \Alg\ref{alg:atomic_filter_check}.

\begin{algorithm}[t]
  \caption{Updated formalism for cross-report optimization (diff with \Alg\ref{alg:e2e_setup})}
  \label{alg:cross-report-setup}
  \begin{algorithmic}[1]

    \State \graycomment{Collect, aggregate and noise reports to answer $Q$}
    \Function{$\operatorname{AnswerQuery}$}{$D, Q, \lambda, b$}
    \State $(\rho_r^b)_{r \in R} \gets Q$ \graycomment{Get report
    identifiers from $Q$}

    \For{$r \in R$}
    \If{$\alpha_r$ is not defined}
    \State \graycomment{Initialize attribution object}
    \State $\alpha_r \gets \operatorname{MeasureConversion}(D, \rho_r^b, \lambda) $
    \EndIf
    \State $\rho_r^b(D), \alpha_r \gets \operatorname{GetReport}(\alpha_r, \rho_r^b, \lambda, b)$
    \EndFor
    \State Sample $X \sim \cL(\lambda)$
    \State \Return $\sum_{r \in R} \rho_r^b(D) + X$
    \EndFunction

  \end{algorithmic}
\end{algorithm}

\begin{algorithm}[h]
  \caption{\sysname algorithm (on-device) with cross-report optimization for histogram reports}
  \label{alg:cross-report-on-device}
  \begin{algorithmic}[1]
    \Input
    \State Filter and quota capacities $\ec$, $\enc$, $\eqimp$, $\eqconv$

    \EndInput

    \Function{$\operatorname{MeasureConversion}$}{$D, \rho, \lambda, \bfb$}
    \State Read $\rho$ to get device $d$, epoch $E$,  \convsite $c$, \impsites $\mathbf{i}$, histogram attribution function $A_{a_\rho,S_\rho, H_\rho}$ with scalar attribution function $a$, support impressions $S$ and histogram bin mapping $H$.
    \For{$e \in E$}
    \State $x \gets (d,e,D_d^e)$
    \If{$\cF_{x}$ is not defined}
    \State $\cF_{x} \gets \operatorname{InitializeFilters}(\ec,\enc,$
    \Statex \hspace{\algorithmicindent}$\eqimp,\eqconv)$
    \EndIf

    \State $\epsilon_x \gets  2a^{\max} / \lambda $
    \State $\epsilon_\mathbf{x}^{\mathbf{i}} \gets \{i : 2a^{\max} / \lambda, i \in \mathbf{i}\}$

    \If{$\operatorname{AtomicFilterCheckAndConsume2}(\cF_x, c,
    \mathbf{i}, \epsilon_x, \epsilon_\mathbf{x}^{\mathbf{i}}) = \false$}
    \State $F_e \gets \emptyset$ \graycomment{Empty the epoch if any
    filter check fails} \label{line:cross-report-zero-out}
    \EndIf
    \EndFor
    \State \Return $\alpha = (a_{\bfF}, \bfF, \lambda, \emptyset)$ \graycomment{Start with $S_\alpha=\emptyset$}
    \EndFunction

    \State
    \Function{$\operatorname{GetReport}$}{$\alpha, \rho, \lambda$}
    \State Read $\alpha$ to get attribution function $a_\alpha$, impressions $\bfF_\alpha$, noise scale $\lambda_\alpha$.
    \State Read $\rho$ to get device $d$, querier $b$,
    target epochs $E$,
    histogram attribution function $A_{a,S, H}$ with scalar attribution function $a$, support impressions $S$ and histogram bin mapping $H$.
    \If{$S \cap S_\alpha \neq \emptyset \vee \lambda \neq \lambda_\alpha \vee a_\rho \neq a_\alpha$}
    \State $\rho \gets A(\emptyset, \dots, \emptyset)$ \graycomment{Null report if inconsistent with $\alpha$}
    \State\Return $\rho$
    \EndIf
    \For{$e \in E$}
    \State $x \gets (d,e,{\bfF_\alpha}_e)$
    \State $\epsilon_x \gets
    \operatorname{EpochBudget}(x,\rho, \lambda)$ \graycomment{Querier budget only}
    \If {$\cF_x^{\persiteFilterVarName[b]}.\operatorname{tryConsume}(\epsilon^t_x) = \false$}
    \State $F_e \gets \emptyset$
    \Else
    \State $F_e \gets {\bfF_\alpha}_e$
    \EndIf
    \EndFor
    \State $\alpha \gets (a_\alpha, \bfF, \lambda, S_\alpha \sqcup S_\rho)$ \graycomment{Update impression support}
    \State $\rho \gets A_{a,S,H}((F_e)_{e \in E})$ \graycomment{Clipped
    attribution report}
    \State\Return $\rho, \alpha$
    \EndFunction
  \end{algorithmic}
\end{algorithm}

\subsection{Histogram reports definition and properties}
\label{sec:histogram-reports}

The following definitions (\Def\ref{def:scalar-attribution-function} and \ref{def:histogram-report}) are adapted from \cite[Thm. 18]{TKM+24}.
Histogram reports distribute a positive value across impressions, map each impression to a bucket, and then sum up the attributed value in each bucket. \Lem\ref{lem:histogram-sensitivity} gives the global sensitivity of such histogram reports, which is bounded by the maximum attributable value.

\begin{definition}[Scalar attribution function]
  \label{def:scalar-attribution-function}
  Fix a number of epochs $k > 0$.
  A {\em scalar attribution function} is a function $a: \mathcal{P}(\mathcal{I})^k \times \mathcal{I} \rightarrow \mathbb{R}_+$
  that attributes a positive value $a_{\mathbf{F}}(f)$ to each impression $f \in \mathcal{I}$,
  depending on all the impressions in $k$ epochs $\mathbf{F} \in \mathcal{P}(\mathcal{I})^k$.

  For a scalar attribution function $a$, we define its {\em maximum attributable value} $a^{\max}$ as follows:

  \begin{align}
    \label{eq:max-attributable-value}
    a^{\max} := \max_{\mathbf{F} \in \mathcal{P}(\mathcal{I})^k} \sum_{j = 1}^k\sum_{f \in \bfF_j} a_{\bfF}(f)
  \end{align}
\end{definition}

\begin{definition}[Histogram report]
  \label{def:histogram-report}
  Consider a scalar attribution function $a: \mathcal{P}(\mathcal{I})^k \times \mathcal{I} \rightarrow \mathbb{R}_+$,
  a support set of impressions $S \subset \cI$,
  an output dimension $m > 0$,
  and a one-hot encoding function $H$ that maps each event $f$ to one of $m$ buckets. That is, $H: \mathcal{I} \rightarrow \{0,1\}^m$ such that $\forall f \in \mathcal{I}, ||H(f)||_1  = 1$.

  First, we define $A_{a, S, H}: \mathcal{P}(\mathcal{I})^k \to \R^m$ as follows:

  \begin{align}
    A(\mathbf{F}) = \sum_{j = 1}^k\sum_{f \in \bfF_j} \mathds{1}[f \in S] a_{\mathbf{F}}(f) \cdot H(f)
  \end{align}

  $A_{a, S, H}$ is a well-defined attribution function (in the sense of \Def\ref{def:attribution-function}).

  Second, for a device $d$ and a set of epochs $E$, we define the {\em histogram report} associated with $A_{a, S, H}$ as in \Def\ref{def:attribution-report}:

  \begin{align}
    \rho: D \mapsto A_{a, S, H}(D_d^E)
  \end{align}

\end{definition}

Next, \Lem\ref{lem:histogram-sensitivity} and \ref{lem:total-histogram-contribution} give two preliminary properties of histogram reports that will be used in \Thm\ref{thm:cross-report-privacy-loss-optimization-adaptive}.

\begin{lemma}[Histogram sensitivity]
  \label{lem:histogram-sensitivity}
  Consider a histogram report $\rho$ with associated attribution function $A_{a, S, H}$.
  We have:

  \begin{align}
    \Delta(\rho) \le 2a^{\max}
  \end{align}
\end{lemma}
\begin{proof}
  Take a report $\rho$ with scalar attribution function $a$, device $d$ and epochs $E$.
  Consider two neighboring databases $D, D'$ and denote $\bfF := D_d^E$ and $\bfF' := {D'}_d^E$.
  We have:
  \begin{align}
    & \|\rho(D) - \rho(D')\|_1 = \|A_{a, S, H}(\bfF) - A_{a, S, H}(\bfF')\|_1                                                                    \\
    & = \left\| \sum_{j = 1}^k\sum_{f \in \bfF_j} a_{\bfF}(f) \cdot H(f) - \sum_{j = 1}^k\sum_{f \in \bfF'_j} a_{\bfF'}(f) \cdot H(f) \right\|_1 \\
    & \le \sum_{j = 1}^k\sum_{f \in \bfF_j} a_{\bfF}(f) \| H(f)\| + \sum_{j = 1}^k\sum_{f \in \bfF'_j} a_{\bfF'}(f) \| H(f)\|                    \\
    & \le 2a^{\max}
  \end{align}
  Even though this bound holds even for non-neighboring databases, \cite[Thm. 18]{TKM+24} provides mild conditions under which the bound is tight.
\end{proof}

\subsection{Privacy proof}
\label{sec:cross-report-proofs}

\begin{lemma}[Correlated histogram sensitivity]
  \label{lem:total-histogram-contribution}
  Fix a device-epoch $x = (d,e,F)$ and a database $D$.
  Fix a report identifier $r \in \Z$ corresponding to a histogram attribution object $\alpha_r$.
  Fix a sequence of reports $\rho_1, \dots, \rho_n$ that request a report from $r$, ordered lexicographically by time and querier $(t,b)$. In particular, MeasureConversion is called for $\rho_1$ and then reused for subsequent reports.
  $\sum_{i=1}^n \|\rho_i(D) - \rho_i(D+x)\|/\lambda_i$ represents the total contribution over reports computed from $\alpha_r$, each with its own requested noise scale $\lambda_i$.
  Denote by $\pass_1$ the output of the 2PC for $x$ in \Alg\ref{alg:cross-report-on-device}.

  We have:

  \begin{align}
    \sum_{i=1}^n & \|\rho_i(D) - \rho_i(D+x)\|/\lambda_i \nonumber \\
    & \le \pass_1  \cdot 2a^{\max}/ \lambda_1
  \end{align}

\end{lemma}

\begin{proof}

  If $\pass_1 = 0$, then $\rho_i(D) = \rho_i(D+x)$ because in both cases the data for $x$ is zeroed out ($F_e = \emptyset$ at \Line\ref{line:cross-report-zero-out}), which completes the proof in this case.

  Now, suppose that $\pass_1 =1$.
  Take a report $\rho_i$ with scalar attribution function $a_i$ and support impressions $S_i$.
  These only depend on past results $v_{<t_i,b_i}$.
  If $S_i \cap (S_1 \cup \dots \cup S_{i-1}) \neq \emptyset$, $\lambda_i \neq \lambda_1$ or $a_i \neq a_1$, then $\rho_i(D) = A(\emptyset, \dots, \emptyset) = \rho_i(D+x)$.
  Also, if $x = (d,e,F)$ is not queried by $\rho_i$ ($d \neq d_i$ or $e \not \in E_i$), then $\rho_i(D) = \rho_i(D+x)$.

  Denote by $I$ the set of remaining reports satisfying $S_i \cap (S_1 \sqcup \dots \sqcup S_{i-1}) = \emptyset$, $\lambda_i = \lambda_1$, $a_i = a_1$, $d_i = d$, and $e \in E_i$.
  We have:

  \begin{align}
    & \sum_{i =1}^n  \|\rho_i(D) - \rho_i(D+x)\|/\lambda_i                                            \\
    & = \sum_{i \in I}   \|\rho_i(D) - \rho_i(D+x)\|/\lambda_1                                        \\
    & = \sum_{i \in I} \| \sum_{j=1}^k \sum_{f \in \bfF_j} \mathds{1}[f \in S_i] a_{\bfF}(f) H_i(f) - \\
    & \sum_{j=1}^k \sum_{f \in \bfF_j'} \mathds{1}[f \in S_i] a_{\bfF'}(f) H_i(f) \| / \lambda_1
  \end{align}
  \begin{align}
    & \le \sum_{i \in I} \sum_{j=1}^k \sum_{f \in \bfF_j} \mathds{1}[f \in S_i](a_{\bfF}(f) + a_{\bfF'}(f)) / \lambda_1    \\
    & \le \sum_{j=1}^k \sum_{f \in \bfF_j} \mathds{1}[f \in \sqcup_{i \in I} S_i] (a_{\bfF}(f) + a_{\bfF'}(f)) / \lambda_1 \\
    & \le 2a^{\max} / \lambda_1
  \end{align}

  by \Def\ref{def:scalar-attribution-function} and using the fact that the $S_i$ are disjoint so each impression is counted at most once.

\end{proof}

\begin{theorem}
  \label{thm:cross-report-privacy-loss-optimization-adaptive}
  Consider $x \in \cX$ with \c capacity $\ec$.
  For simplicity, because we are considering a global guarantee, we consider a non-adaptive data generation process in \Alg\ref{alg:cross-report-setup}, which defines a mechanism $\cM$.
  Then, $\cM$ satisfies individual device-epoch $\ec$-DP for $x$ under public information $\cC$.
\end{theorem}

\begin{proof}
  Denote by $x_{\cC} = (d,e,F\cap \cC)$ the device-epoch obtained by keeping only public events $\cC$ from $x$, where public events are the set of all conversions.

  Take $v \in \operatorname{Range}(\cM)$.
  As in \Thm\ref{thm:privacy-guarantee}, we want to show that:
  \begin{align}
    \left| \ln \left( \frac{\Pr[\cM(D + x_{\cC}) = v]}{\Pr[\cM(D + x) = v]} \right) \right| \leq \ec.
  \end{align}

  Using Bayes' rule and $\rho(D+x_\cC) = \rho(D)$, as in \Thm\ref{thm:privacy-guarantee}, we have:

  \begin{align}
    &
    \left| \ln \left( \frac{\Pr[\cM(D + x_{\cC}) = v]}{\Pr[\cM(D + x) = v]} \right) \right|                                                                                                                                                      \\
    =   &
    \left| \ln \left( \frac{\prod_{t=1}^{t_{\max}} \prod_{b \in S} \Pr[\sum_{r \in R_t^b} \rho_{r,t}^b(D) + X_t^b = v_t^b]}{\prod_{t=1}^{t_{\max}}  \prod_{b \in S} \Pr[\sum_{r \in R_t^b} \rho_{r,t}^b(D + x) + X_t^b = v_t^b]} \right) \right| \\
    \le & \sum_{t=1}^{t_{\max}} \sum_{b \in S} \left|\ln \left(\frac{\Pr[\sum_{r \in R_t^b} \rho_{r,t}^b(D) + X_t^b = v_t^b]}{\Pr[\sum_{r \in R_t^b} \rho_{r,t}^b(D + x) + X_t^b = v_t^b]} \right) \right| \label{eq:loss-sum-bayes}
  \end{align}

  where each query $Q_t^b = \{\rho_{r,t}^b, r \in R_t^b\}$ is chosen adaptively, potentially based on previous results $v_1^b, \dots, v_{t-1}^b$.
Since the filters and attribution functions are identical for $D$ and $D+x$ when we condition on past results, we write $\rho_{r,t}^b(D)$ for simplicity instead of $\rho_{r,t}^b(D ; \cF_{v_{<t}},\alpha_{v_{<t}}))$.

Fix $t \in [t_{\max}]$ and $b \in \cS$.
Without the optimization, as in \Eq\ref{ineq:sensitivity-guarantee-on-global-filter}, the device would pay for each report sent to any querier, which would give:
\begin{align}
  & \left|\ln \left(\frac{\Pr[\sum_{r \in R_t^b} \rho_{r,t}^b(D) + X_t^b = v_t^b]}{\Pr[\sum_{r \in R_t^b} \rho_{r,t}^b(D + x) + X_t^b = v_t^b]} \right) \right| \\
  & \leq \sum_{r \in R_t^b} \Delta_x (\rho_{r,t}^b)\pass_r^b/\lambda_t^b
\end{align}

Instead of upper-bounding the difference $\|\rho_{r,t}^b(D) - \rho_{r,t}^b(D+x)\|$ for each report by $\Delta_x(\rho_r)$ separately, which takes a maximum over all $D$ right away, we keep information about $D$ a bit longer.
This will allow us to leverage the fact that reports $\rho_{r,t}^b$ across different time steps and queriers tied to the same identifier $r$ are correlated:

\begin{align}
  & \left|\ln \left(\frac{\Pr[\sum_{r \in R_t^b} \rho_{r,t}^b(D) + X_t^b = v_t^b]}{\Pr[\sum_{r \in R_t^b} \rho_{r,t}^b(D + x) + X_t^b = v_t^b]} \right) \right| \\
  & \leq \sum_{r \in R_t^b} \|\rho_{r}^b(D) - \rho_{r}^b(D+x)\|/\lambda_t^b \label{eq:report-local-sensitivity}
\end{align}

For a report identifier $r \in \Z$, we now define $\cT_r$, which keeps track of all queriers that requested a report from $r$ and the time steps at which they requested it:

\begin{align}
  \cT_r := \{(t,b) \in [t_{\max}] \times \cS : r \in R_t^b \}
\end{align}

This notation allows us to swap the sums, after putting \Eq\ref{eq:report-local-sensitivity} into \Eq\ref{eq:loss-sum-bayes}:
\begin{align}
  &
  \left| \ln \left( \frac{\Pr[\cM(D + x_{\cC}) = v]}{\Pr[\cM(D + x) = v]} \right) \right|                              \\
  & \le \sum_{t=1}^{t_{\max}} \sum_{b \in S} \sum_{r \in R_t^b} \|\rho_{r}^b(D) - \rho_{r}^b(D+x)\|/\lambda_t^b       \\
  & = \sum_{r \in \Z} \sum_{(t, b) \in \cT_r} \|\rho_{r}^b(D) - \rho_{r}^b(D+x)\|/\lambda_t^b \label{eq:swapped-sums}
\end{align}

Fix a report identifier $r \in \Z$ corresponding to a histogram attribution object $\alpha_r$.
Denote by $(t_0, b_0)$ the first time-step/querier pair that requests a report from $r$, thereby calling MeasureConversion.
By \Lem\ref{lem:total-histogram-contribution}, we have:

\begin{align}
  \sum_{(t, b) \in \cT_r} \|\rho_{r}^b(D) - \rho_{r}^b(D+x)\|/\lambda_t^b \le \pass_{r, t_0}^{b_0}  \cdot \Delta_x(\alpha_r) / \lambda_{t_0}^{b_0}
\end{align}

Finally, since $\pass_{r, t_0}^{b_0}$ implies that $\epsilon_{r,t_0}^{b_0} = \Delta_x(\alpha_r) / \lambda_{t_0}^{b_0}$ passes the \c, by the definition of \c, \Eq\ref{eq:swapped-sums} becomes:

\begin{align}
  &
  \left| \ln \left( \frac{\Pr[\cM(D + x_{\cC}) = v]}{\Pr[\cM(D + x) = v]} \right) \right| \\
  & \le \sum_{r\in\Z} \pass_{r, t_0}^{b_0}  \epsilon_{r,t_0}^{b_0}                       \\
  & \le \ec
\end{align}
which concludes the proof.
\end{proof}

\heading{Remark.}
\Lem\ref{lem:histogram-sensitivity} and \ref{lem:total-histogram-contribution} show that for histogram reports with $k>1$ epochs, \Alg\ref{alg:cross-report-on-device} spends up to $|\bfb|$ times less budget than \Alg\ref{alg:e2e_functional_view} when $|\bfb|$ queriers request reports from the same conversion with a single report identifier $r$.
This is because the privacy loss in these cases is proportional to $\Delta(\rho) = 2a^{\max}$.
For histogram reports with a single epoch, we can use the individual sensitivity, which renders this optimization unnecessary.

\fi  %

\end{document}